\documentclass[twocolumn]{aastex631}

\usepackage{amsmath}
\usepackage{amssymb}
\usepackage[varg]{txfonts}
\usepackage{graphicx}
\usepackage{multirow}
\usepackage{subfigure}
\usepackage{hyperref}

\begin{document}

\title{The lively accretion disc in NGC 2992. III. Tentative evidence of rapid Ultra Fast Outflow variability.}

\author{Alfredo Luminari}
\affiliation{INAF - Istituto di Astrofisica e Planetologia Spaziali, Via del Fosso del Caveliere 100, I-00133 Roma, Italy; \url{alfredo.luminari@inaf.it}}
\affiliation{INAF - Osservatorio Astronomico di Roma, Via Frascati 33, 00078, Monte Porzio Catone (Roma), Italy}

\author{Andrea Marinucci}
\affiliation{ASI - Italian Space Agency, Via del Politecnico snc, 00133, Rome, Italy}

\author{Stefano Bianchi}
\affiliation{Dipartimento di Matematica e Fisica, Universit\`a degli Studi Roma Tre, via della Vasca Navale 84, 00146 Roma, Italy}

\author{Barbara de Marco}
\affiliation{Departament de F\`isica, EEBE, Universitat Polit\`ecnica de Catalunya, Av. Eduard Maristany 16, E-08019 Barcelona, Spain}

\author{Chiara Feruglio}
\affiliation{INAF - Osservatorio Astronomico di Trieste, Via G. B. Tiepolo 11 I–34143 Trieste, Italy}

\author{Giorgio Matt}
\affiliation{Dipartimento di Matematica e Fisica, Universit\`a degli Studi Roma Tre, via della Vasca Navale 84, 00146 Roma, Italy}

\author{Riccardo Middei}
\affiliation{Space Science Data Center - ASI, Via del Politecnico s.n.c., 00133 Roma, Italy}
\affiliation{INAF - Osservatorio Astronomico di Roma, Via Frascati 33, 00078, Monte Porzio Catone (Roma), Italy}

\author{Emanuele Nardini}
\affiliation{INAF - Osservatorio Astrofisico di Arcetri, Largo Enrico Fermi 5, 50125 Firenze, Italy}

\author{Enrico Piconcelli}
\affiliation{INAF - Osservatorio Astronomico di Roma, Via Frascati 33, 00078, Monte Porzio Catone (Roma), Italy}

\author{Simonetta Puccetti}
\affiliation{ASI - Italian Space Agency, Via del Politecnico snc, 00133, Rome, Italy}

\author{Francesco Tombesi}
\affiliation{Department  of Physics, Tor Vergata University of Rome,Via della Ricerca Scientifica 1, I-00133 Rome, Italy}
\affiliation{INAF - Osservatorio Astronomico di Roma, Via Frascati 33, 00078, Monte Porzio Catone (Roma), Italy}
\affiliation{Department of Astronomy, University of Maryland, College Park, MD 20742, USA}
\affiliation{NASA – Goddard Space Flight Center, Code 662, Greenbelt, MD 20771, USA}
\affiliation{INFN - Sezione di Roma Tor Vergata, Via della Ricerca Scientifica 1, I-001133 Rome, Italy}

\begin{abstract}

We report on the 2019 XMM-{\it Newton}+{\it NuSTAR} monitoring campaign of the Seyfert galaxy NGC 2992, observed at one of its highest flux levels in the X-rays. The time-averaged spectra of the two XMM-{\it Newton} orbits show Ultra Fast Outflows (UFOs) absorbing structures above 9 keV with $> 3 \sigma$ significance.
A detailed investigation of the temporal evolution on a $\sim$5 ks time scale reveals UFO absorption lines at a confidence level $>$95\% (2$\sigma$) in 8 out of 50 XMM-Newton segments, estimated via Monte Carlo simulations. We observe a wind variability corresponding to a length scale of 5 Schwarzschild radii ${\rm r_S}$. Adopting the novel Wind in the Ionised Nuclear Environment (WINE) model, we estimate the outflowing gas velocity (v=0.21-0.45 c), column density (N$_{\rm H}=4-8\cdot 10^{24}$ cm$^{-2}$) and ionisation state ($\log(\xi_0/\rm {erg\ cm\ s^{-1}})=3.7-4.7$), taking into account geometrical and special relativity corrections. These parameters lead to instantaneous mass outflow rates $\dot{{\rm M}}_{\rm out}\simeq0.3-0.8$ ${\rm M}_{\odot}$yr${}^{-1}$, with associated outflow momentum rates  $\dot{{\rm p}}_{\rm out}\simeq20-90$ ${\rm L}_{\rm Bol}$/c and kinetic energy rates $\dot{{\rm E}}_{\rm K}\simeq2-25$ ${\rm L}_{\rm Bol}$. We estimate a wind duty cycle $\approx$ 12\% and a total mechanical power $\approx$ 2 times the AGN bolometric luminosity, suggesting the wind may  drive significant feedback effects between the AGN and the host galaxy. Notably, we also provide an estimate for the wind launching radius and density $\approx 5 {\rm r_S}, 10^{11} {\rm cm}^{-3}$, respectively.

\end{abstract}

\section{Introduction}
\label{introduction}
Variability is one of the best tools to investigate the emission mechanisms at play in Active Galactic Nuclei (AGN). While in many cases significant flux variations can be attributed to variations in the line-of-sight absorbers \citep[e.g. NGC 1365:][]{ris05, wrh14}, some sources have been also observed to vary dramatically in the X-ray intrinsic flux. Recent long XMM-{\it Newton} and {\it NuSTAR} observations of highly variable sources have led to a number of results which shed light on the accretion/ejection mechanisms in their innermost regions, such as PDS 456 \citep{nrg15, rbn18}, IRAS 13349+2438 \citep{pma20}, IRAS 13224-3809 \citep{ppf17}, MCG-03-58-007 \citep{brs21} and NGC 3783 \citep{cdv21} among the others.

NGC 2992 is a nearby \citep[z=0.00771:][]{ward78,keel96} Seyfert 1.5/1.9 galaxy \citep{trippe08}. The large 2-10 keV amplitude variations found in the deep 2005 RXTE monitoring on time scales of days (F=0.8-8.9$\cdot10^{-11}$ erg cm$^{-2}$ s$^{-1}$: \citealp{mkt07}) and its high peak brightness make it the ideal case to study the response of the accretion disc to strong changes of the nuclear continuum, via time-resolved spectroscopy.
In 2010, NGC 2992 was observed eight times with XMM-{\it Newton} and three times with {\it Chandra}, with a 2-10 keV flux ranging from $\sim 5\cdot10^{-12}$ erg cm$^{-2}$ s$^{-1}$ (its historical minimum) to $1.5\cdot10^{-11}$ erg cm$^{-2}$ s$^{-1}$. A narrow, constant iron line component at 6.4 keV was detected \citep{mkt07,mbb18}. The total iron line Equivalent Width (EW) and the reflection component are anti-correlated with the flux, suggesting that at least part of them originate from matter rather distant (light-years) from the black hole.
The source was simultaneously observed with {\it Swift} and {\it NuSTAR} in 2015 and a 2-10 keV flux of $5.8\pm0.3\cdot10^{-11}$ erg cm$^{-2}$ s$^{-1}$ was measured. All the past X-ray features of the source were detected \citep{mbb18}: a broad iron K$\alpha$ line (EW=$250^{+190}_{-120}$ eV), a rather flat intrinsic emission ($\Gamma=1.72\pm0.03$, cutoff energy $E_c>350$ keV) and a Compton reflection continuum (with a ratio $R=0.18\pm0.07$). The rise in brightness is accompanied by X-ray spectral features arising from an Ultra Fast Outflow with velocity v$\rm{_1}$=0.21$\pm$0.01 c, one of the few ever detected with {\it NuSTAR} alone. The total kinetic energy rate of such a wind is $\approx5\%$ L$_{\rm bol}$, sufficient to switch on feedback mechanisms on the host galaxy (\citealp{dsh05}; see also \citealp{zn20} for the dependence on the wind duty cycle). A re-analysis of the 2003 XMM-{\it Newton} bright state confirmed such outflowing absorption structure with an additional wind component detected at v$\rm{_2}$=0.305$\pm$0.005 c, one of the fastest detected so far in a Seyfert galaxy at an accretion rate of only few percent of the Eddington value \citep{tcr10, grt13}. \\
The Swift-XRT monitoring campaigns \citep{middei20} have been performed between late March and mid December 2019 and January to December 2021, with a variable interval between the observations: 2 days during the XMM-{\it Newton} observing windows and 4 days in the remaining months. Large 2-10 keV amplitude variability was found (ranging between 0.3 and 1.1$\cdot10^{-10}$ erg cm$^{-2}$ s$^{-1}$), indicating that the variability timescale is quite short, of the order of days. Simultaneous XMM-{\it Newton} (250 ks) and {\it NuSTAR} (120 ks) observations were hence triggered on 6 May 2019 (\citealp{mbb20}, hereafter Paper I). Several iron K emission transients were detected in the 5-7 keV energy band and their location was estimated from fitting 50 EPIC pn spectra ($\sim5$ ks long each). Two components can be ascribed to a flaring emitting region of the accretion disc located at $\simeq10$-40 r$_g$ from the central black hole (r$_g=$G M$_{BH}$/c$^2$ is the gravitational radius and M$_{BH}$, c are the black hole mass and the speed of light) and one is likely produced at much larger radii ($>50 r_g$).\\
We hereby present novel results from the same XMM-{\it Newton} and {\it NuSTAR} 2019 observations, describing the detection of Ultra Fast Outflows (UFOs) and constraining their properties with the novel Wind in the Ionised Nuclear Environment model (WINE), which couples radiative transfer photoionisation with a Monte Carlo treatment \citep{lpt18,ltp20,lne21,llt21}. The paper is structured as follows: in Sect. 2 we discuss the data analysis procedure and the statistical significance of the UFOs, in Sect. 3  we apply the WINE model to the most significant spectra, in Sect. 4 and 5 we discuss and summarise our findings. 

\section{Data analysis}
\label{dataanalysis}
\subsection{Observations and data reduction}
NGC 2992 was monitored with {\it Swift}-XRT throughout the year 2019 (from March 26 to December 14), to trigger a deep, high flux XMM-{\it Newton} observation of the source. The triggering flux threshold was met on May 6, 2019, with a 2-10 keV flux F$_{2-10}$=7.0$\pm0.4\cdot10^{-11}$ erg cm$^{-2}$ s$^{-1}$ \citep{middei20} and XMM-{\it Newton} promptly started its 250 ks pointing on the following day, for two consecutive orbits. {\it NuSTAR} observed NGC 2992 on May 10, 2019 for 120 ks, simultaneously to the second XMM-{\it Newton} orbit, i.e. after $\sim 185$ ks from the beginning of the pointing. In this paper we consider the same data set presented in Paper I and we adopt the same binning strategy and nomenclature for each spectral slice. We report a journal of the observations in Table \ref{table-journal}.

\begin{table*}
\begin{center}
\begin{tabular}{c c c c c}
\hline
Satellite & Instrument & Obs. ID & Net Exposure (ks) & Start Date \\
\hline
\textit{XMM-Newton} & pn & 0840920201 & 92.6 & 2019-05-07 \\ 
\textit{XMM-Newton} & pn & 0840920301 & 92.8 & 2019-05-09 \\
\textit{NuSTAR} & FPMA & 90501623002 & 57.5 & 2019-05-10 \\
\textit{NuSTAR} & FPMB & 90501623002 & 57.1 & 2019-05-10 \\
\hline
\end{tabular}
\caption{Journal of the observations}
\end{center}
\label{table-journal}
\end{table*}

\begin{figure}
\centering
\includegraphics[width=\columnwidth]{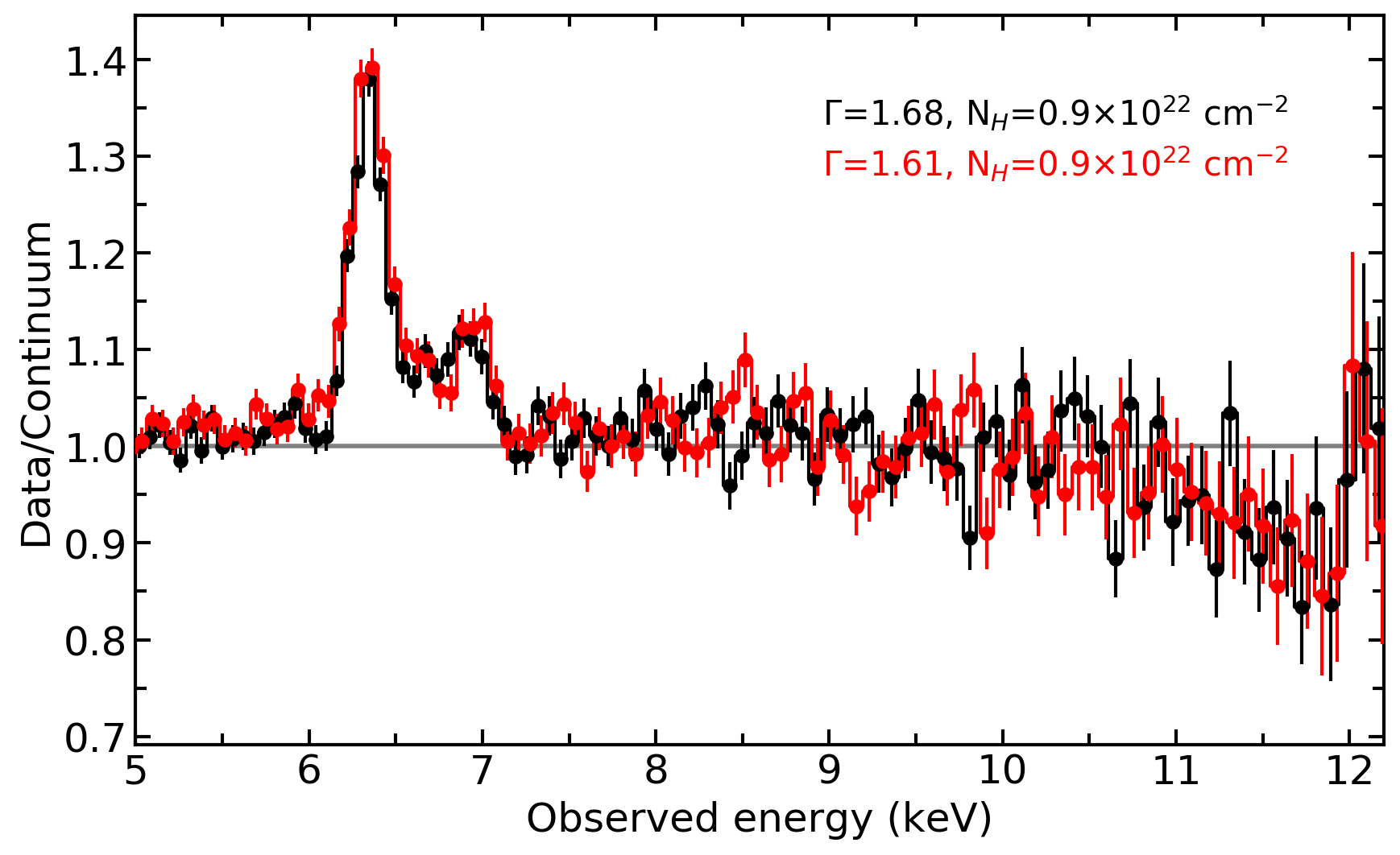} \\
\includegraphics[width=\columnwidth]{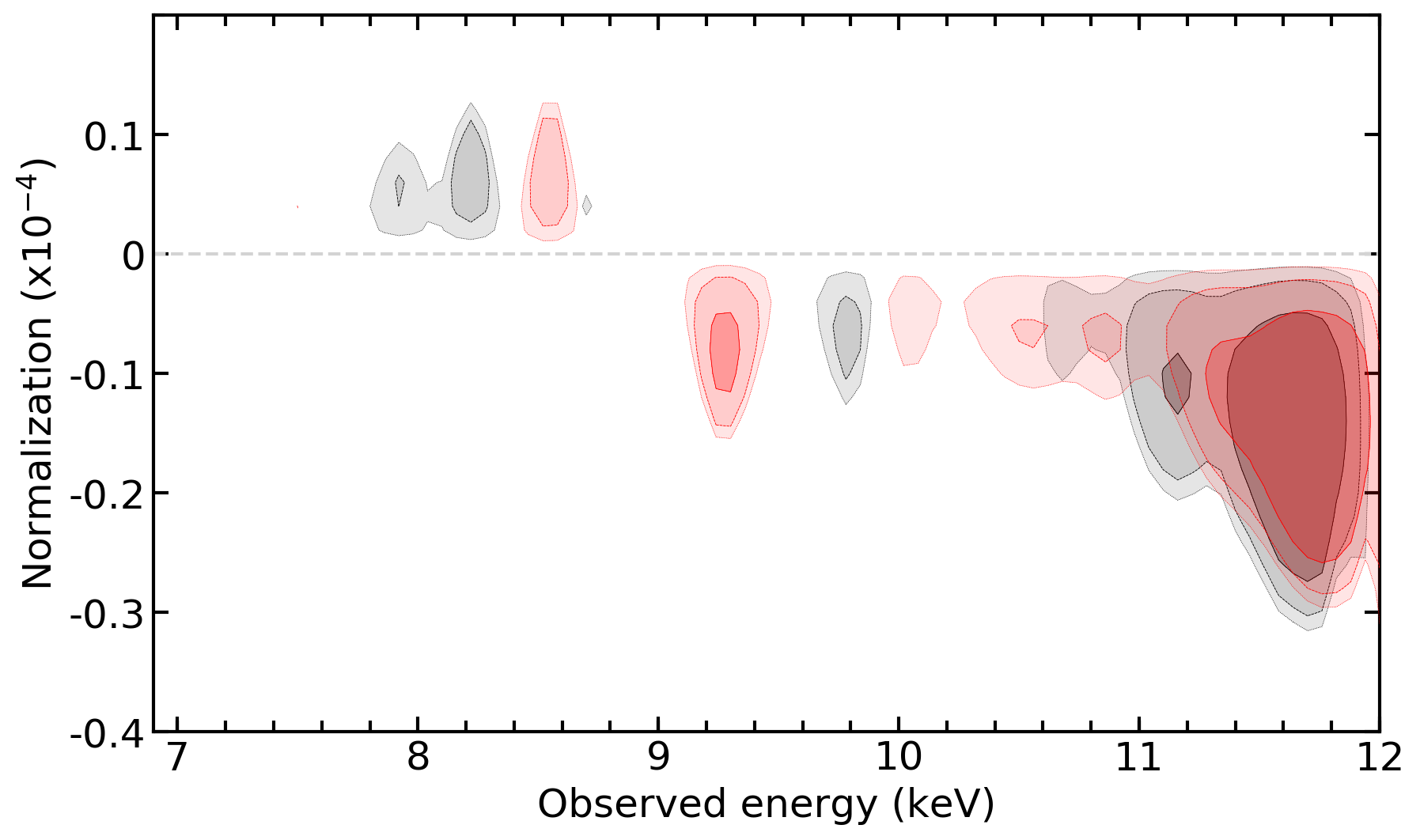}
\caption{{\it Top-panel:} ratio between the XMM-{\it Newton} time-averaged data from the first and second orbit and the associated best-fitting continuum models (black and red lines, respectively). The continuum is composed of an absorbed power law fitted between 3–5 plus 8–12 keV; the best-fitting values for the column density and photon index are reported in the top-right corner. {\it Bottom-panel:} contour plot between the normalisation and the observed energy of a variable Gaussian line in the 7-12 keV range. Black and red shaded regions are used for time-averaged spectra from the first and second orbits (darker to lighter colours indicate 99$\%$, 90$\%$ and 68$\%$ confidence levels).}
\label{ratios}
\end{figure}

\begin{figure*}[!ht]
\begin{center}
\includegraphics[width=0.65\columnwidth]{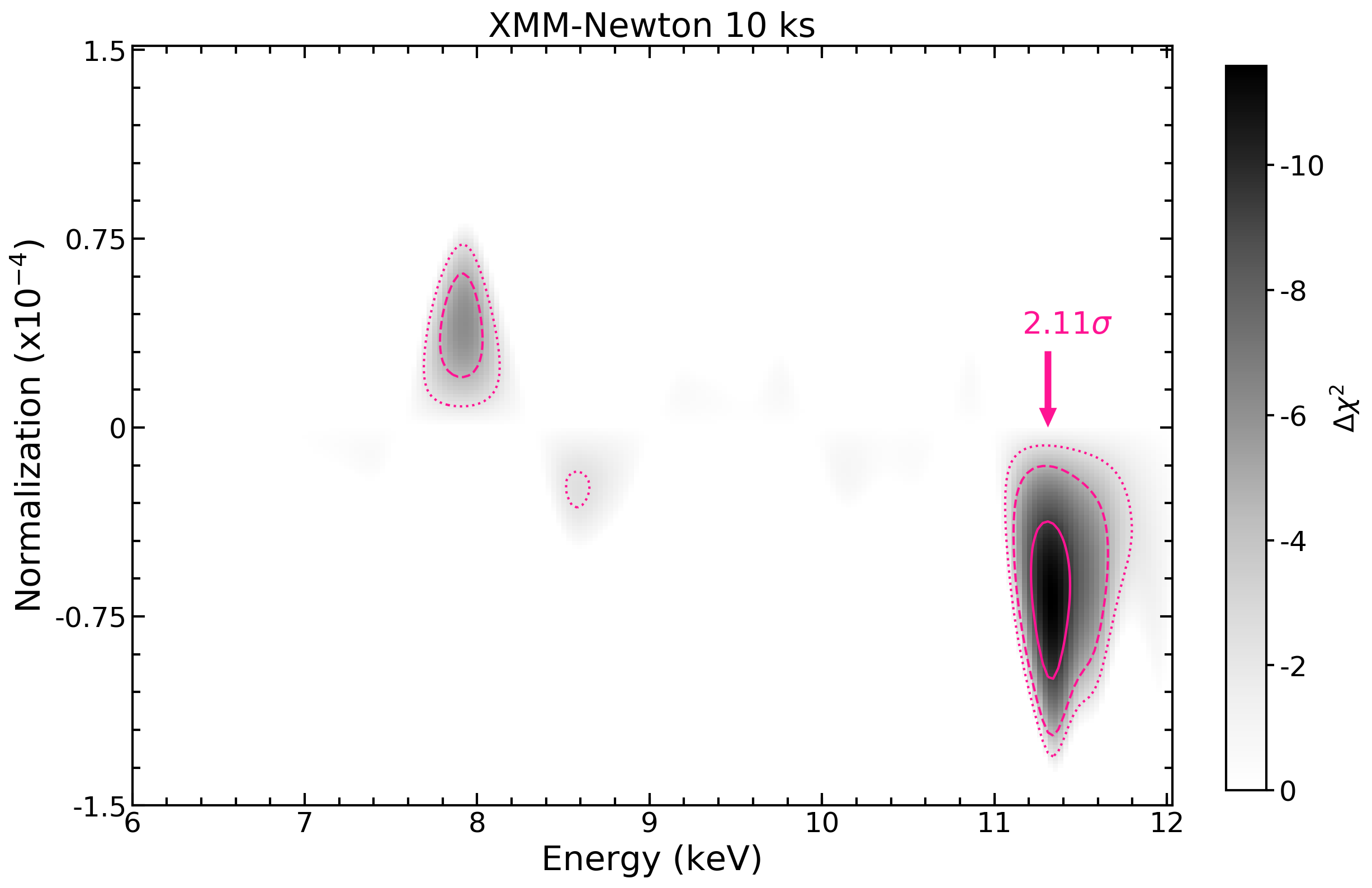}
\includegraphics[width=0.65\columnwidth]{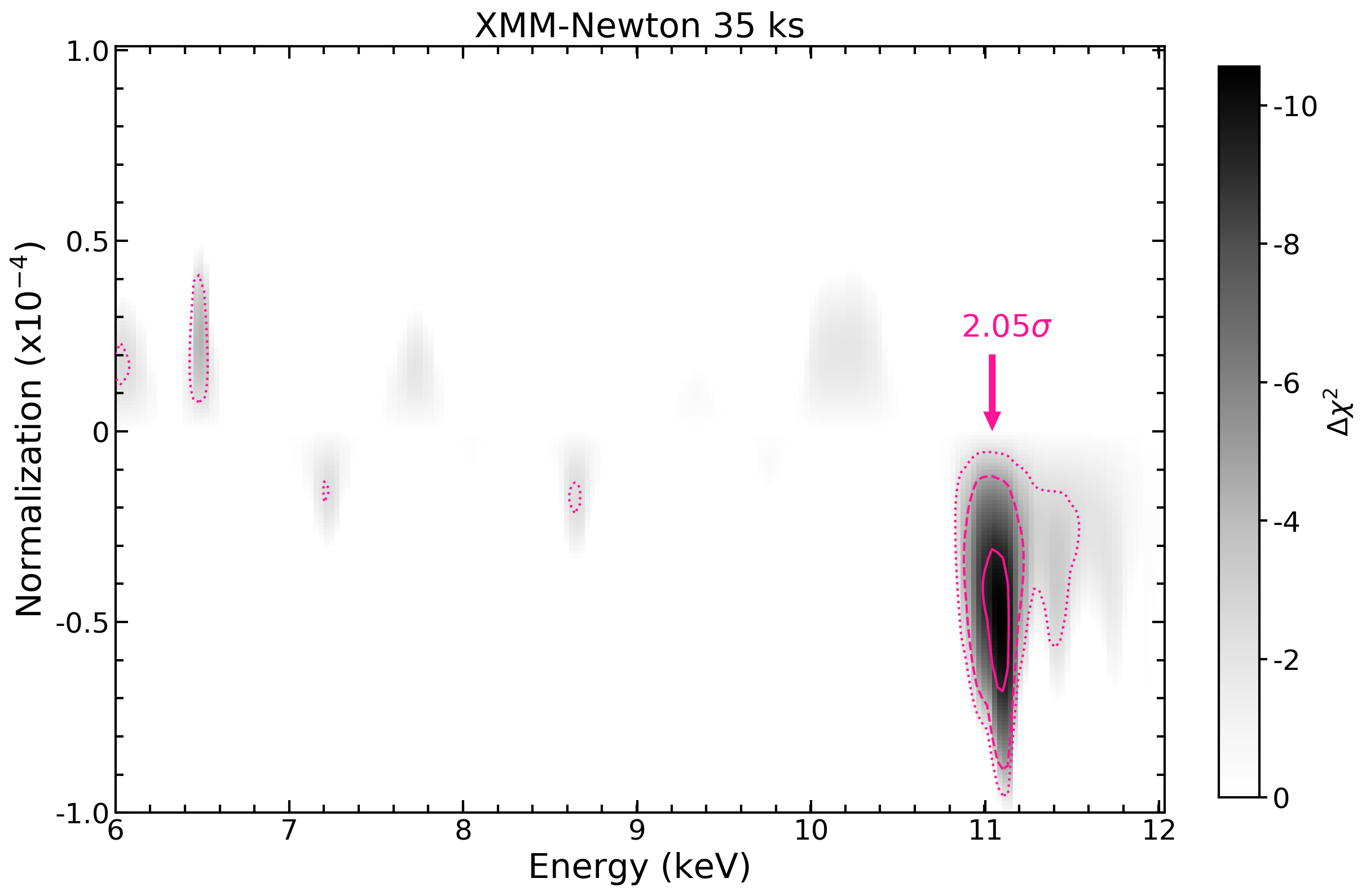}
\includegraphics[width=0.65\columnwidth]{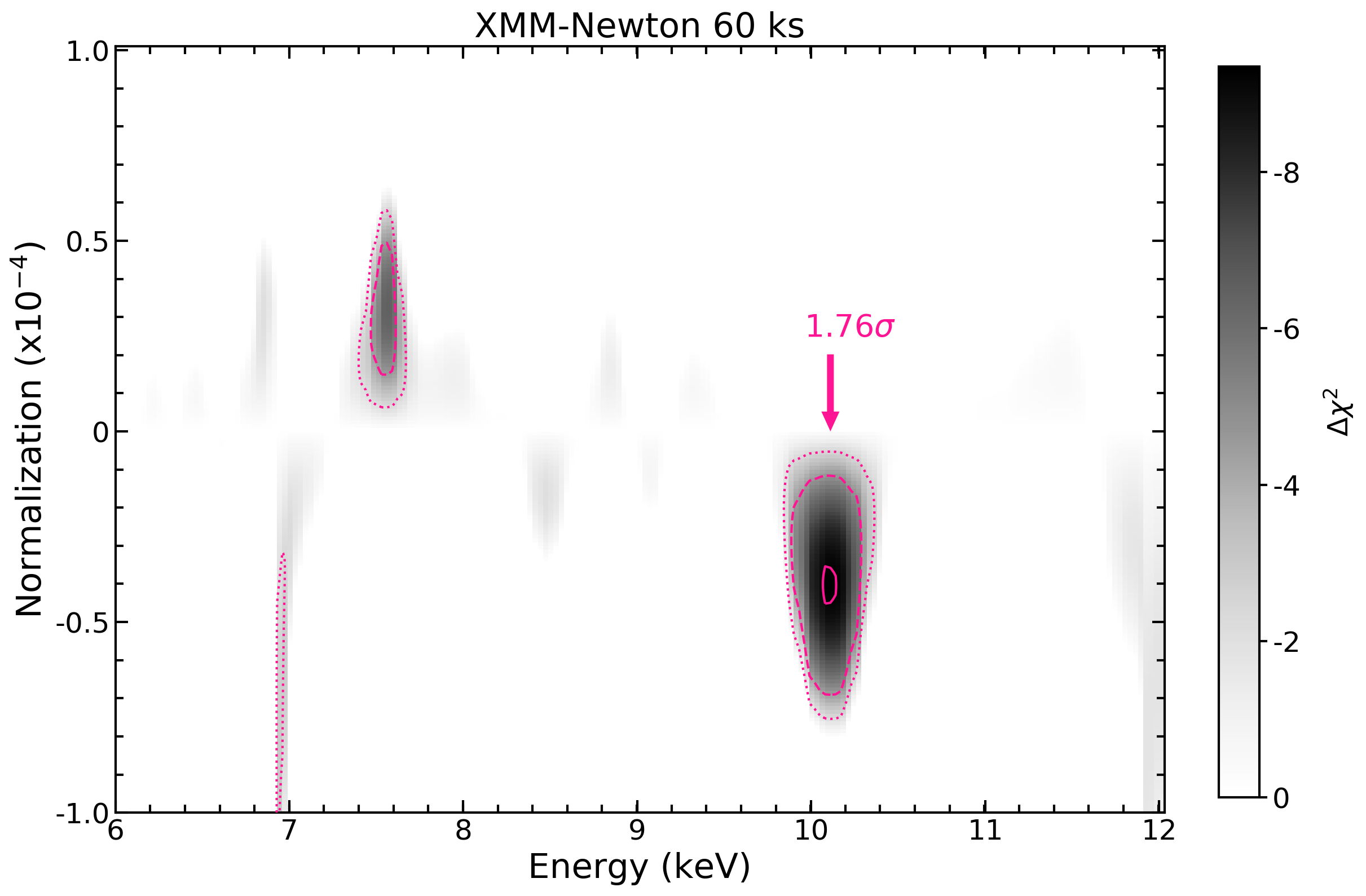}
\includegraphics[width=0.65\columnwidth]{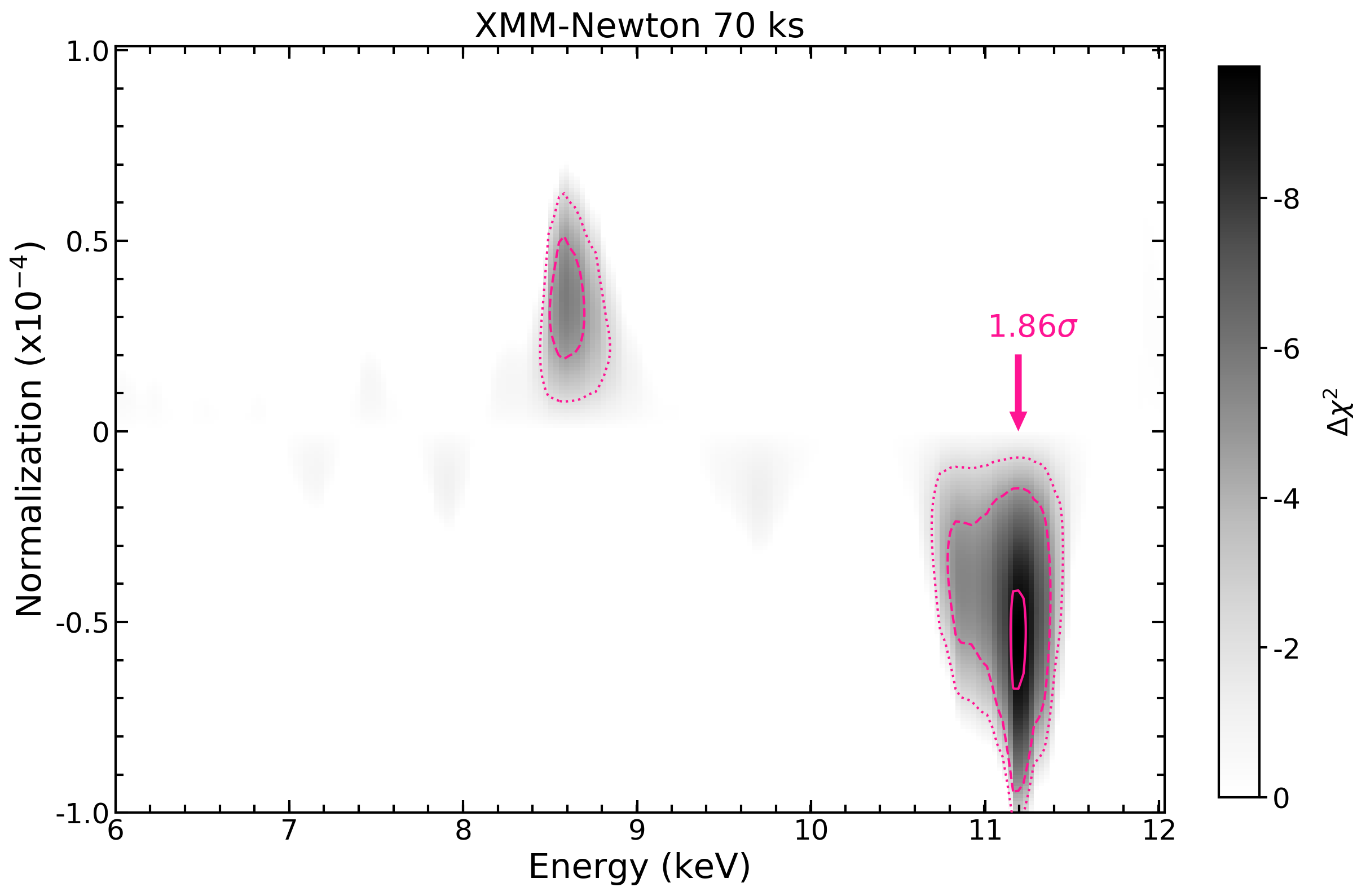}
\includegraphics[width=0.65\columnwidth]{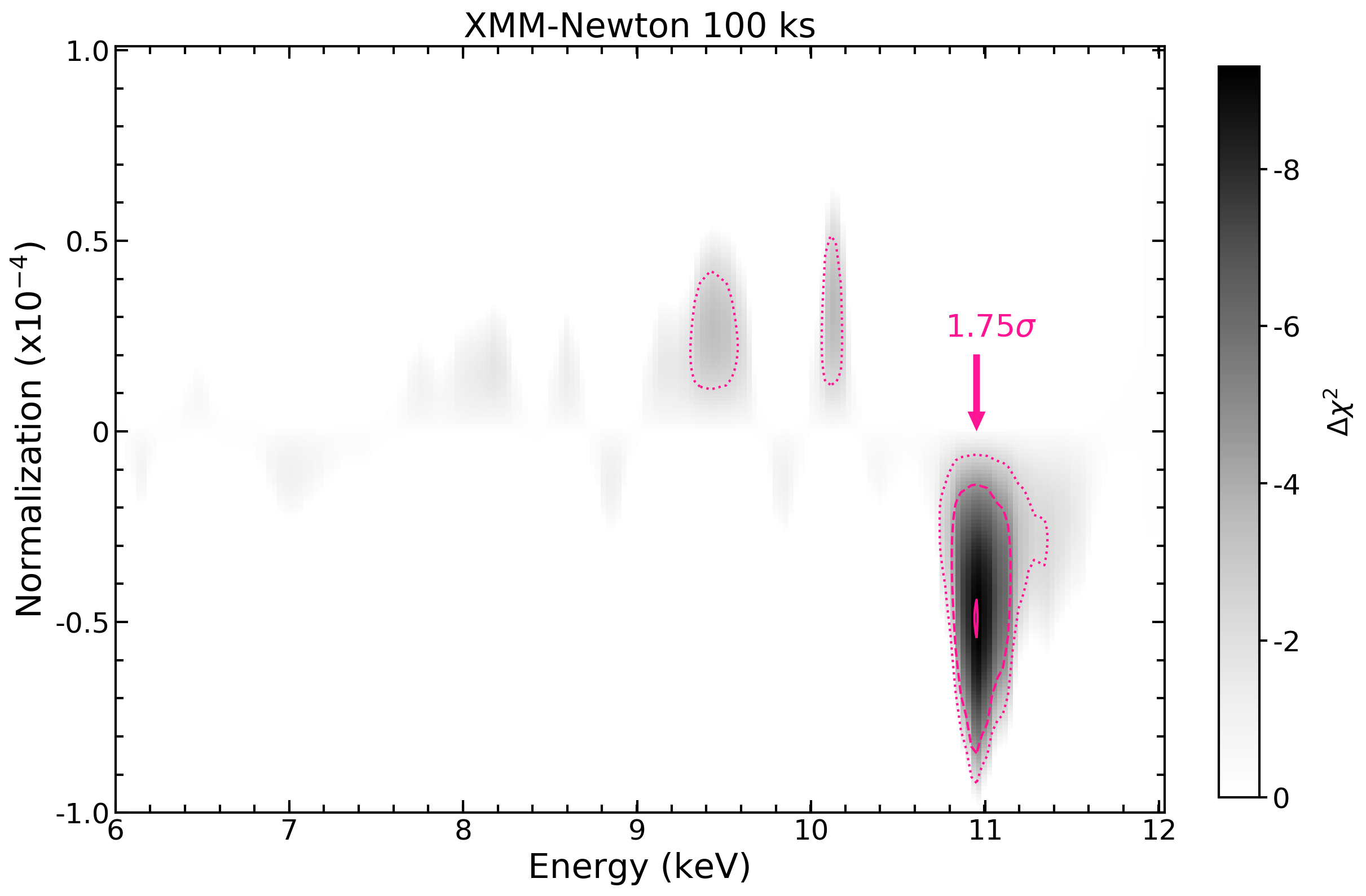}
\includegraphics[width=0.65\columnwidth]{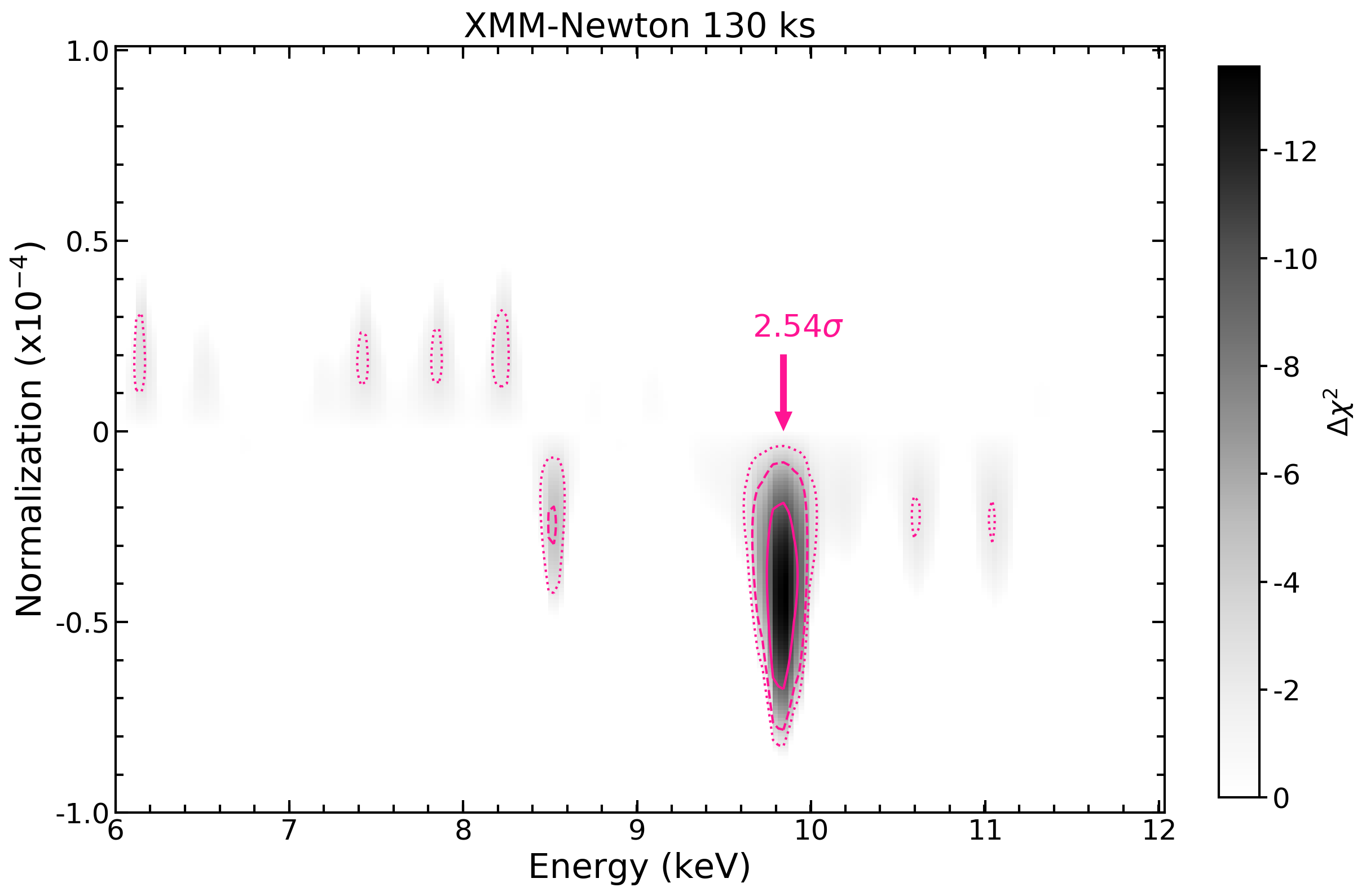}
\includegraphics[width=0.65\columnwidth]{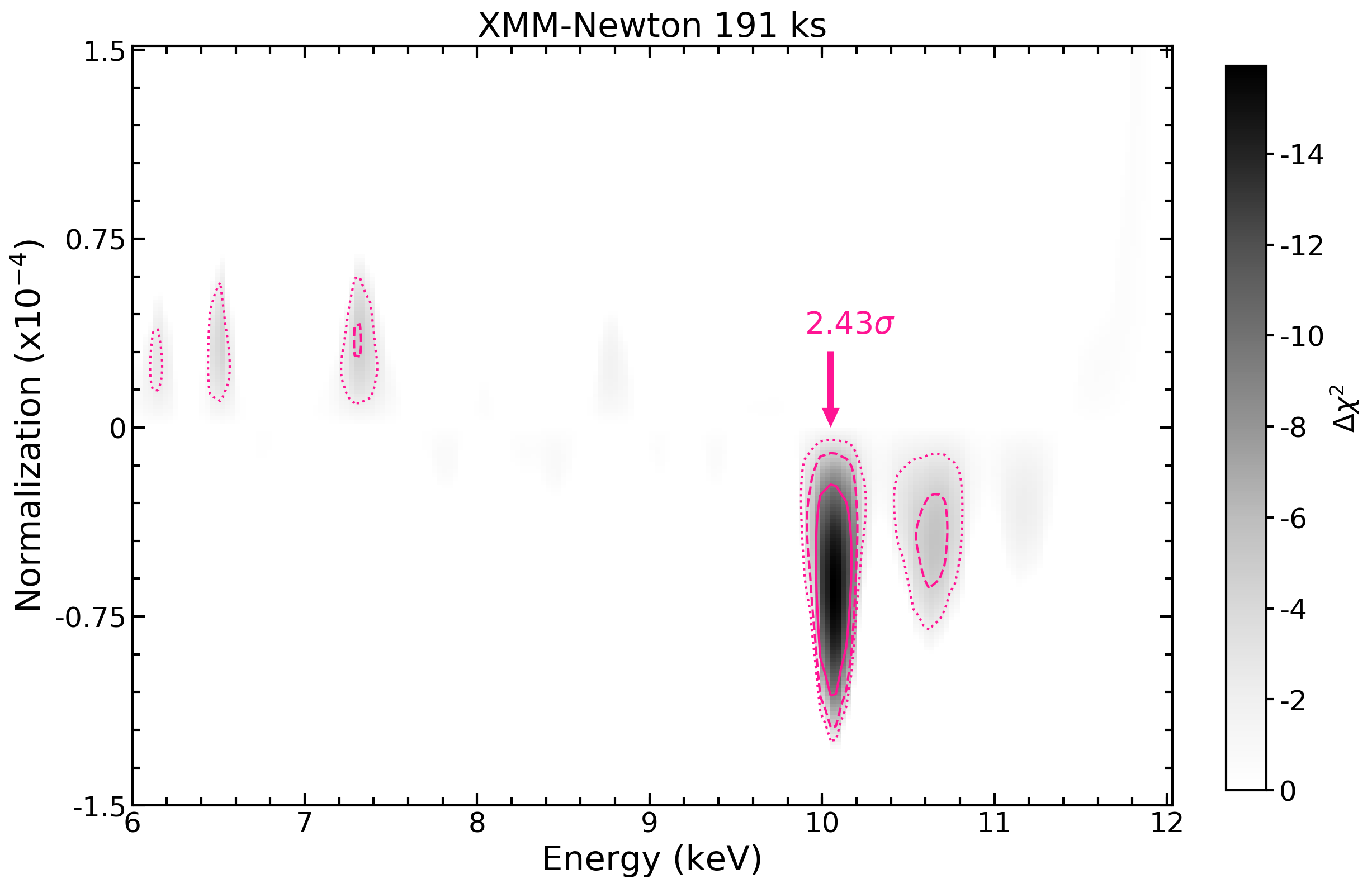}
\includegraphics[width=0.65\columnwidth]{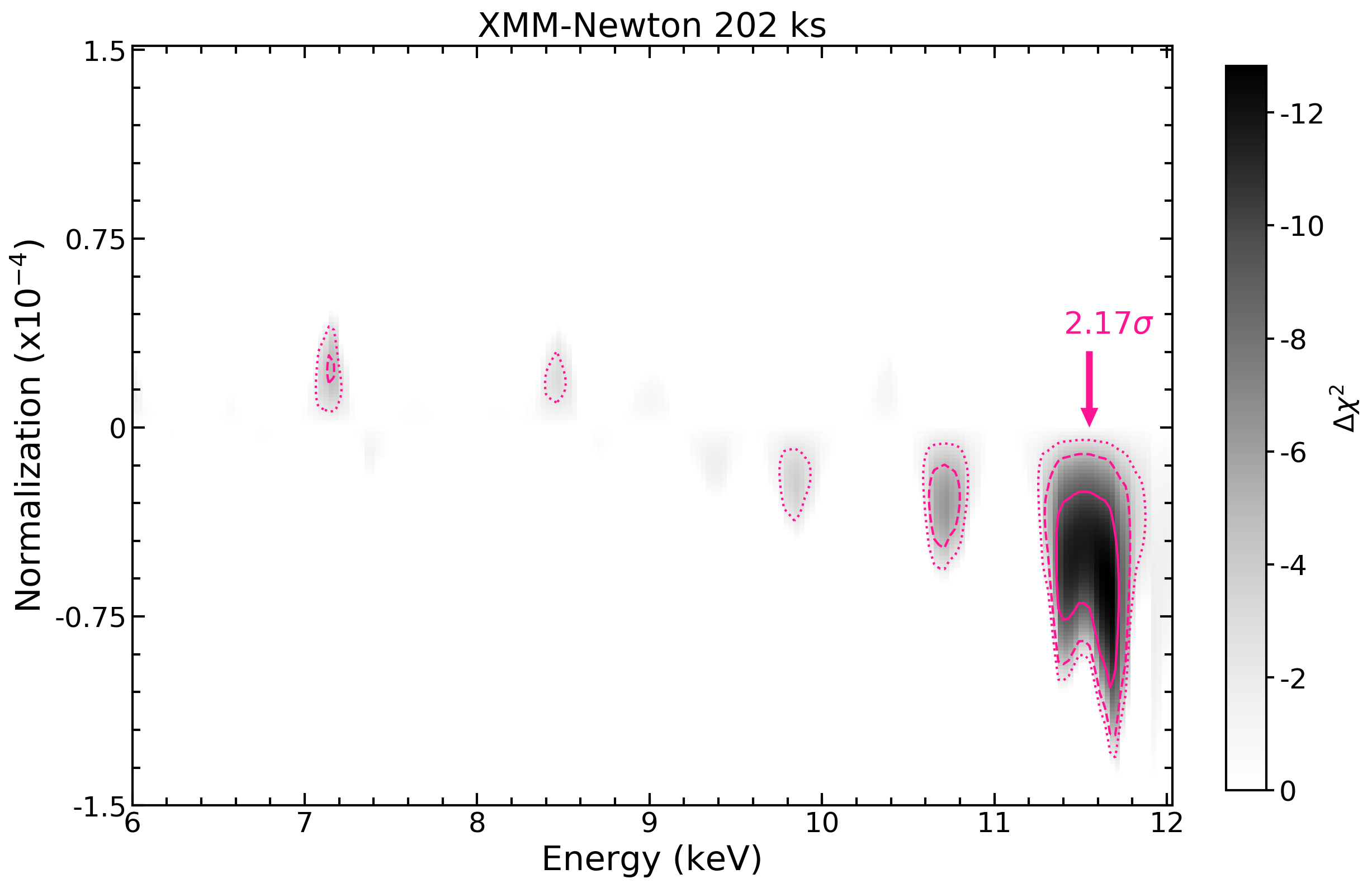}
\includegraphics[width=0.65\columnwidth]{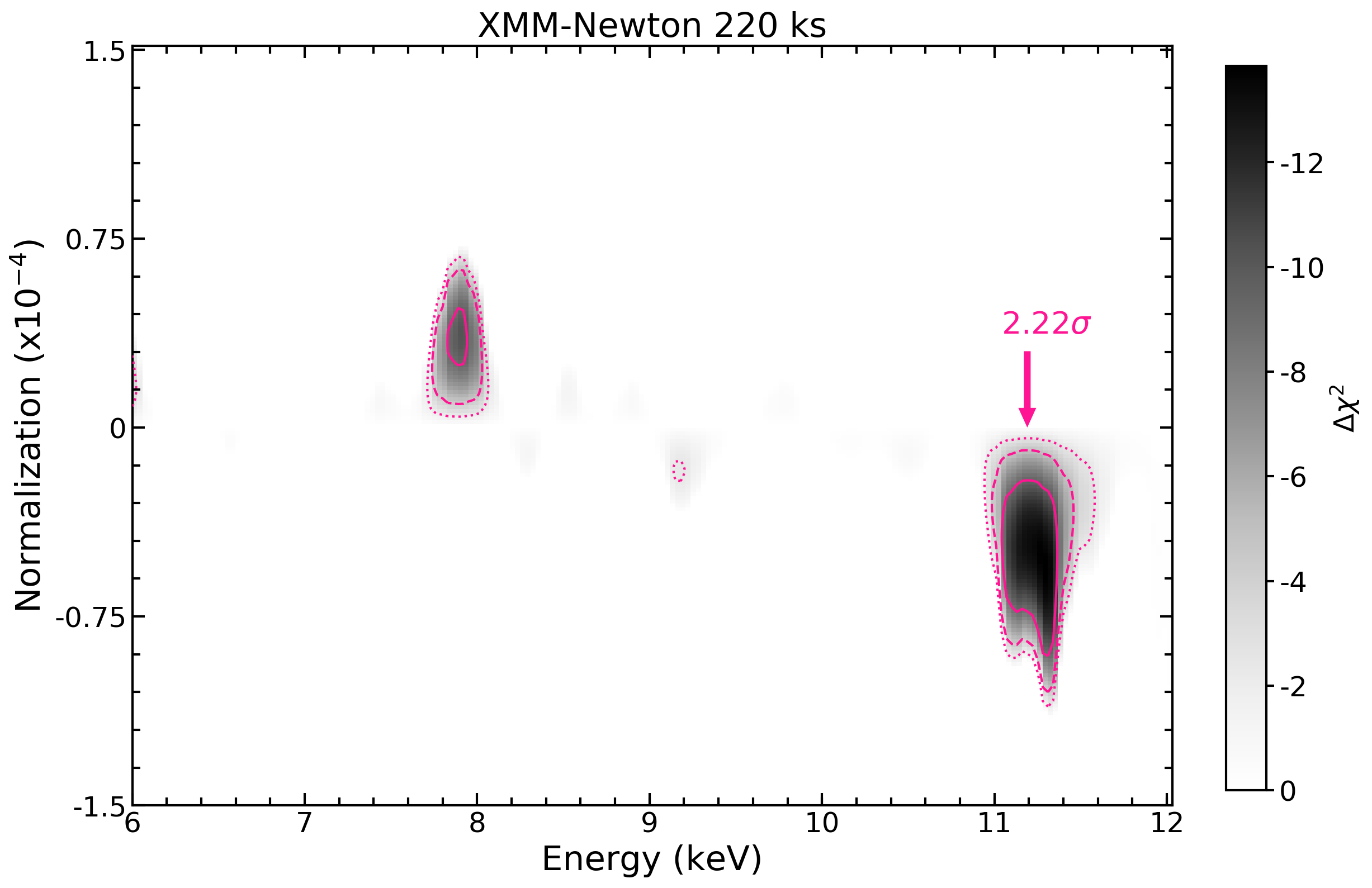}
\includegraphics[width=0.65\columnwidth]{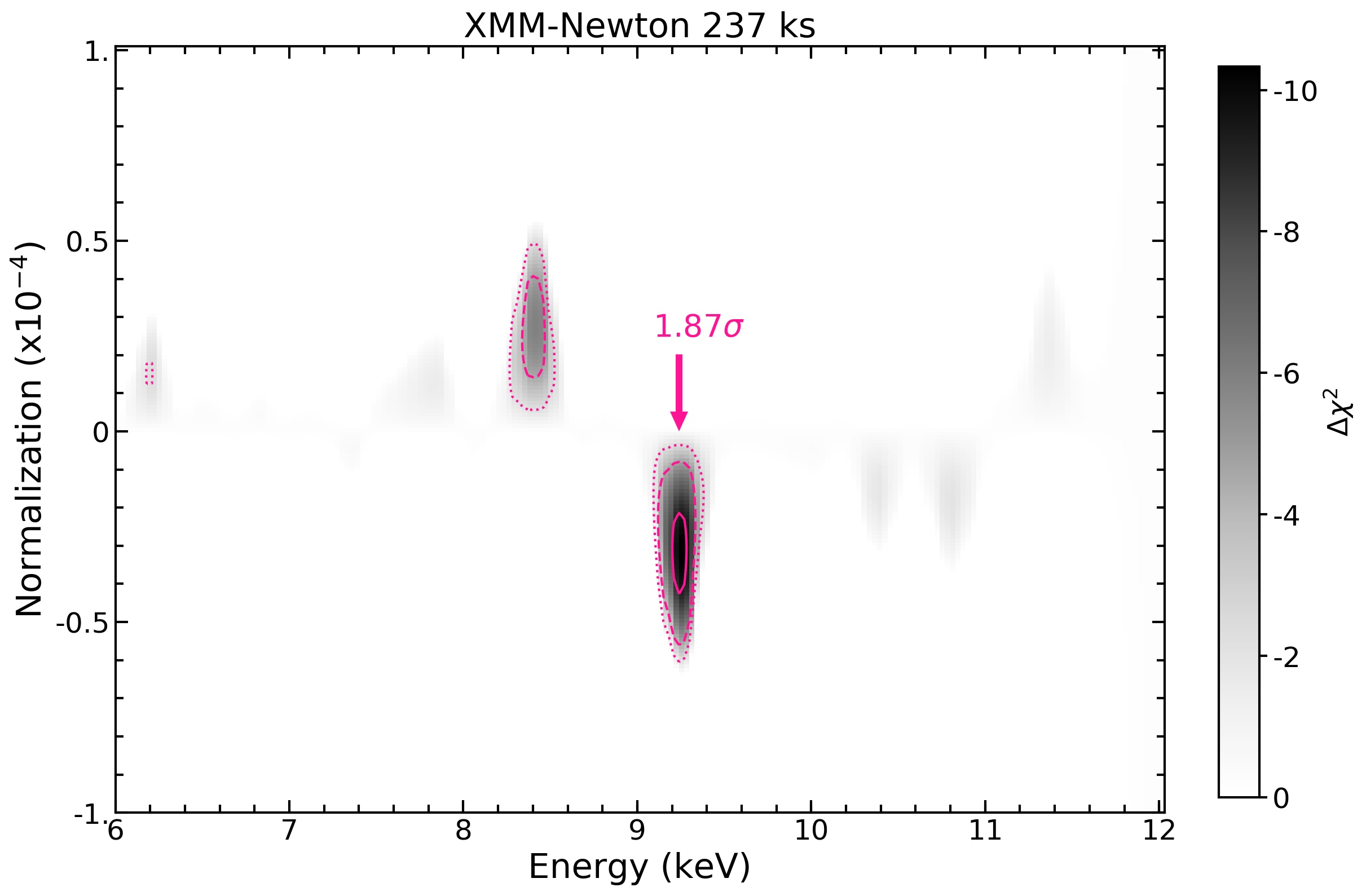}
\includegraphics[width=0.65\columnwidth]{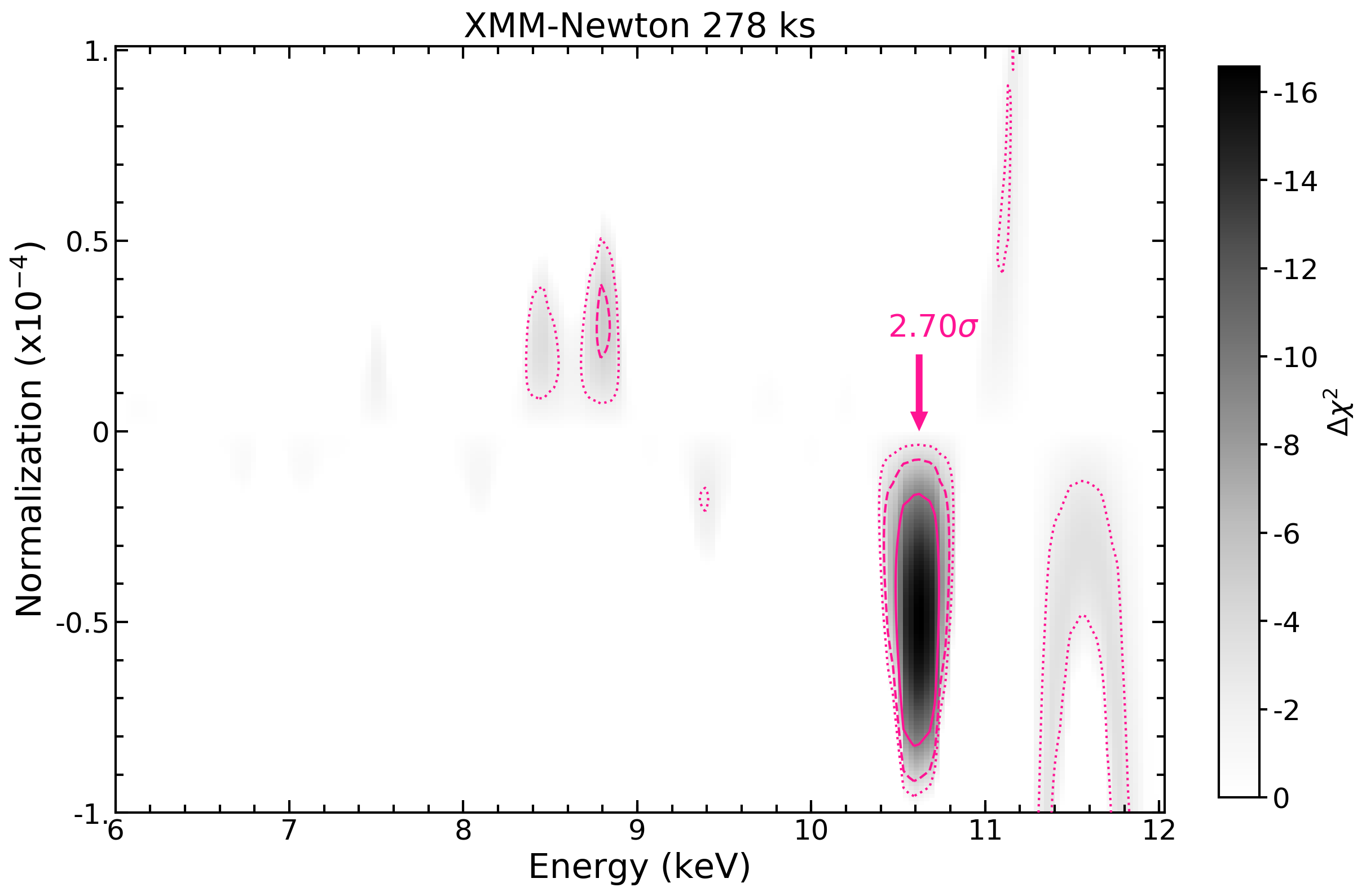}
\includegraphics[width=0.65\columnwidth]{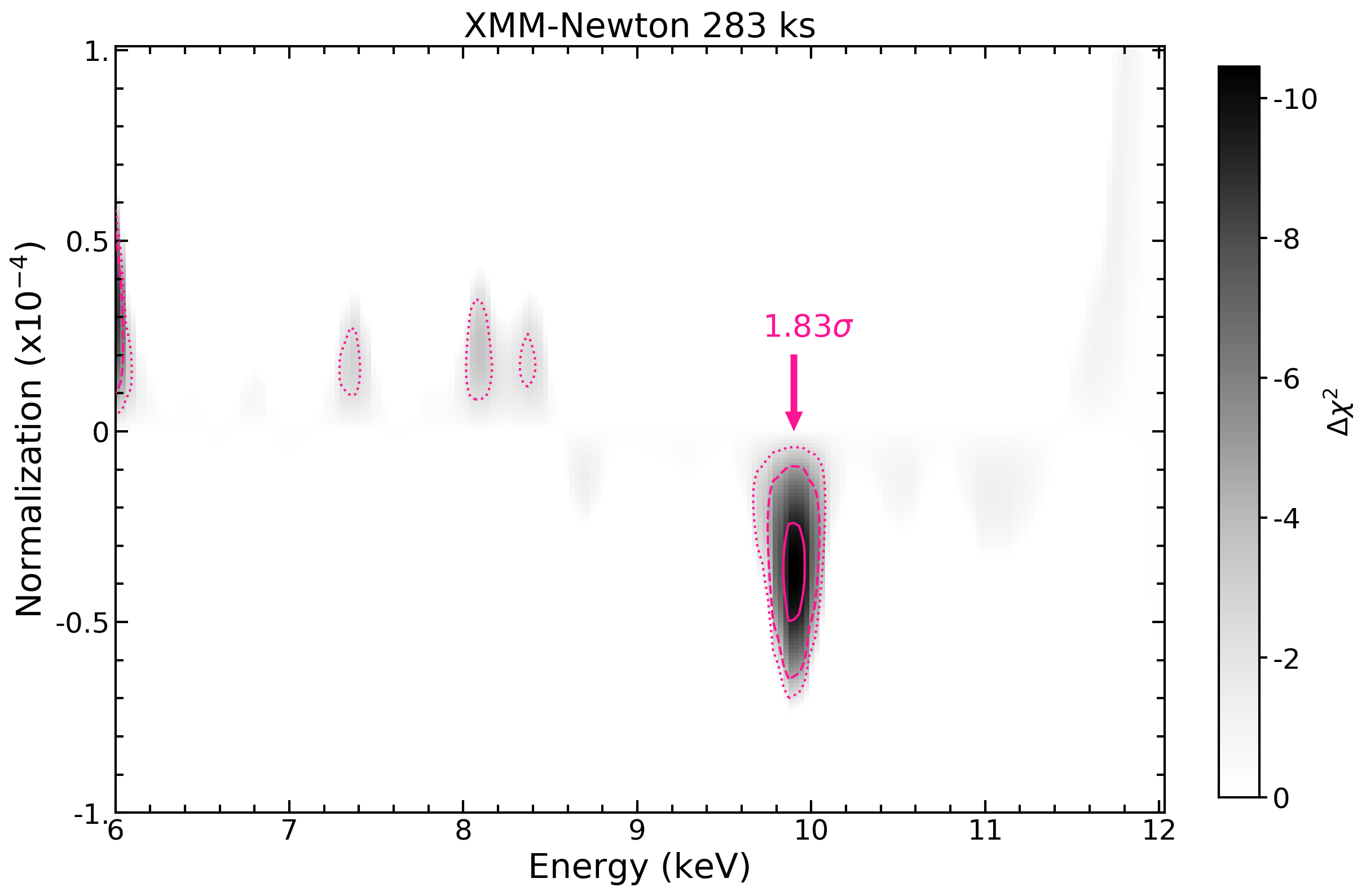}
\includegraphics[width=0.65\columnwidth]{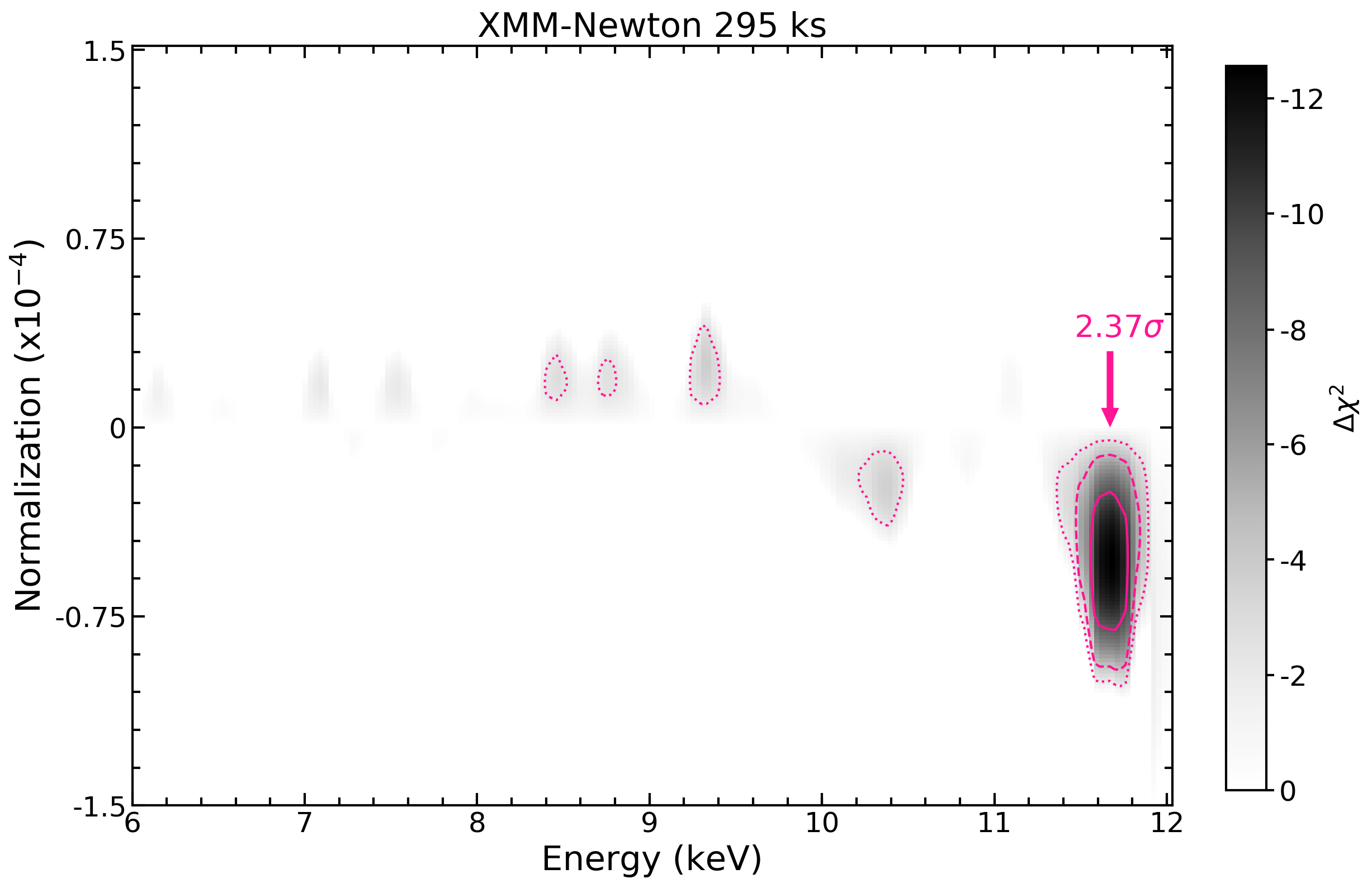}    
\end{center}
\caption{XMM-{\it Newton} contour plots between the normalisation and the observed energy of a variable Gaussian line between 6 and 12 keV. We adopt 200 steps in both normalisation and energy and consider a model composed of an absorbed power law and several emission lines, see Sect. \ref{stat}. Solid, dashed and dotted magenta lines indicate 99$\%$, 90$\%$ and 68$\%$ confidence levels, corresponding to $\Delta\chi^2$=-9.21,-4.61 and -2.3, respectively. For simplicity, we only show  results from time intervals with a $\Delta\chi^2$<-9.21. The significance of the absorption lines, inferred via Monte Carlo simulations, is indicated in magenta.}
\label{blind}
\end{figure*}

\begin{figure*}
\begin{center}
\includegraphics[width=0.65\columnwidth]{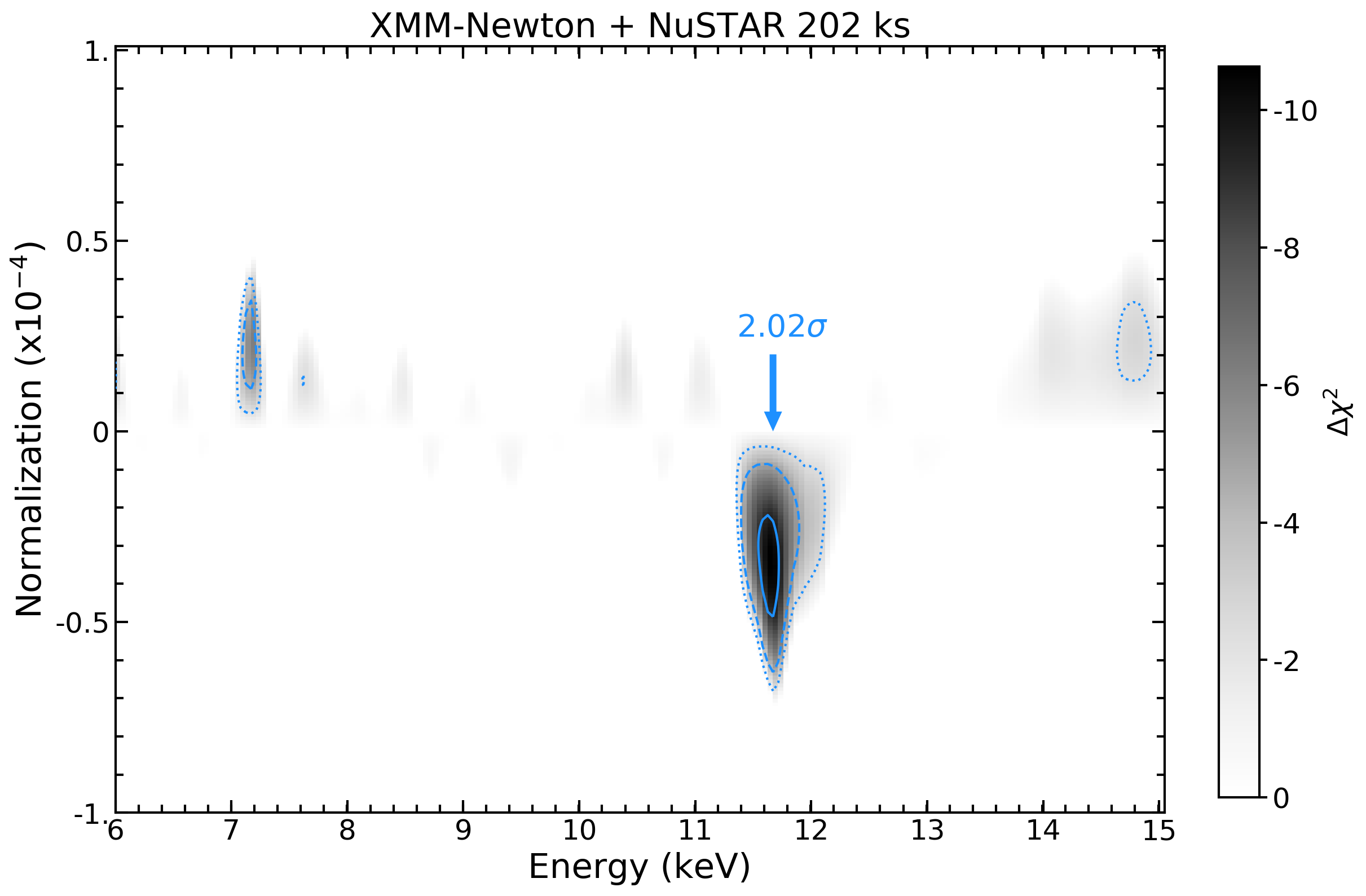}
\includegraphics[width=0.65\columnwidth]{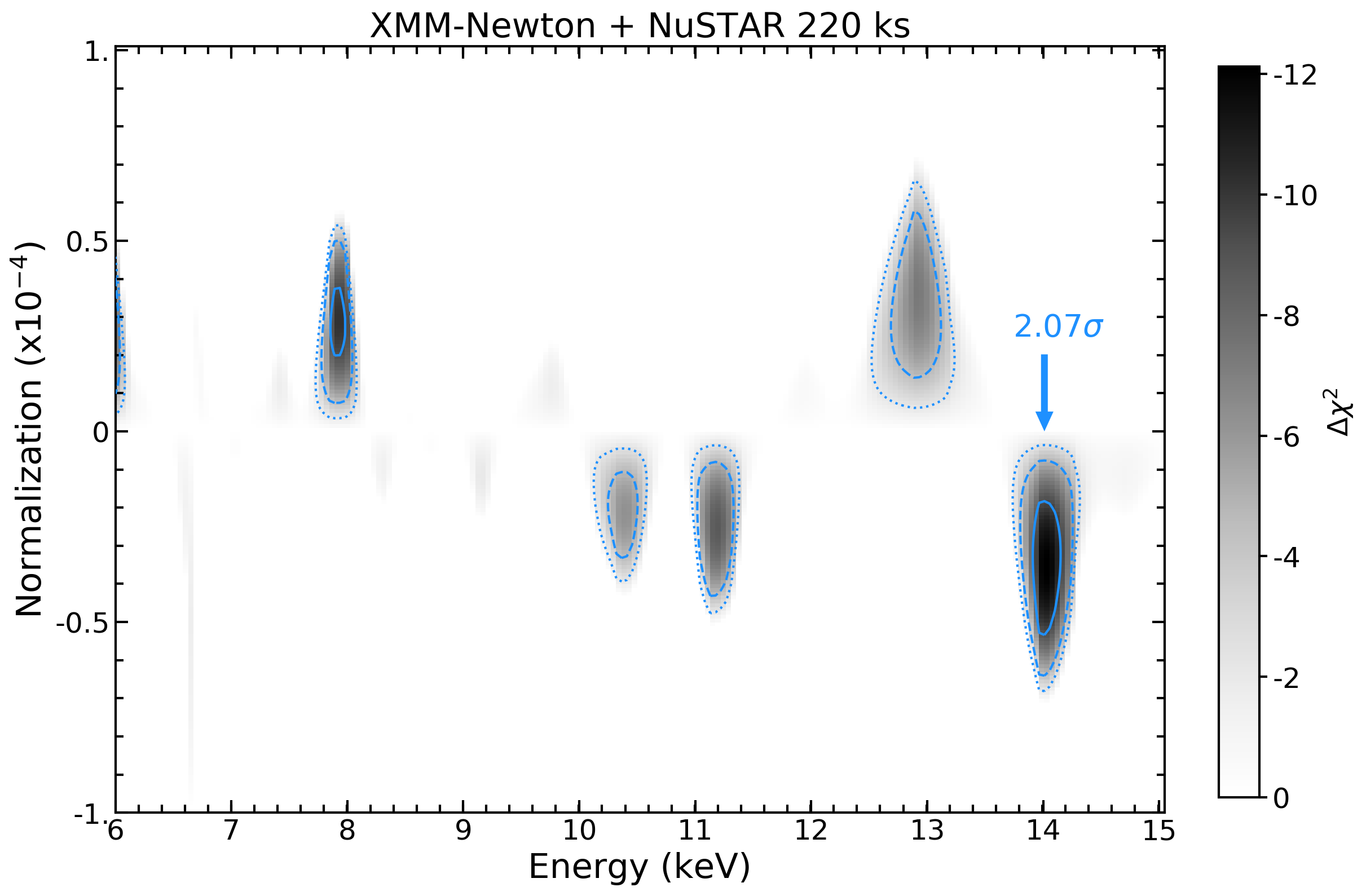}
\includegraphics[width=0.65\columnwidth]{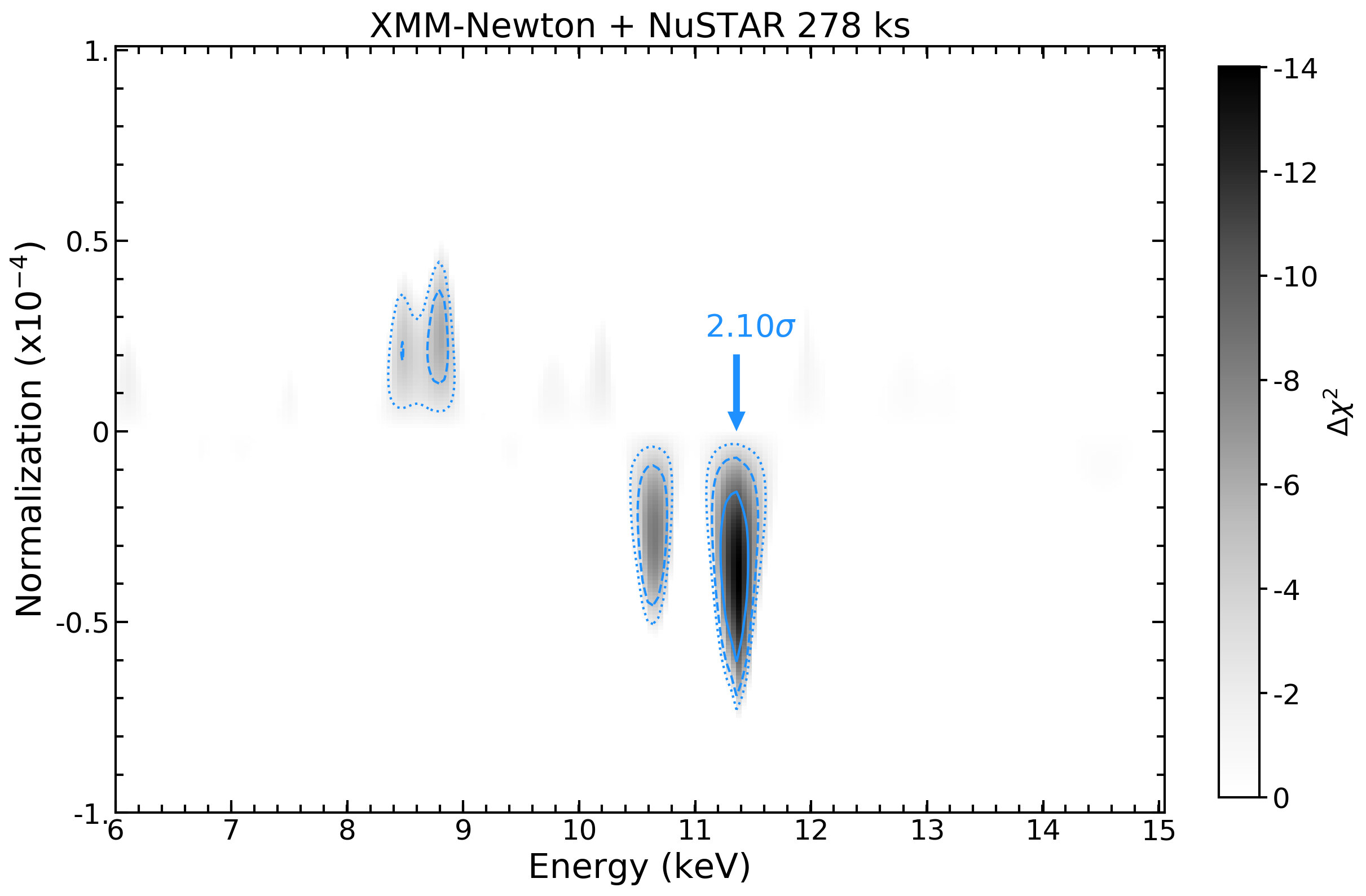}
\includegraphics[width=0.65\columnwidth]{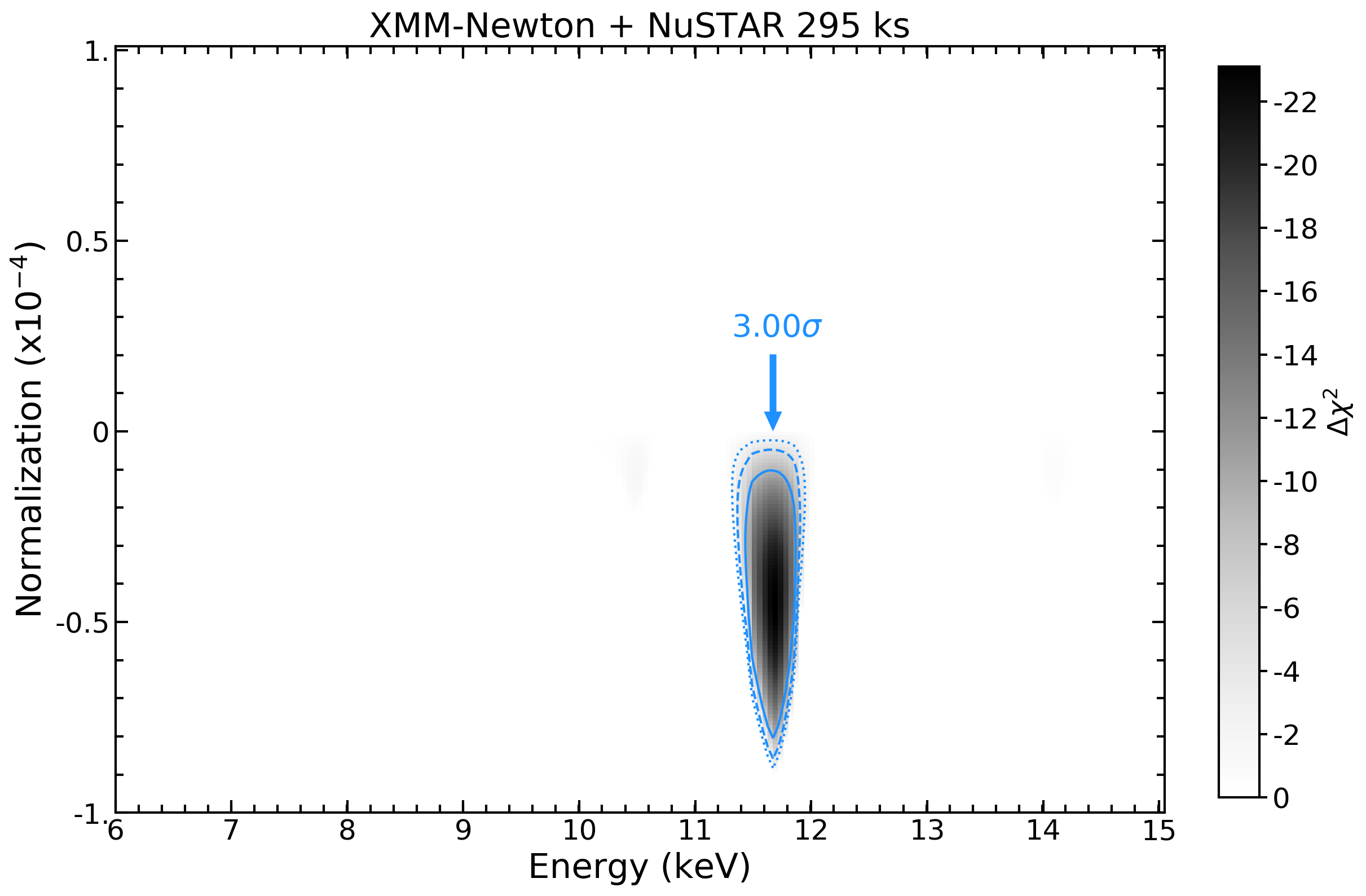}
\end{center}
\caption{XMM-{\it Newton} and {\it NuSTAR} contour plots between the normalisation and the observed energy of a variable Gaussian line between 6 and 15 keV. As in Fig. \ref{blind} solid, dashed and dotted light blue lines indicate 99$\%$, 90$\%$ and 68$\%$ confidence levels, and we only show  results from time intervals with a $\Delta\chi^2$<-9.21. The significance of the absorption lines, inferred via Monte Carlo simulations, is indicated in light blue.}
\label{blind2}
\end{figure*}

For the XMM data, we oversample the instrumental resolution by at least a factor of 3 and require to have no less than 30 counts in each background-subtracted spectral channel. The first XMM orbit is divided in bins of 5 ks each, while the second one, together with the {\it NuSTAR} observation, in bins of $5.8$ ks. Different background regions do not affect the outcomes of the spectral analysis of the XMM-{\it Newton} time-averaged spectra, as discussed in Appendix \ref{AppendixBACK}. {\it NuSTAR} spectra are binned in order to oversample the instrumental resolution by at least a factor of 2.5 and to have a SNR greater than 3$\sigma$ in each spectral channel.
We adopt the cosmological parameters H$_0$=70 km s$^{-1}$ Mpc$^{-1}$, $\Omega_\Lambda=0.73$ and $\Omega_m=0.27$, i.e. the default ones in \textsc{Xspec 12.11.1} \citep{Xspec}. Errors correspond to the 90\% confidence level for one interesting parameter ($\Delta\chi^2=2.7$), if not stated otherwise.

\subsection{Ultra Fast Outflows detection and statistical significance}
\label{stat}
We fit the time-averaged spectra of the two XMM-{\it Newton} orbits between 2 and 12 keV with a model composed of an absorbed power law ({\sc zwabs}$\times${\sc pow} in {\sc XSPEC}) multiplied by a Galactic absorption component ({\sc TBabs}) with N$_{\rm H} \equiv 4.8 \cdot 10^{20}$ cm$^{-2}$ \citep{kalberla05} and removing the energy range dominated by the Fe K lines (5–8 keV). The ratios between the time-averaged data and the best-fitting continuum models are plotted in Fig. \ref{ratios} (top panel), once the 5–8 keV band is included, and clear absorption features above 9 keV can be seen. We add to the absorbed power law baseline model ({\sc tbabs}$\times${\sc zwabs}$\times${\sc pow}) five narrow Gaussian lines, of which two reproduce the neutral Fe K$\alpha$ and K$\beta$ and the other three are associated to the $\sim5.5$ keV, $\sim6.7$ keV and $\sim7.0$ keV transient emission lines, indicated as {\it red flare}, {\it blue flare I} and {\it blue flare II}, respectively, in Paper I. No strong residuals are present below 9 keV and we obtain best fit statistics $\chi^2/$dof=200/154 and $\chi^2/$dof=216/154 for the first and second orbit spectra, respectively.

To blindly search for any absorption signature we add an additional Gaussian line left free to vary in the 6-12 keV range, with a normalisation which can take both positive and negative values. The Gaussian width is fixed to zero to represent a narrow, unresolved line. We fit the two time-averaged XMM-{\it Newton} orbits, leaving the baseline parameters free to vary as well. Fig. \ref{ratios} (bottom panel) shows the corresponding contour plots between the normalisation and the line energy centroid. 
For the first orbit (black data points and contour plots in Fig. \ref{ratios}) the inclusion of a Gaussian line leads to a fit improvement of $\Delta\chi^2=-18.48$ for two additional degrees of freedom. The best fit energy of the line is $11.78_{-0.15}^{+0.08}$ keV, with a normalisation N=$-1.6\pm0.6\cdot10^{-5}$ ph cm$^{-2}$ s$^{-1}$. 

To evaluate the statistical significance of narrow, unresolved emission/absorption lines standard likelihood ratio tests could lead to inaccurate results \citep{prot02}. Following the procedures described in \citet{tcr10} and \citet{wmp16}, we create Monte Carlo routines to retrieve the statistical significance of the absorption line. We create 10000 fake data sets of the XMM-{\it Newton} EPIC pn spectrum using the {\sc fakeit} command in {\sc Xspec} with responses, background files, exposure times and energy binning of the real data according to the following procedure.
We first load the best fitting model, composed of an absorbed power law and five Gaussian lines, and produce a new list of free parameters (i.e. column density of the cold absorber, power law photon index and normalisation, energy and normalisation of the emission lines) by drawing from a multivariate Normal distribution based on the covariance matrix via the {\sc Xspec} command {\sc 'tclout simpars'}. This allows us to take into account uncertainties in the continuum model, too. Then, this new continuum, without absorption lines and with randomly sampled free parameters, is used to simulate a fake spectrum. 
Finally, the fake spectrum is fitted, first with the input model (i.e. without absorption lines) and, then, including an additional unresolved Gaussian line to the model, with normalisation and energy centroid left free to vary in the range [-1.0:+1.0]$\cdot10^{-4}$ ph cm$^{-2}$ s$^{-1}$ and [6:12] keV, respectively. Being $N$ the number of spectra in which the fit improvement is equal or higher than that of our observed line (i.e. $\Delta\chi^2=-18.48$) and $S$ the total number of simulated spectra, then the estimated statistical significance of the detection is $1-N/S$. We obtain 5 spurious detections out of the total 10000 trials, implying a statistical significance of 3.5$\sigma$.

We then apply the same procedure to the XMM spectrum of the second orbit (red data points and contour plots in Fig. \ref{ratios}). We find two absorption lines at $9.28\pm0.07$ keV, with a normalisation N=$-8.5\pm4.0\cdot10^{-6}$ ph cm$^{-2}$ s$^{-1}$ ($\Delta\chi^2=-10.18$) and at $11.75\pm0.15$ keV, with a normalisation N=$-1.5\pm0.6\cdot10^{-5}$ ph cm$^{-2}$ s$^{-1}$ ($\Delta\chi^2=-17.65$). The significance of these two detections, estimated as above, is 2.2$\sigma$ and $3.2\sigma$, respectively. We note that any broadening of such unresolved absorption lines is likely due to the superposition of different spectral features from several time intervals, as we will show in the next sections.

As a further step we take into account the phenomenological model adopted in Paper I, in which the EPIC pn spectra from the two orbits (250 ks in total) are divided in 50 bins ranging between 2 and 12 keV. {\it NuSTAR} FPMA/B spectra, when available, are simultaneously fitted in the energy interval between 3 and 79 keV.
We fit the 50 EPIC pn spectra leaving the baseline parameters free to vary.  Following the same procedure for the blind search of absorption lines described above, we show in Fig. \ref{blind} the contour plots between the normalisation and the energy centroid of the Gaussian line. We adopt the best fitting models of Paper I and, in addition, we leave the energy centroid of the absorption line  free to vary between 6 and 12 keV. We only show the time intervals where we find a detection at a confidence level greater than 99$\%$ ($\Delta\chi^2$ improvement larger than 9.21, for two parameters of interest). The Gaussian width is fixed to 0 keV to represent a narrow, unresolved line. Solid, dashed and dotted magenta lines indicate 99$\%$, 90$\%$ and 68$\%$ confidence levels, corresponding to $\Delta\chi^2$=-9.21,-4.61 and -2.3, respectively. We detect an absorption line at a confidence level greater than 99$\%$ in 13 out of 50 spectra (24$\%$). Best fit values of the baseline parameters are instead fully consistent with the values in Paper I and, thus, we do not report them here. The same procedure is then repeated for the 20 time intervals in which simultaneous EPIC pn and FPMA/B spectra are available, including cross-calibrations constants between the three detectors and leaving the Gaussian energy centroid to vary between 6 and 15 keV. Fig. \ref{blind2} reports the contour plots for the 4 out of 20 spectra (20$\%$) showing an absorption line with a confidence level $>$99$\%$.

\begin{table*}
\begin{center}
\begin{tabular}{ccccccccc}
 Time & Energy & normalisation & EW & $\Delta\chi^2$ & v$_{\rm out}$/c & v$_{\rm out}$/c & MC sign. & $\chi^2/\nu$ \\
  & (keV) & ($10^{-5}$ ph cm$^{-2}$ s$^{-1}$) & (eV) & & (Fe XXV He-$\alpha$)&  (Fe XXVI Ly-$\alpha$)  & $\sigma$ \\
   \hline
 \multicolumn{9}{c}{XMM-{\it Newton} only} \\
10 ks&$11.40^{+0.08}_{-0.10}$ & $-6.5\pm3.2$&$-165\pm80$&-11.79& $0.486^{+0.005}_{-0.007}$& $0.456^{+0.005}_{-0.007}$& 2.11&126/131 \\
35 ks&$11.15^{+0.08}_{-0.13}$ & $-5.1\pm2.6$&$-155\pm80$&-11.10&$0.469^{+0.006}_{-0.009}$ & $0.438^{+0.006}_{-0.009}$& 2.05 &100/123 \\
60 ks&$10.15\pm0.15$ & $-4.1\pm2.1$&$-90\pm50$&-9.62&$0.39\pm0.01$ &$0.360\pm0.013$ & 1.76 &119/129 \\
70 ks&$11.25^{+0.12}_{-0.10}$ & $-5.6\pm2.8$&$-130\pm70$&-10.09&$0.476^{+0.008}_{-0.007}$ &$0.446^{+0.008}_{-0.007}$ & 1.86 &105/128 \\
100 ks&$11.05\pm0.10$ & $-5.6\pm5.0$&$-130\pm70$&-9.59& $0.462\pm0.007$& $0.431\pm0.007$& 1.75 &131/126 \\
130 ks&$9.90^{+0.07}_{-0.06}$ & $-4.5\pm2.0$&$-100\pm40$&-14.40& $0.372^{+0.006}_{-0.005}$& $0.337^{+0.006}_{-0.005}$& 2.54 &137/126 \\
191 ks&$10.15\pm0.06$ & $-4.3\pm1.8$&$-110\pm45$&-15.83& $0.393\pm0.005$&$0.359\pm0.005$ & 2.43 &151/127 \\
202 ks&$11.73^{+0.07}_{-0.28}$ & $-6.5\pm3.5$&$-220\pm120$&-13.13&$0.508^{+0.004}_{-0.018}$ &$0.478^{+0.004}_{-0.019}$ & 2.17 &151/125 \\
220 ks&$11.35^{+0.07}_{-0.18}$ & $-5.5\pm2.7$&$-170\pm80$&-14.15&$0.483^{+0.005}_{-0.012}$ & $0.453^{+0.005}_{-0.013}$& 2.22 &107/124 \\
237 ks&$9.32^{+0.05}_{-0.08}$ & $-3.1\pm1.7$&$-75\pm40$&-10.03& $0.319^{+0.005}_{-0.008}$& $0.283^{+0.005}_{-0.008}$& 1.87 &118/125 \\
278 ks&$10.70\pm0.08$ & $-4.9\pm2.0$&$-125\pm55$&-15.76& $0.437\pm0.006$&$0.405\pm0.006$ & 2.70 &117/128 \\
283 ks&$9.96^{+0.10}_{-0.08}$ & $-3.7\pm1.8$&$-90\pm50$&-9.94&$0.377^{+0.008}_{-0.007}$ & $0.343^{+0.009}_{-0.007}$& 1.83 &122/128 \\
295 ks&$11.75^{+0.10}_{-0.11}$ & $-5.4\pm2.5$&$-165\pm80$&-12.58&$0.509^{+0.007}_{-0.006}$ & $0.480^{+0.007}_{-0.006}$& 2.37 &160/141 \\
 \hline
 \multicolumn{9}{c}{XMM-{\it Newton} + {\it NuSTAR}} \\
 202 ks&$11.73^{+0.10}_{-0.12}$ & $-3.7\pm1.7$&$-130\pm60$&-10.78& $0.508^{+0.006}_{-0.008}$&$0.478^{+0.006}_{-0.008}$ & 2.02 &430/375 \\
\\
220 ks & $11.24^{+0.12}_{-0.10}$ & $-2.7\pm1.4$&$-90\pm50$& -8.48&$0.476^{+0.008}_{-0.007}$ & $0.445^{+0.008}_{-0.007}$& 2.07 & 384/386 \\
 & $14.15^{+0.08}_{-0.14}$ & $-3.6\pm1.7$&$-170\pm80$&-12.20&$0.634^{+0.004}_{-0.006}$ &$0.610^{+0.004}_{-0.006}$ & &\\
 \\
278ks &$10.71^{+0.08}_{-0.11}$ & $-2.7\pm1.6$&$-70\pm45$& -8.87&$0.437^{+0.006}_{-0.008}$ & $0.405^{+0.006}_{-0.008}$& 2.10 & 369/376 \\
 & $11.46\pm0.13$ & $-3.8^{+1.6}_{-1.3}$&$-120^{+50}_{-30}$&-13.19&$0.490^{+0.008}_{-0.009}$ & $0.460^{+0.008}_{-0.009}$& &\\ 
\\
295 ks&$11.65^{+0.08}_{-0.05}$ & $-4.6\pm1.5$&$-150\pm50$&-22.99&$0.503^{+0.006}_{-0.005}$ &$0.473^{+0.007}_{-0.005}$ & 3.00 & 461/409 \\
\hline
\end{tabular}
\end{center}
\caption{\label{bestfitPar} Best fit parameters of the time-resolved XMM-{\it Newton} and XMM-{\it Newton}+{\it NuSTAR} analysis (top and bottom box, respectively). Energies are in keV and in the rest-frame of the source (z=0.00771). The statistical significance of each absorption line is determined via Monte Carlo simulations, see text for details.}
\end{table*}

Table \ref{bestfitPar} reports the best fitting energies and fluxes of the absorption lines, the $\Delta\chi^2$ improvement and the overall $\chi^2/\nu$ value. For the XMM+{\it NuSTAR} spectra corresponding to 220 ks and 278 ks we include also the absorption lines detected with XMM-{\it Newton} alone, despite their lower statistical significance (at 11.24$^{+0.12}_{-0.10}$ and 10.71$^{+0.08}_{-0.11}$ keV, respectively).
The non-detection of these lines in {\it NuSTAR} spectra can be explained in terms of their shorter exposure times and lower spectral resolution at these energies with respect to XMM-{\it Newton}. As a consistency check, we fit the unbinned pn spectra together with the FPMA/B spectra with a fixed 200 eV energy binning, leaving the normalisation of the lines free to vary between the three detectors and using the Cash statistics \citep{cash76}. The inferred normalisations and upper limits are always consistent with each other.

The statistical significance of each detected absorption line listed in in Tab. \ref{bestfitPar} is then estimated via Monte Carlo simulations, using 1000 fake data sets for each spectrum and the procedure outlined above. We report the inferred significances in Tab. \ref{bestfitPar}, ranging between 1.76$\sigma$ and 3.00$\sigma$, in Fig. \ref{blind} and \ref{blind2} for the XMM-{\it Newton} and joint XMM-{\it Newton}+{\it NuSTAR} observations, respectively.

\begin{figure*}
\begin{center}
\includegraphics[width=0.65\columnwidth]{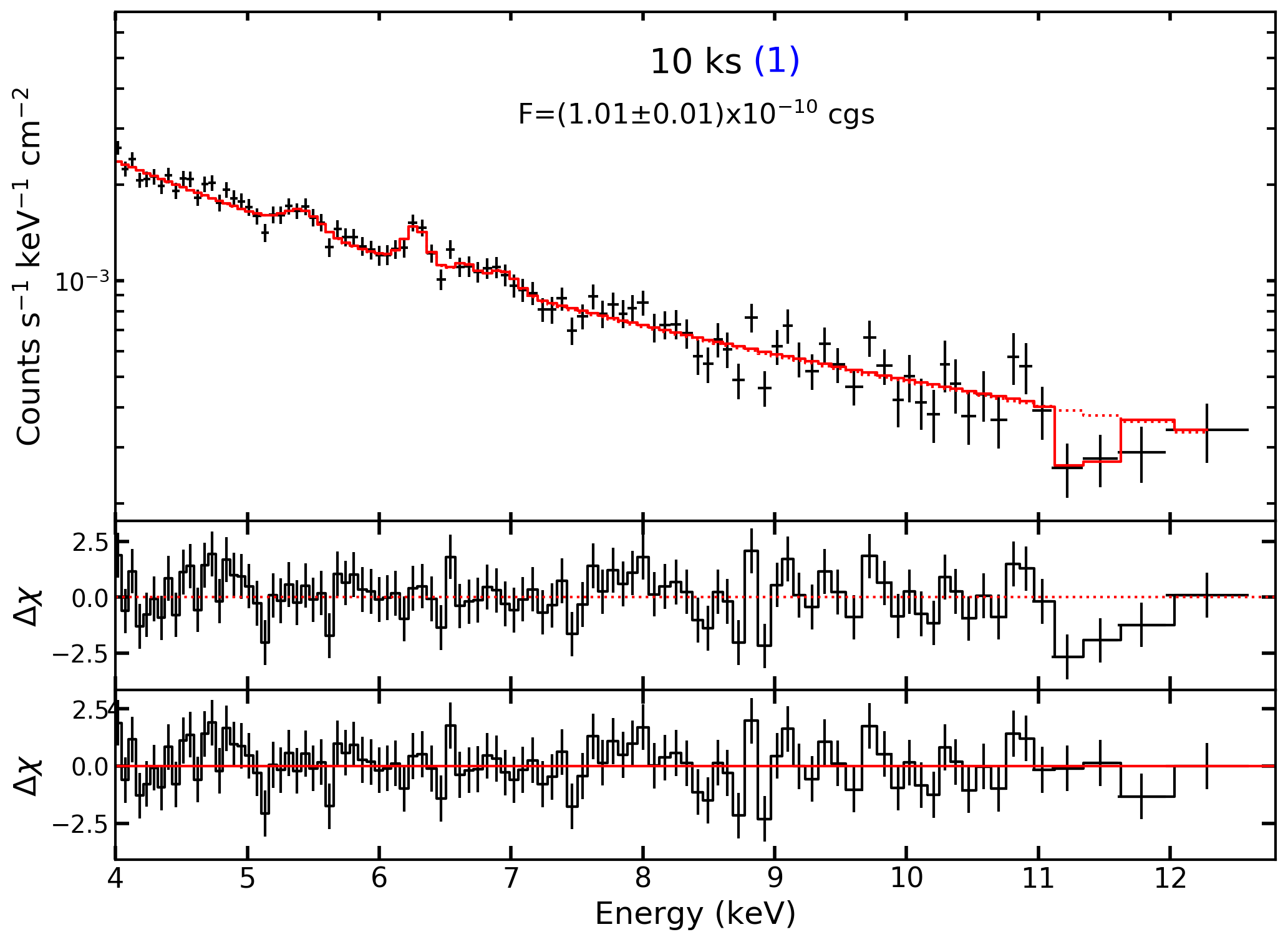}
\includegraphics[width=0.65\columnwidth]{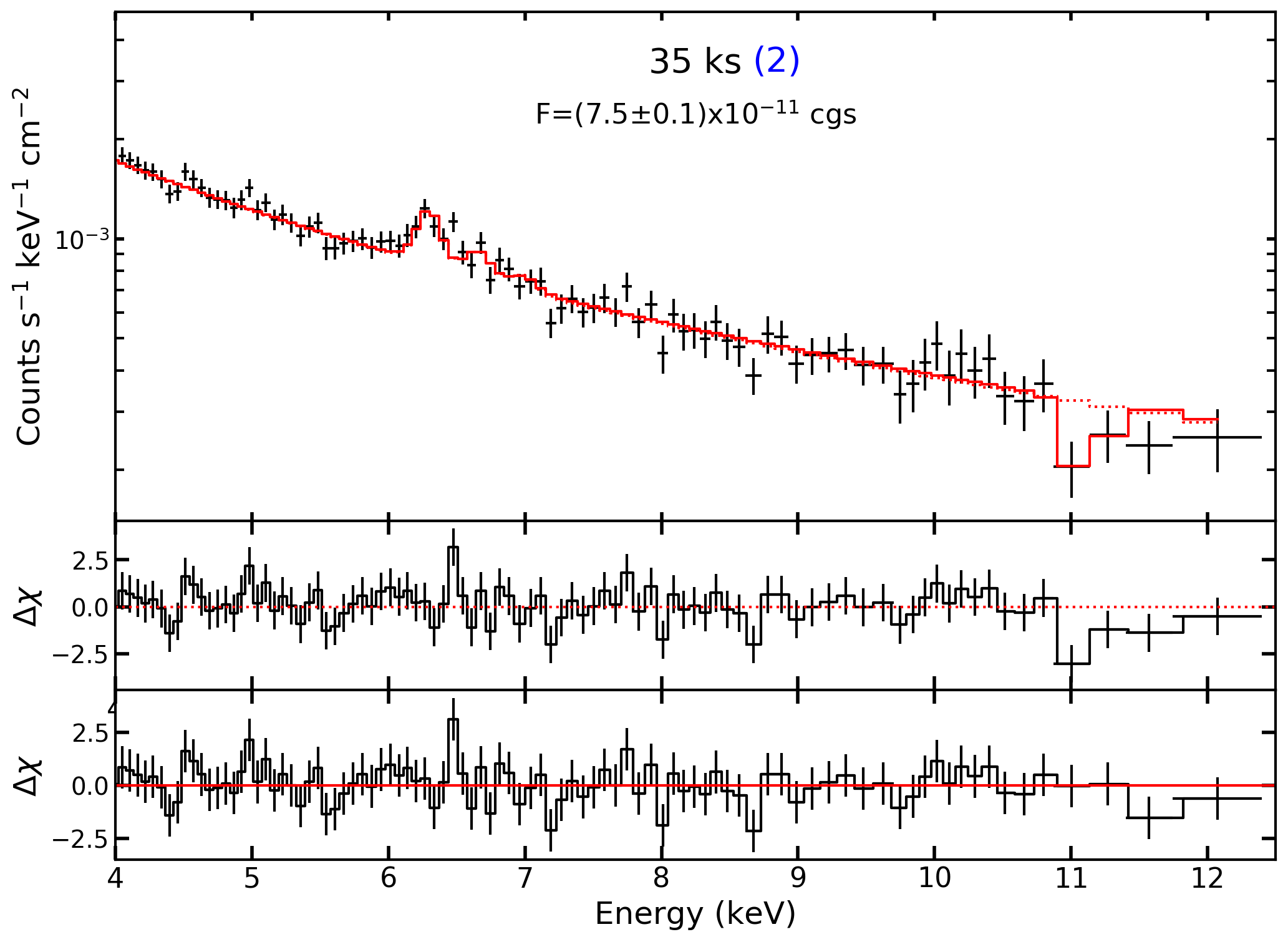}
\includegraphics[width=0.65\columnwidth]{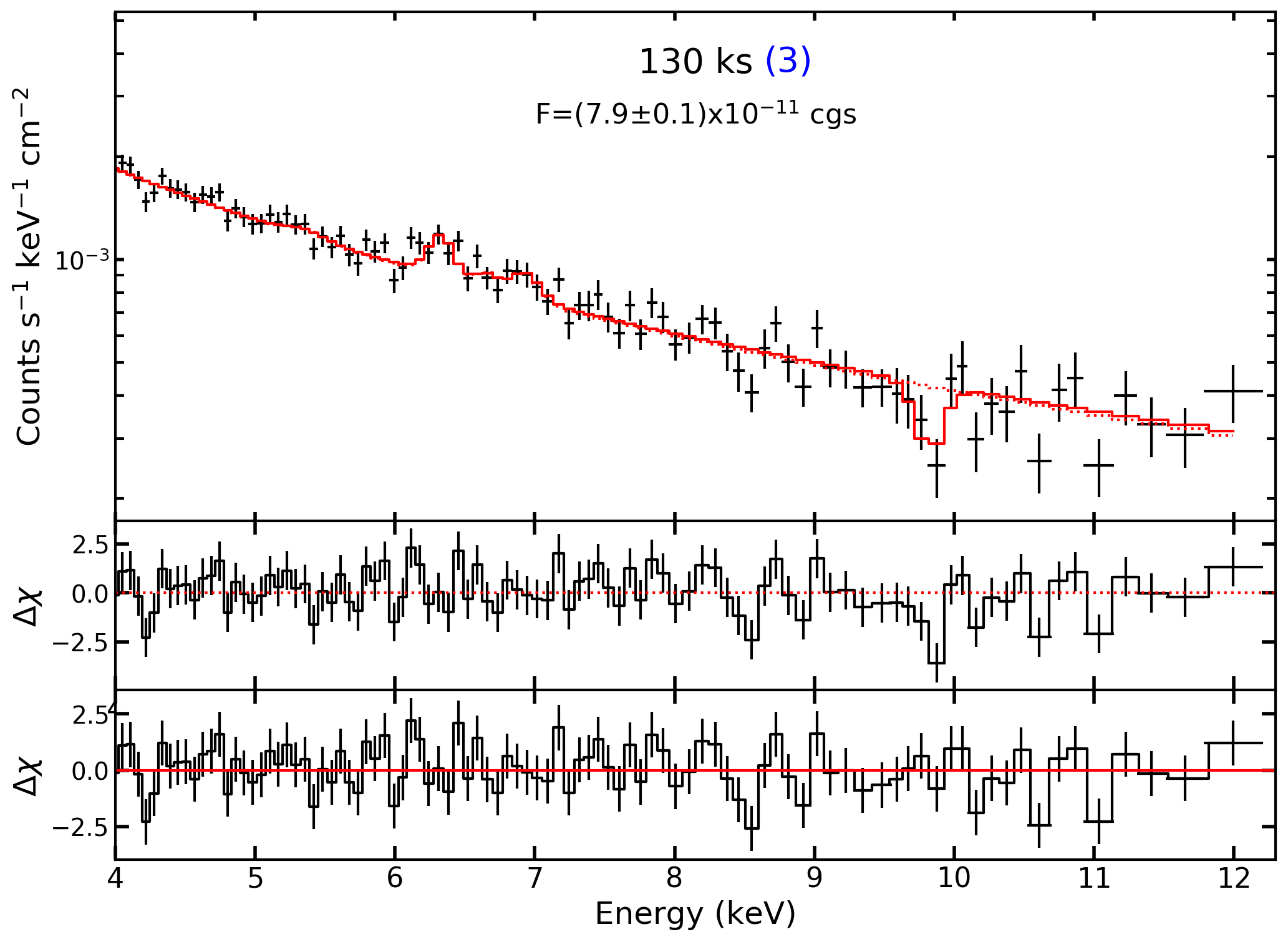}
\includegraphics[width=0.65\columnwidth]{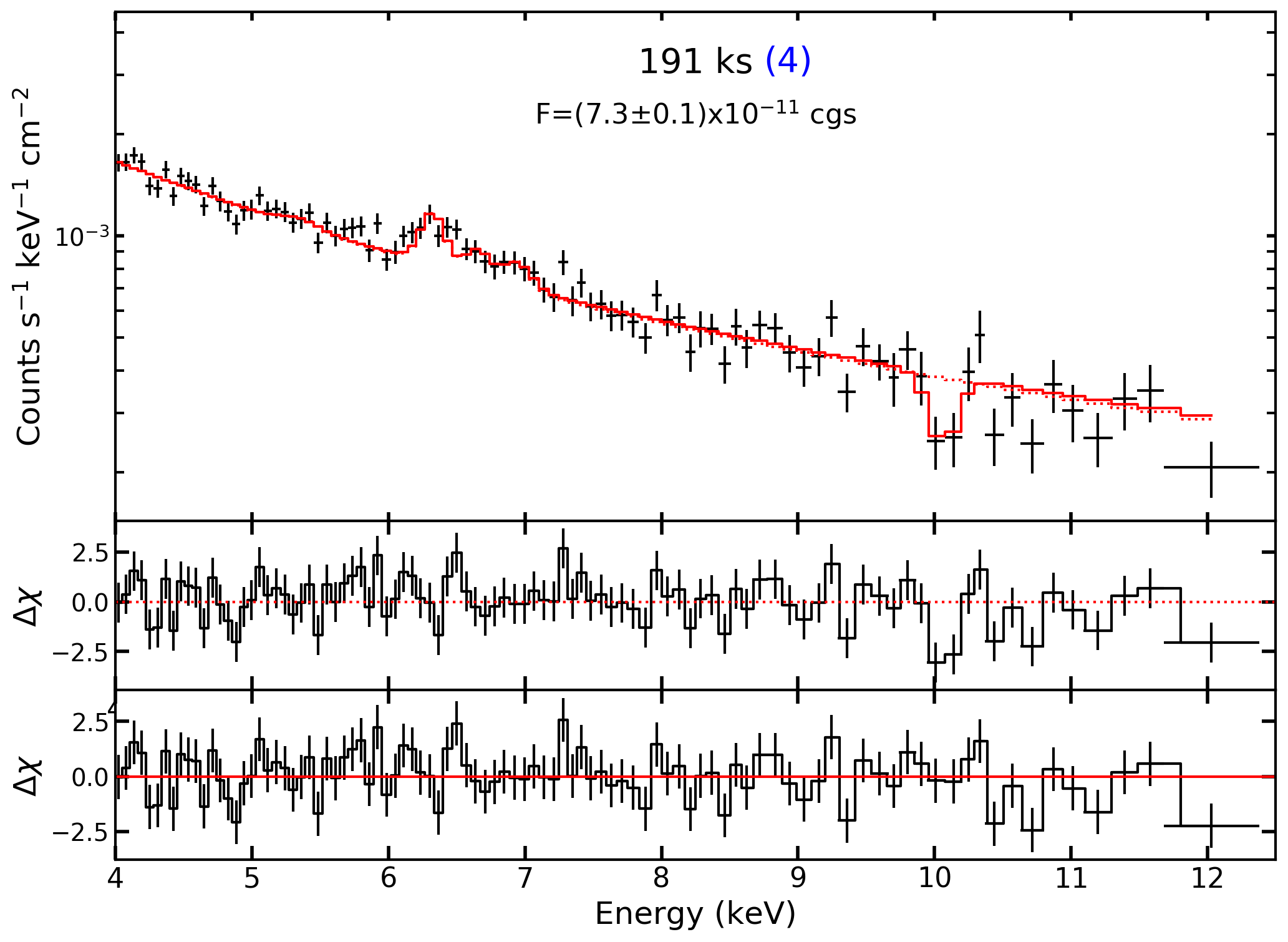}
\includegraphics[width=0.65\columnwidth]{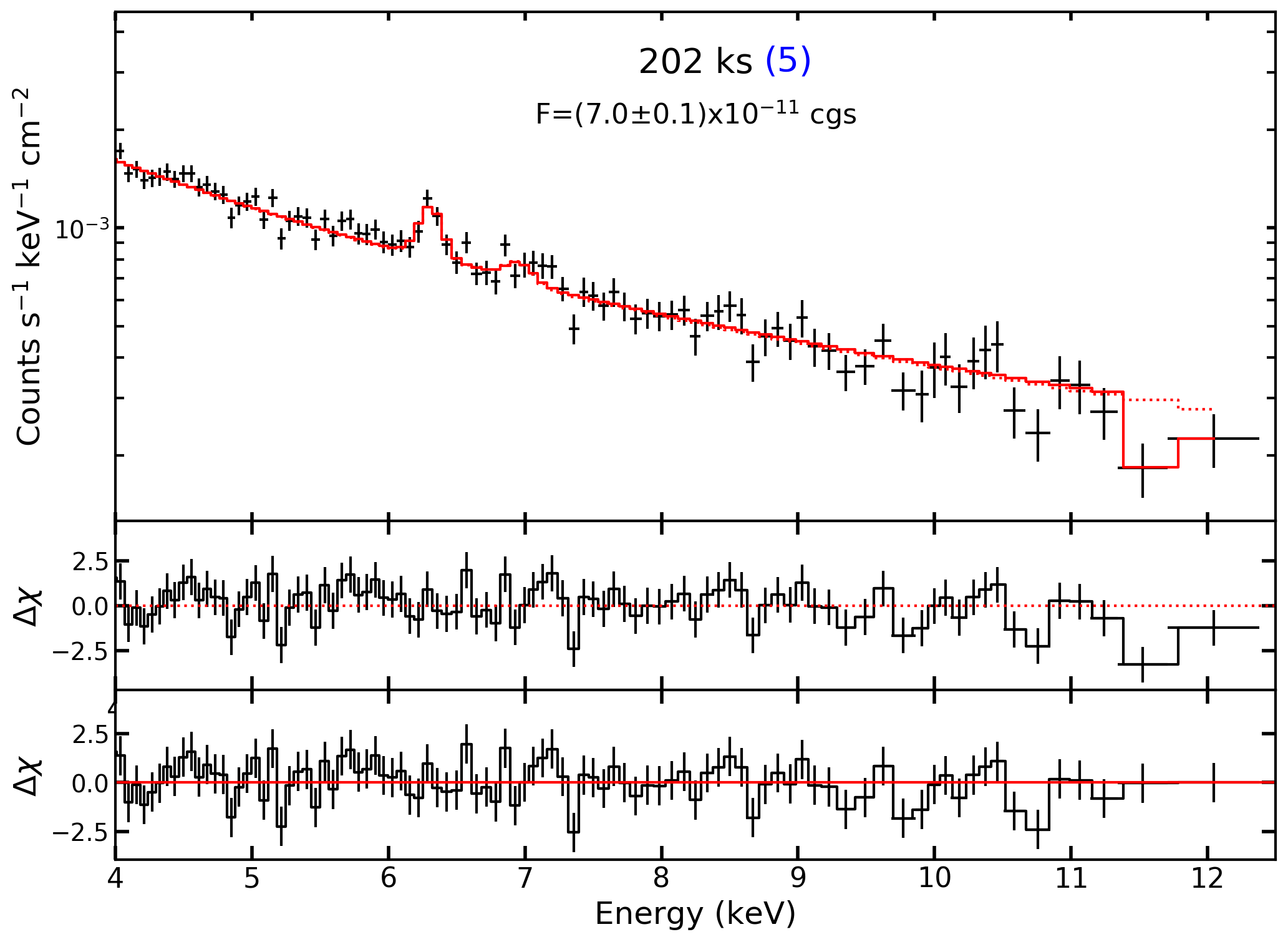}
\includegraphics[width=0.65\columnwidth]{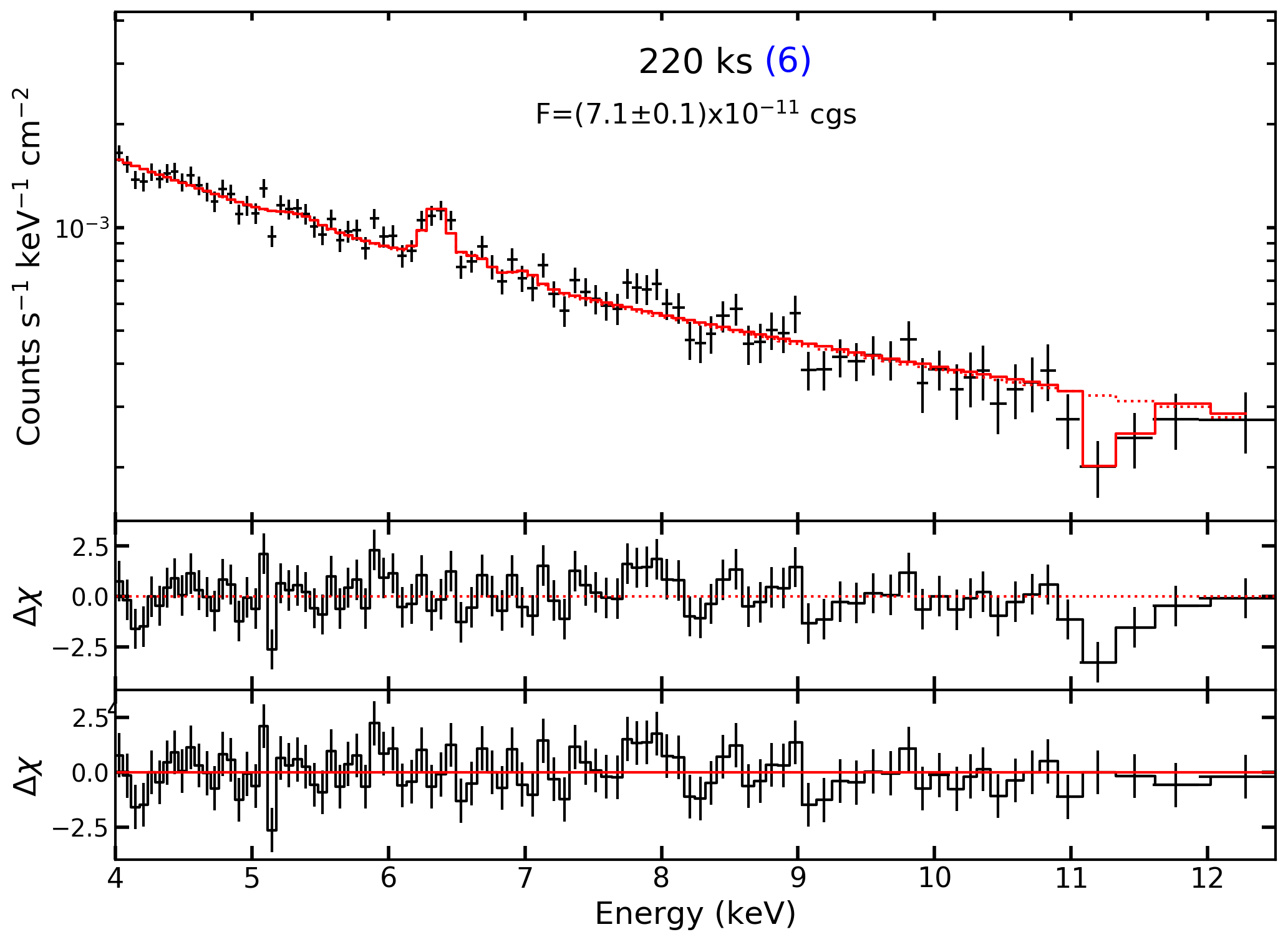}
\includegraphics[width=0.65\columnwidth]{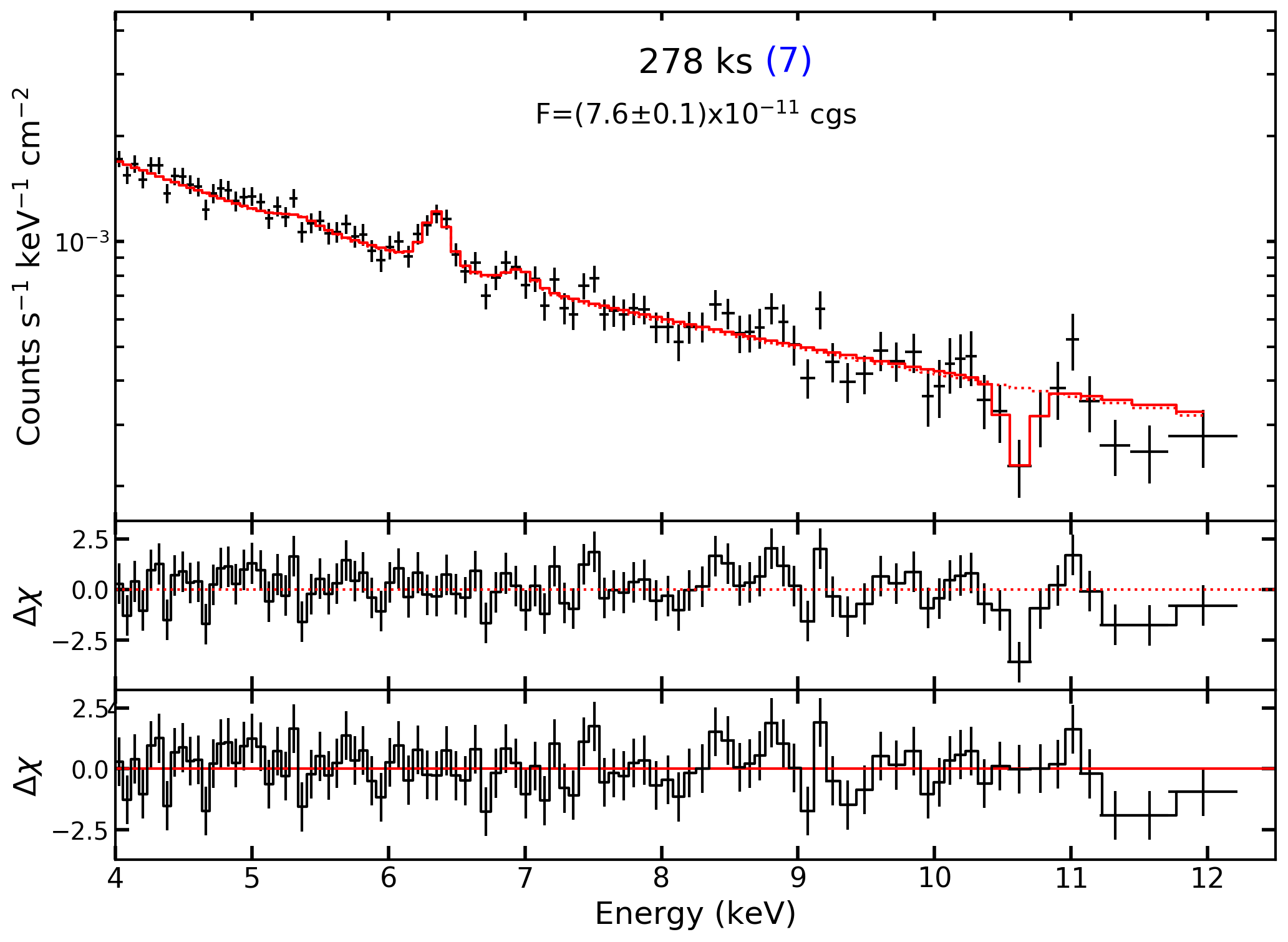}
\includegraphics[width=0.65\columnwidth]{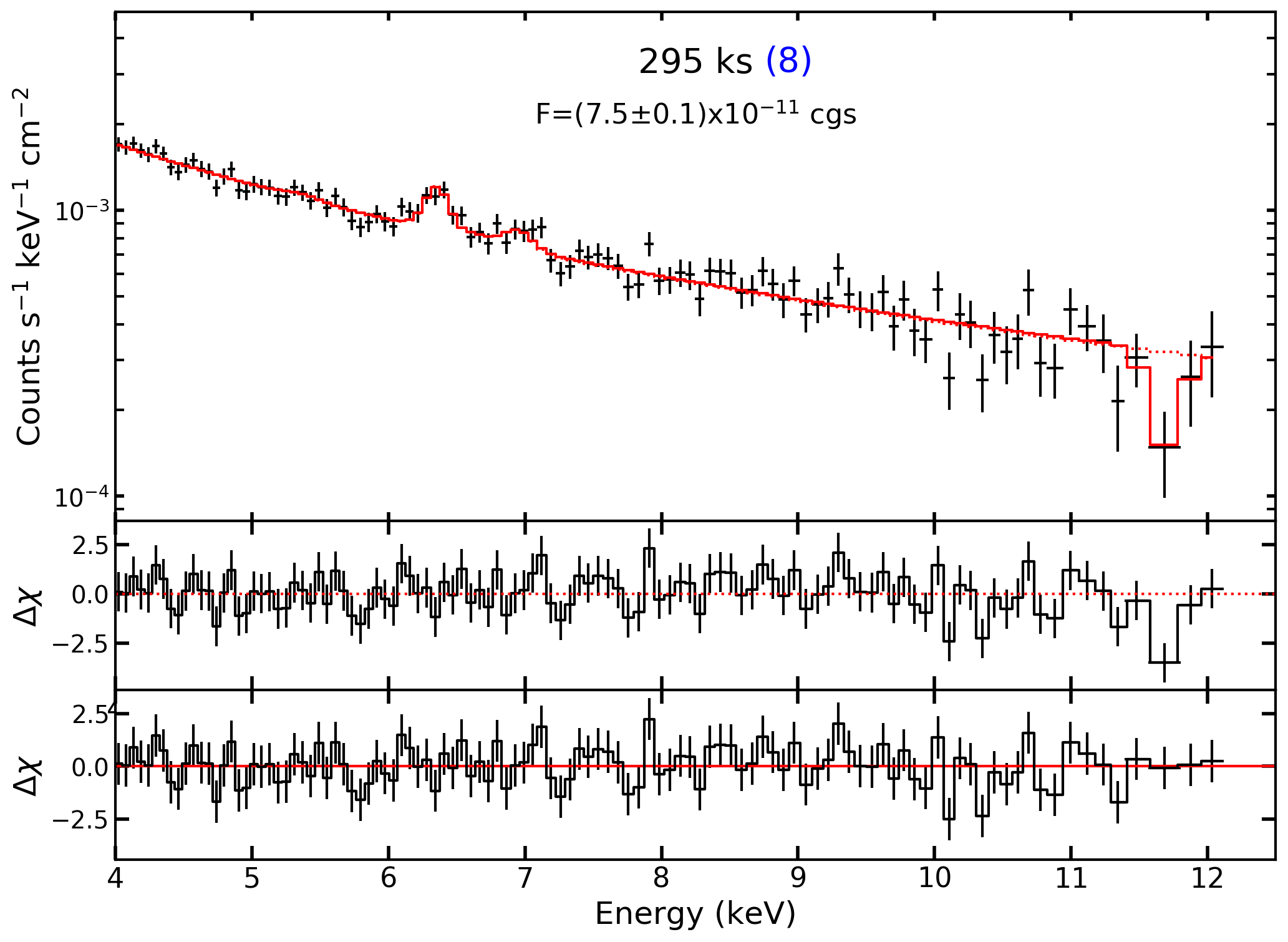}
\includegraphics[width=0.65\columnwidth]{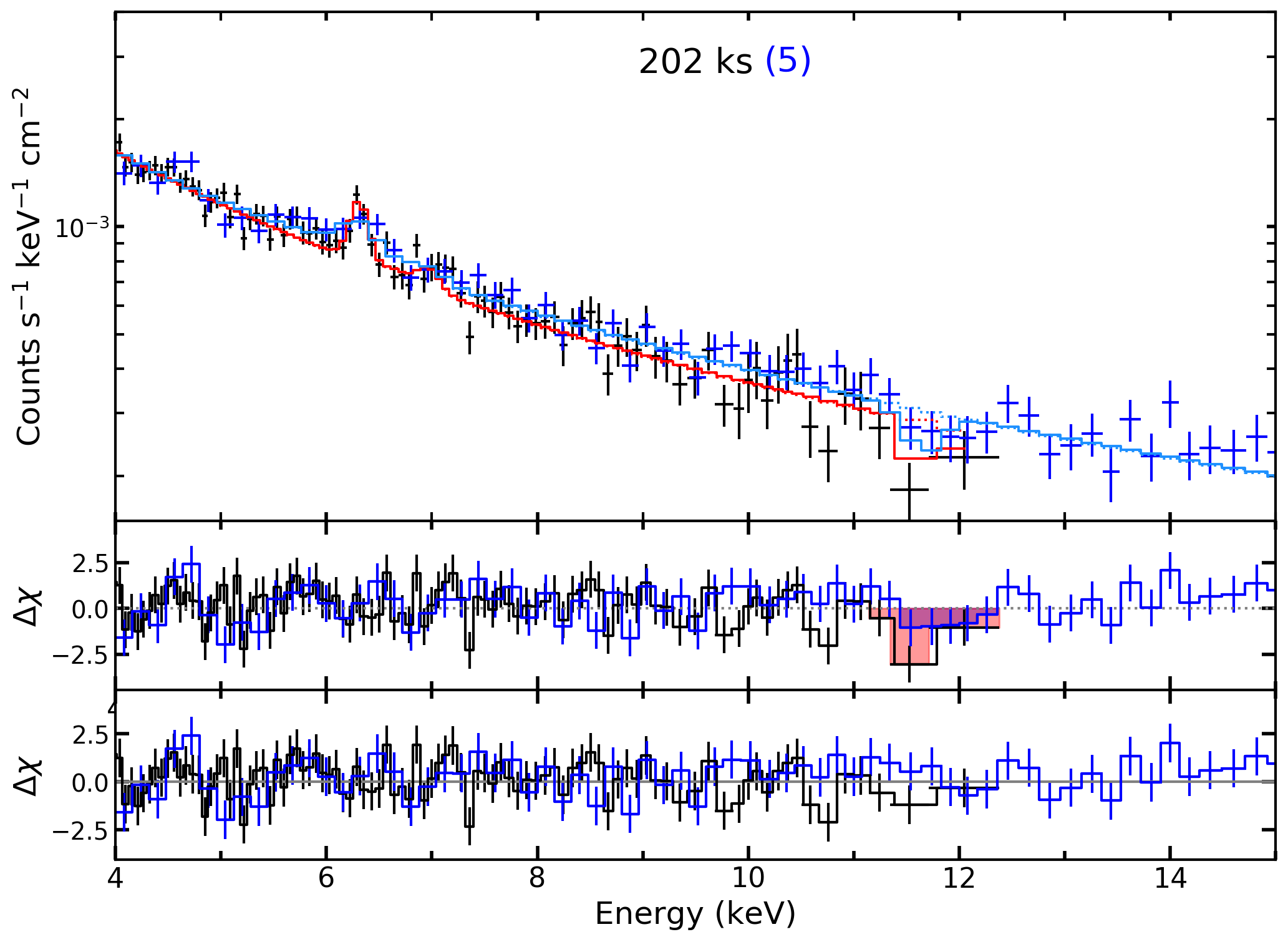}
\includegraphics[width=0.65\columnwidth]{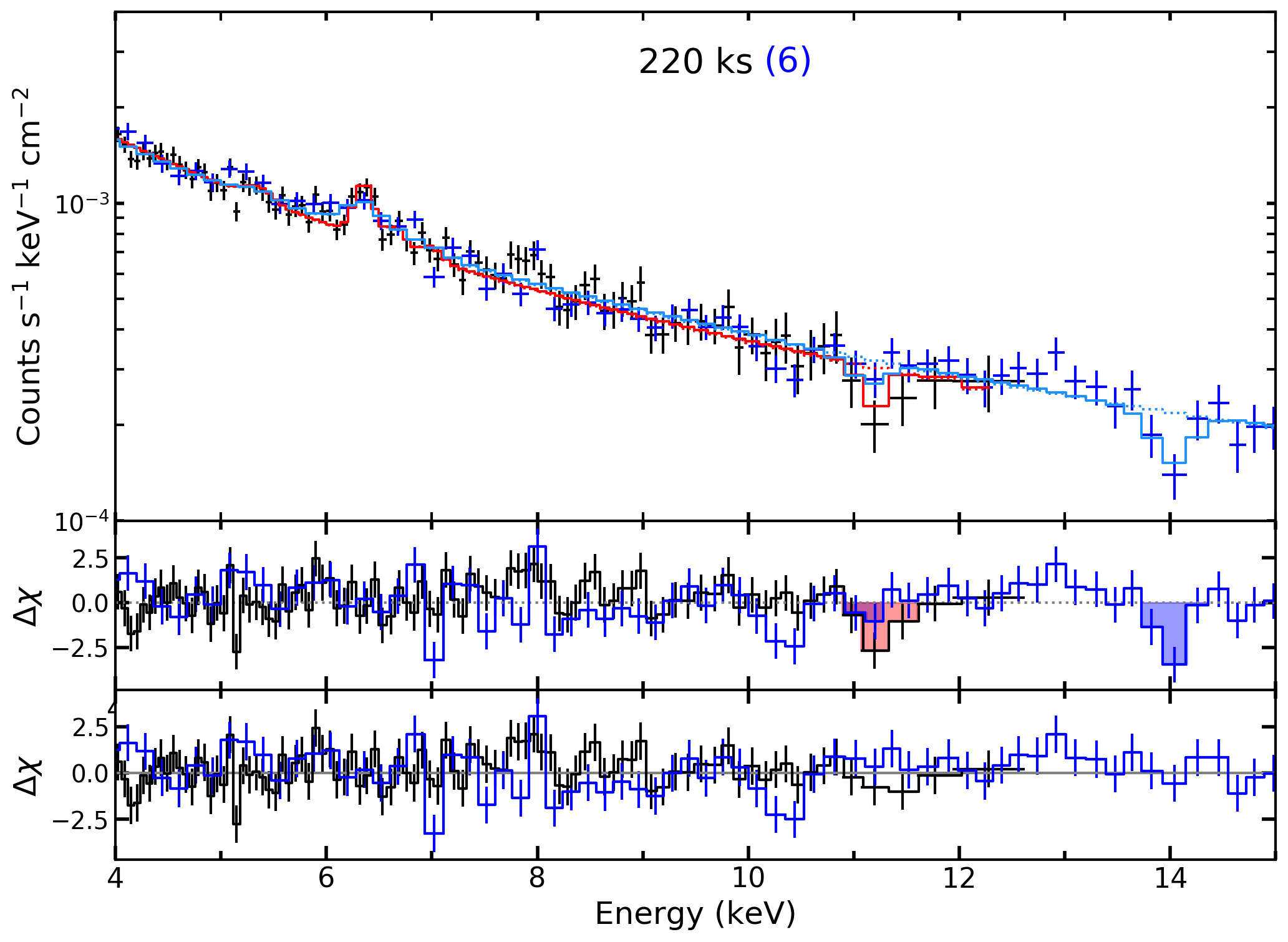}
\includegraphics[width=0.65\columnwidth]{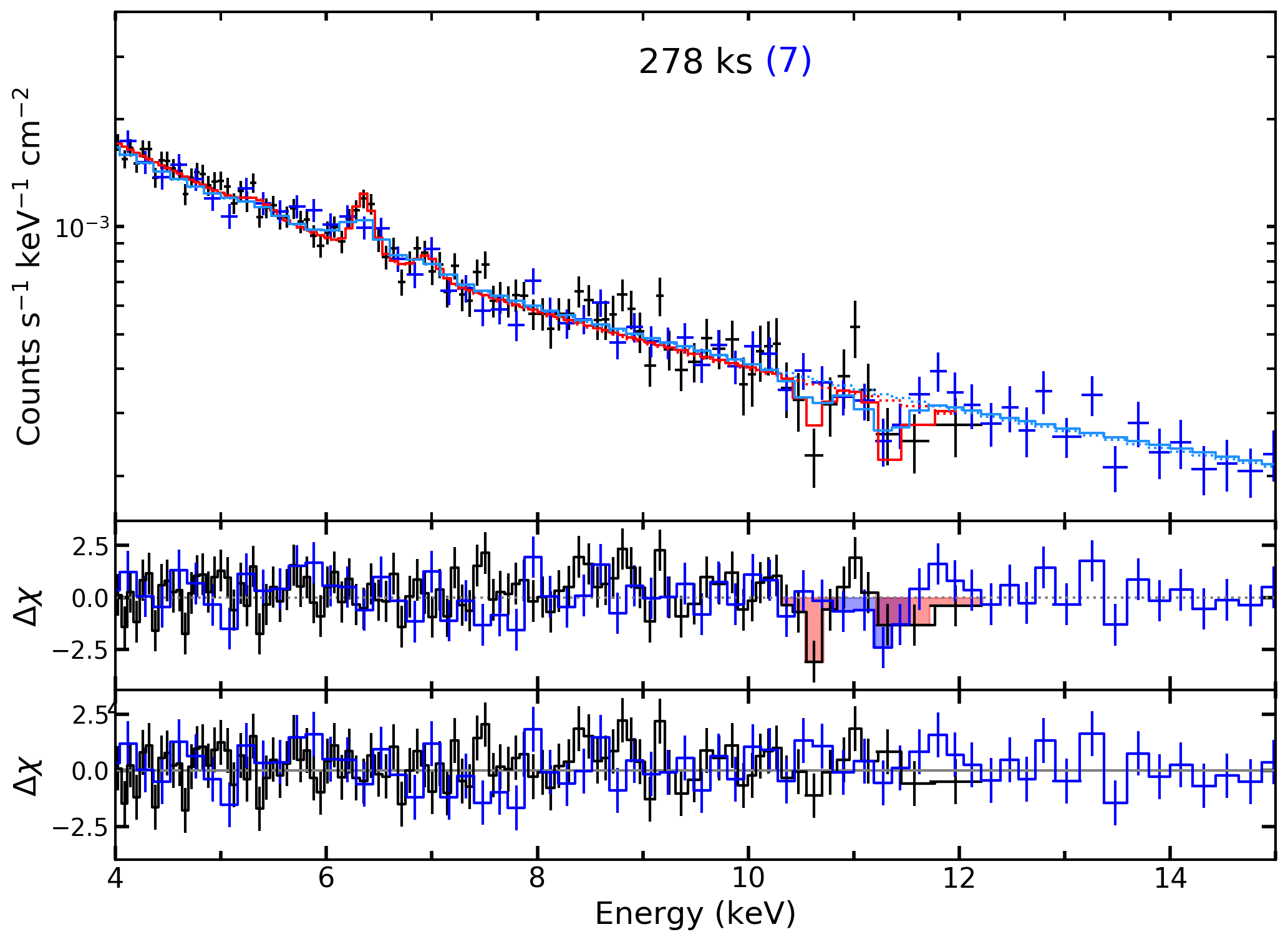}
\includegraphics[width=0.65\columnwidth]{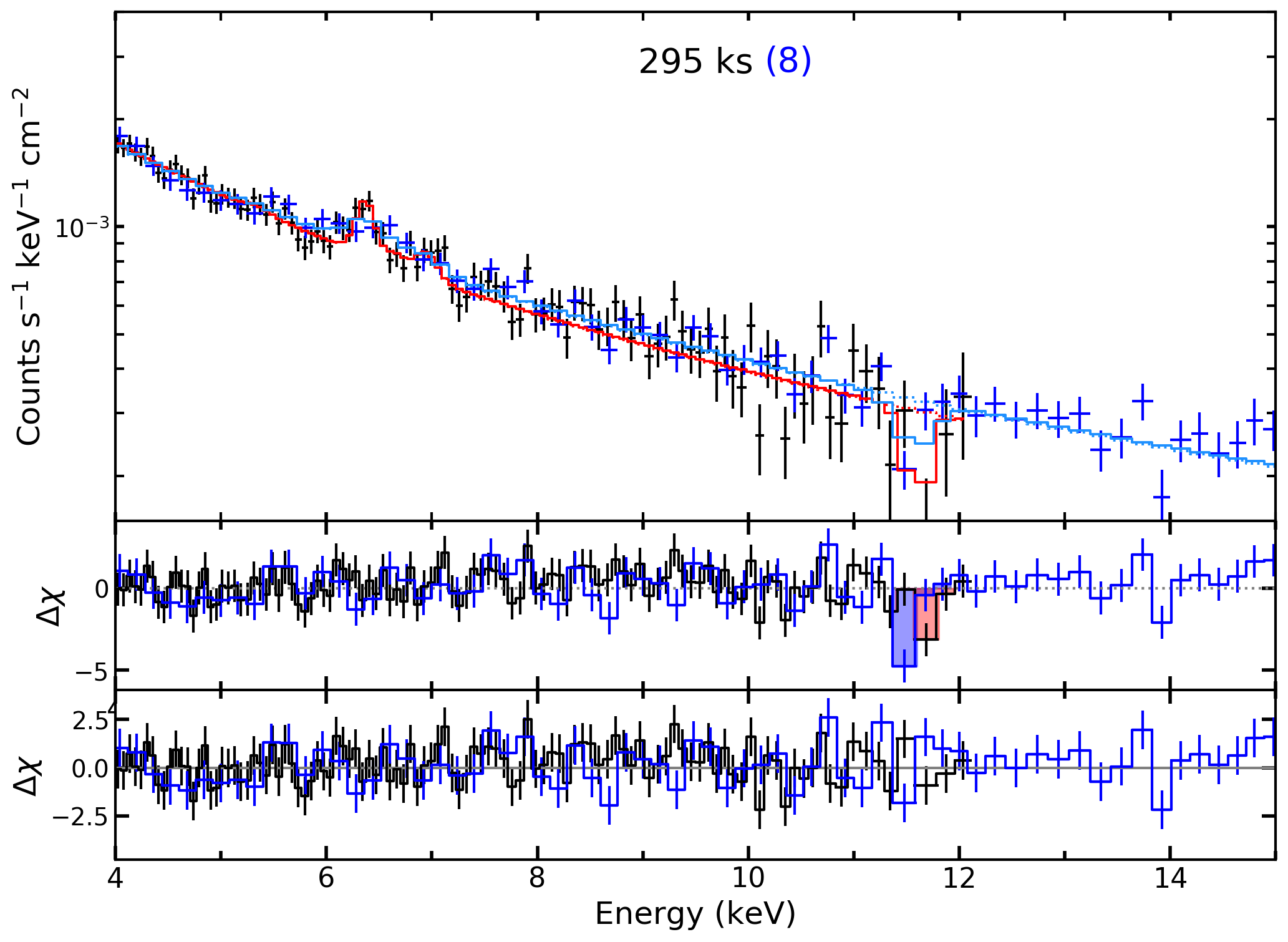}
\includegraphics[width=0.7\paperwidth]{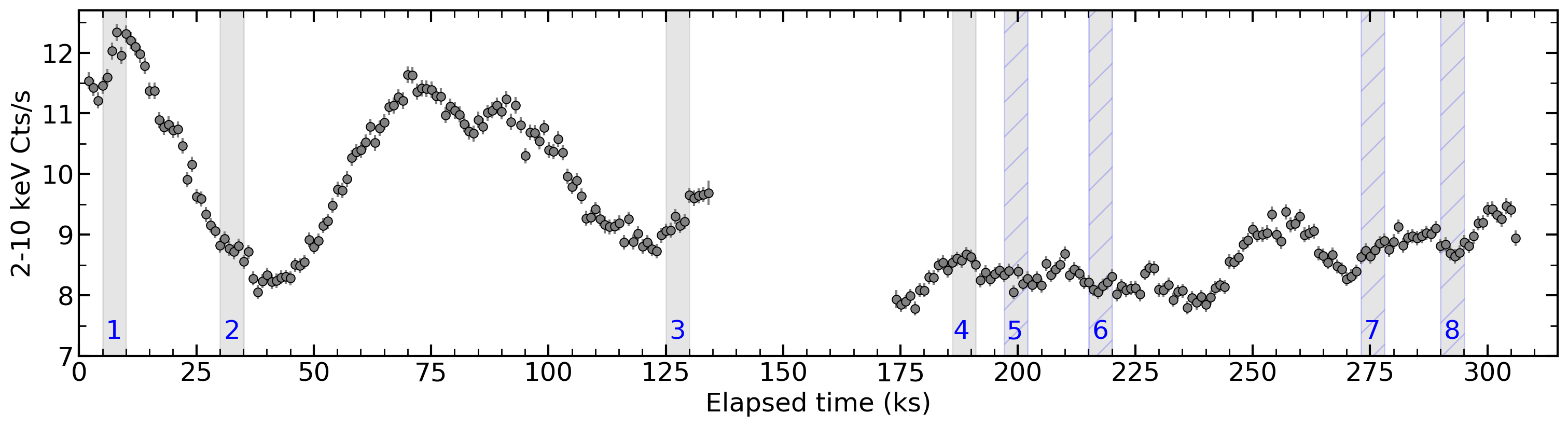}
\end{center}
\caption{{\it From top to bottom and left to right:} First seven boxes: top panels show the EPIC pn spectra with a $>2 \sigma$ significance absorption line. Middle and bottom panels include residuals to a fit using the continuum model and the continuum model plus an absorption line, indicated with dotted and solid red lines, respectively. Last four boxes: EPIC pn and the combined FPMA/B spectra (black and blue lines) with a $>2\sigma$ significance absorption line. {\it Bottom panel:} 2-10 kev EPIC pn light curve, with 1 ks time binning. The eight time intervals with $\sigma>2$ absorption features are numbered and plotted as grey shaded regions. Blue stripes indicate the presence of {\it NuSTAR} spectra as well.}
\label{bfiti}
\end{figure*}

We show in the top panels of Fig. \ref{bfiti} the eight EPIC pn spectra with an absorption line with a significance $> 2 \sigma$ together with the residuals with respect to a spectral model without and with the absorption line, respectively. Similarly, in the four bottom panels we show the fits to the joint XMM-{\it Newton}+{\it NuSTAR} observations with $> 2 \sigma$ significance absorption lines, indicating in red and blue the residuals due to absorption lines in the EPIC pn and FPMA/B datasets, respectively. For visual clarity we plot the combined FPMA/B spectra ({\sc setplot group} command in {\sc Xspec}). {As a summary of our findings, we plot in Fig. \ref{bfiti} (bottom panel) the eight time intervals in which an absorption line is detected on top of the 2-10 keV EPIC pn light curve.}

To estimate the statistical significance of the absorption lines in the global set of spectra, rather than in each single one, we simulate the entire set of 50 XMM-{\it Newton} time slices via Monte Carlo routines, using the same procedure outlined above. For each time slice we set input parameters drawn from a Normal distribution around the best fit values excluding the Gaussian absorption component. Then, we fit again all the time slices and check how many spurious absorption components are detected. We simulate the entire set 1000 times and we find that, on average, 2.4 spurious absorption lines are detected at a significance level equal or higher than our Monte Carlo-derived 2 $\sigma$ threshold, corresponding to a fit improvement $\Delta \chi^2<-11$ (see Table \ref{bestfitPar}). We note that 5 out of 8 detections in the observed spectra have a significance $\Delta \chi^2<-13$, while for such significance only 1.2 spurious lines, on average, are detected in the simulated set. Fig. \ref{setsim} (Appendix \ref{appsim}) reports the full set of results and the average number of spurious detections for different$\Delta \chi^2$ thresholds.

Most of the variable absorption lines are detected above 10 keV, an energy range in which the effective area of the EPIC-pn detector significantly decreases. To assess the impact of possible calibration effects we analyse the EPIC-pn spectrum of the Blazar 3C 273, which is well known for showing a simple, featureless continuum (see Appendix \ref{AppendixBACK}). The spectrum does not show any deviation from a simple powerlaw continuum or absorption features, further demonstrating that the absorption features in NGC 2992 cannot be ascribed to instrumental effects.

\section{The {\sc WINE} model}
\label{winesection}
\subsection{Overview of the code}
\label{code overview}
To get a complete characterisation of the outflow we apply WINE to all the time intervals with UFO signatures detected with a significance $>2 \sigma$.

WINE is a self-consistent, physically motivated model for wind absorption and emission profiles. We give here a brief overview of the model and we refer to \cite{llt21} and Luminari et al., in prep.) for a comprehensive description.
We represent the wind as a series of thin slabs and we start radiative transfer from the innermost one using the {\sc XSTAR} photoionisation code \citep{xstar} and providing the required parameters, i.e. the inner radius $r_0$, the slab column density $\delta N_H$ and $\xi_0$, the ionisation parameter at $r_0$. The density profile of the wind $n(r)$ is parameterised through the exponent $\alpha$ such that $n(r)=n_0(r_0/r)^{\alpha}$, while $n_0$ can be derived by inverting the definition of the ionisation parameter:
\begin{equation}
\xi_0 \equiv L'_{ion}/(n_0 r_0^2)
\label{xi_definition}
\end{equation} where $L'_{ion}$ is the incident ionising luminosity in the energy range between 13.6 eV and 13.6 keV in the gas reference frame which, together with the incident spectrum, is a proxy of the AGN luminosity in {\sc XSTAR}. Similarly to the density, we implement a powerlaw behaviour for the wind velocity as $\rm{v}(r)=\rm{v_0}(r_0/r)^{\zeta}$, where $\zeta$ is a free parameter of the model. Then, we propagate the simulation to the second slab, using the transmitted spectrum and luminosity as the incident ones and calculating analytically $\xi, r, n, v$. We iterate the procedure up to the $k$-th slab, so that the total wind column density $N_H=k \cdot \delta N_H$ is reached. 

This slicing allows us to reproduce the scaling of the wind properties, including its velocity profile. Given that {\sc XSTAR}, as well as many other photoionisation codes (e.g. Cloudy, \citealp{fcg17} and SPEX, \citealp{spex}), assumes a null gas outflowing velocity v, we implemented a procedure to take into account v in the radiative transfer calculations. This procedure is carried out in a fully special relativity framework and represents a major novelty of WINE. Given the mildly relativistic velocities usually displayed by UFOs, relativistic effects lead to sizeable effects on the appearance of both emission and absorption profiles, which must be properly accounted for to correctly estimate the wind properties, particularly its $N_H$ (see \citealp{ltp20} for a detailed description). As a result of these effects, $L'_{ion}$ will be in general different (i.e., lower) than the rest frame measured luminosity, $L_{ion}$. In particular, in the case of a powerlaw incident spectrum with photon index $\Gamma$, the luminosity in the gas frame can be written as $L'_{ion} = \Big( \frac{1-\rm{v}}{1+\rm{v}} \Big)^{\frac{2+\Gamma}{2}} \cdot L_{ion}$, where v is in units of c. Moreover, line emissivities calculated by {\sc XSTAR} are convolved for each slab with Monte Carlo profiles to accurately represent the wind emission spectrum as a function of its geometry, as well as its ionisation, column density and velocity.
This allows us to constrain the presence of wind emission components with higher accuracy with respect to XSTAR or other general-purpose photoionisation codes. However, we find that the inclusion of emission features is not statistically supported in any of the time slices, possibly due to a small wind covering factor and/or the limited signal-to-noise ratio of the spectra.
For the same reason we do not implement a detailed velocity profile, but we rather set $\zeta \equiv 0$, i.e. a constant velocity. As a result, the free parameters of the WINE model, which will then be constrained by fitting the model to the data, are:
\begin{itemize}
\item $\xi_0$, the ionisation parameter at the inner boundary of the wind
\item $N_H$, the wind column density
\item v$_0$, the outflowing velocity
\end{itemize}
The remaining free parameters of the model, i.e. $r_0, \alpha$, are fixed to r$_0$=5 r${\rm_S}$, $\alpha$=0, as explained in Sect. \ref{winetables}. Finally, the AGN ionising SED is described through a powerlaw with $\Gamma=1.7$ and 2-10 keV luminosity $L_{2-10}=1.0 \cdot 10^{43}$ erg s$^{-1}$ (which imply $L_{ion}=2.67 \cdot 10^{43}$ erg s$^{-1}$), which are the average values for the time slices analysed (see Paper I).

\begin{figure*}
\centering
\includegraphics[width=0.8\paperwidth]{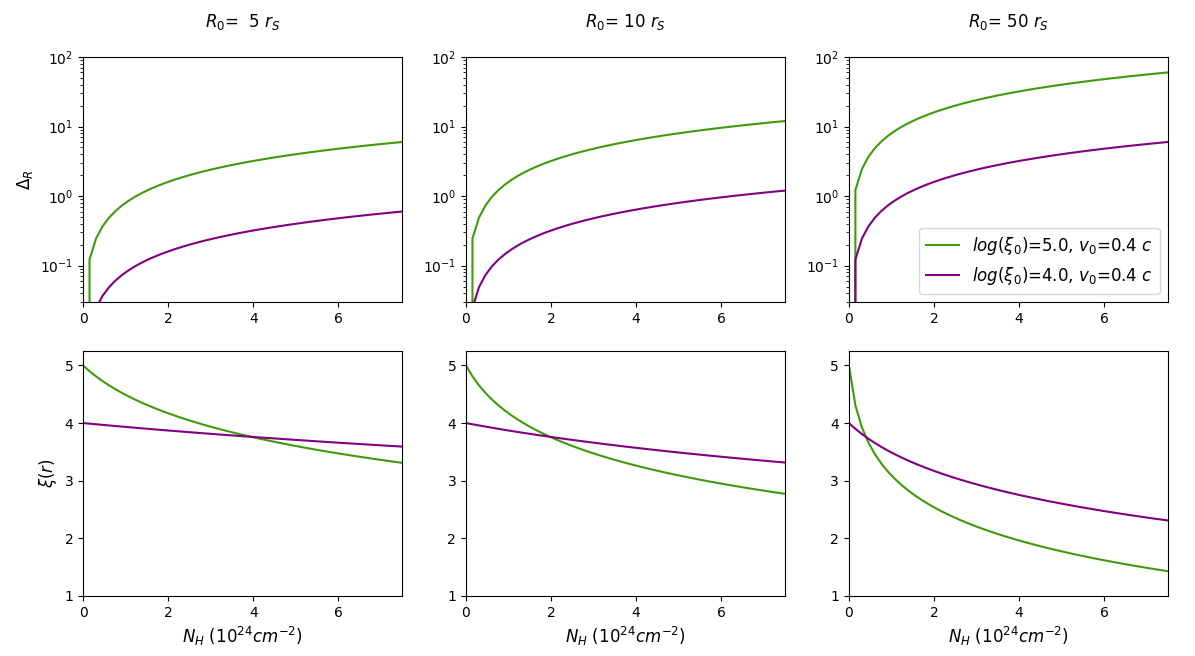}
\caption{Wind radial thickness $\Delta_r$ ({\it top}) and ionisation parameter profile $\xi(r)$ ({\it bottom}) as functions of the wind column $N_H$, for increasing $r_0$ (from left to right, $r_0=5,10,50 r_S$).}
\label{teofig}
\end{figure*}

\subsection{Absorption tables for NGC 2992}
\label{winetables}
The specific set of {\sc Xspec} tables used in this work has been tailored on the (rather extreme) properties of the outflow in NGC 2992. In order to fully understand the behaviour of the wind as a function of its $\xi_0, N_H, v_0$ it is instructive to examine the wind radial thickness. We define the thickness as $\Delta_r = \frac{r(N_H)-r_0}{r_0}$, i.e., the difference between the radius enclosing a column density $N_H$ and $r_0$, normalised by $r_0$. According to $\alpha$, $\Delta_r$ can be written as a function of the initial parameters as follows:
\begin{equation}
\Delta_r = \begin{cases} ( 1- \frac{N_H}{n_0 r_0 (\alpha-1)} \big)^{\frac{1}{1-\alpha}} -1 & \mbox{$\alpha \neq 1$} \\
\exp \Big( {\frac{N_H}{n_0 r_0}} \Big) - 1 & \mbox{$\alpha=1$} \\
\end{cases}
\label{expr_deltar}
\end{equation}
It can be seen that, as expected, $\Delta_r$ increases for increasing $\alpha$. In particular, for a constant density profile ($\alpha=0$), the thickness can be written as:
\begin{equation}
\begin{split}
\frac{N_H}{n_0 r_0} & = \frac{N_H \xi_0 r_0}{L'_{ion}} = \frac{N_H \xi_0 r_0}{\Big( \frac{1-\beta}{1+\beta} \Big)^{\frac{2+\Gamma}{2}} L_{ion}} \\ 
& \approx 2.35\cdot 10^{-33} \Big( \frac{1+\beta}{1-\beta} \Big)^{\frac{2+\Gamma}{2}} N_H \xi_0 r_{0,S}\ \frac{1}{ \lambda_{ion}} \\
& = 0.17\Big( \frac{1+\beta}{1-\beta} \Big)^{\frac{2+\Gamma}{2}} \cdot \frac{N_H}{10^{24}} \cdot \frac{\xi_0}{10^5} \cdot \frac{r_{0,S}}{5}
\label{dr}
\end{split}
\end{equation}
where $L'_{ion}$ is the relativistic-corrected luminosity in the wind reference frame (see Sect. \ref{code overview}), $r_{0,S}$ is the launching radius in units of the Schwarzschild radius ${\rm r_S= 2 r_g = 2 G M_{BH}/c^2}$, $\beta=\rm{v}_0/c$ and $N_H, \xi_0$ are expressed in units of ${\rm cm^{-2},\ erg\ cm\ s^{-1}}$, respectively. In the last step we assume $\lambda_{ion}=7 \cdot 10^{-3}$, as appropriate for NGC 2992 (see Paper I). The relativistic correction is expressed by the term $\Big( \frac{1+\beta}{1-\beta} \Big)^{\frac{2+\Gamma}{2}}$, which for v$_0$=0.4 c, $\Gamma=1.7$ corresponds to 4.8.

In Fig. \ref{teofig} we show $\Delta_r$ and $\xi(r)$ (top and bottom panel, respectively) as a function of $N_H$, up to $7.5 \cdot 10^{24} {\rm cm^{-2}}$, for v$_0$=0.4 c, $\log \Big( \frac{\xi_0}{{\rm erg\ cm\ s^{-1}}} \Big)=4.0, 5.0$; these values are representative of those reported in Table 1 and of the best fit values obtained with WINE (see later). From left to right $r_0$=5,10,50 r$_S$. The very low normalised luminosity of NGC 2992, $\lambda_{ion}=7 \cdot 10^{-3}$, together with the high $N_H$ and v$_0$, contribute to significantly increase the radial extension of the flow. This, in turn, produces a rapidly decreasing ionisation profile through the wind column, due to geometric dilution of the incident radiation flux. As a result, to reproduce the observed high $\xi, N_H$ we need a very low $r_0$, of the order of 5 r$_S$. Although this low value may rise questions about its physical meaning, it must be primarily intended as a numerical strategy to represent a very geometrically thin wind, and thus to minimise the decrease of the ionisation parameter and reproduce the high $\xi_0$ of the observations. We discuss this point further in Sect. \ref{r0_determination} using the results from the fits described in Sect. \ref{resolved_results} below. We also note that the wind properties are variable between one time slice and the following one; the observations have a duration of $\approx$ 5 ks, which can be translated in a dynamical length, for a velocity =0.4c, of around 5 r$_S$ (using M$_{{\rm BH}}=3 \cdot 10^7 {\rm M_{\odot}}$, as estimated in Paper I), in agreement with our $r_0$. We also note that an isothermal density profile ($\alpha=2$), as would be expected for a wind expanding in spherical symmetry, would not be able to reproduce such high column densities, since the maximum value would be limited to $N_H= 2 n_0 r_0 = 2 L'_{ion}/ (\xi_0 r_0)=3.2 \cdot 10^{24}$ cm$^{-2}$ for $\xi_0=10^5 {\rm erg\ cm\ s^{-1}}, \rm{v}_0=0.4 {\rm c}, r_0=5 {\rm r_S}$.
The parameter space of the {\sc Xspec} tables is as follows:
\begin{itemize}
\item $N_H$=[0.5,20.0]$\cdot 10^{24}$ cm$^{-2}$ with a 0.5$\cdot 10^{24}$ cm$^{-2}$ step
\item $\log({\xi_0/{\rm erg\ cm \ s^{-1}}})$=[3.00,6.50] with a 0.25 step
\item v$_0$=[0.15,0.55] c with a 0.05 c step
\end{itemize}
Finally, we tried several values for the turbulent broadening $\sigma_v$ of the absorption lines. Although the data are weakly sensitive to $\sigma_v$, we find that $\sigma_v=2500$ km s$^{-1}$ best reproduces the observed lines and we use this value hereafter.

\begin{table*}
\centering
\begin{tabular}{ c c c c c }
 & \multicolumn{2}{c} {1st orbit} & \multicolumn{2}{c} {2nd orbit} \\
\hline
\multicolumn{5}{c}{\sc zwabs}  \\
N$_{\rm H}$ ($10^{22}\ {\rm cm^{-2}}$) & $0.87\pm0.04$ & $0.84\pm0.2$ & $1.02\pm0.03$ & $1.04\pm0.03$\\
\hline
\multicolumn{5}{c}{\sc powerlaw}  \\
$\Gamma$ & $1.67\pm0.01$ & $1.663\pm0.008$ & $1.655\pm0.007$ & $1.658\pm0.007$\\
norm ($10^{-2}$) & $2.17\pm0.04$ & $2.15\pm0.03$ & $1.81\pm0.02$ & $1.82\pm0.02$ \\
\hline
 & {\sc WINE abs} & XSTAR & {\sc WINE abs} & XSTAR \\
$\log\big(\frac{\xi_0}{\rm erg\ cm\ s^{-1}}\big)$ & $4.1^{+0.6}_{-0.2}$ & $4.4^{+0.4}_{-0.6}$ & $4.5\pm0.4$ & $4.02_{-0.08}^{+0.28}$\\
v$_0$ ($c$) & $0.337^{+0.007}_{-0.006}$ & $0.350\pm0.006$ & $0.372_{-0.008}^{+0.007}$ & $0.415\pm0.006$ \\
N$_{\rm H}$ ($10^{24}\ {\rm cm^{-2}}$) & $3.1\pm0.8$ & $4.0^{+4.9}_{-2.5}$ & $3.8^{+1.9}_{-1.6}$ & $1.2^{+1.1}_{-0.4}$\\
\hline
$\chi^2$/dof & 190/151 & 201/151 & 692/466 & 707/466 \\
\hline
\end{tabular}
\caption{Best fit values for the time-averaged spectra, corresponding to the first XMM-{\it Newton} orbit (columns 2,3) and the joint second XMM-{\it Newton} orbit + NuSTAR (columns 4,5). Columns 2 and 4 correspond to the fits using WINE, while 3 and 5 to those using XSTAR.}
\label{fit_avg}
\end{table*}

\begin{table*}
\centering
\begin{tabular}{c c c c c c c c c}
Time frame (ks): & 10 & 35 & 130 & 60+191 & 202 & 220 & 278 & 295 \\
\hline
\multicolumn{9}{c}{\sc zwabs}  \\
N$_{\rm H}$ ($10^{22}\ {\rm cm^{-2}}$)  & $1.1 \pm 0.2$ & 1.0 $\pm$ 0.2 & $0.8 \pm 0.2$ & $0.9 \pm 0.1$ & $1.1 \pm 0.1$ & $1.2 \pm 0.1$ & $1.1 \pm 0.1$ & $1.1\pm0.1$ \\
\hline
\multicolumn{9}{c}{\sc powerlaw}  \\
$\Gamma$ & $1.81\pm0.04$ & 1.69 $\pm$ 0.06 & $1.63\pm0.06$ & $1.64\pm0.04$ & $1.70\pm0.03$ & $1.69 \pm0.03$ & $1.67^{+0.03}_{-0.02}$  & $1.68\pm0.03$ \\
norm $^a$ & $3.2\pm0.2$ & 1.9 $\pm$ 0.2 & $1.9\pm0.2$ & $1.9\pm0.1$ & $1.88\pm0.09$ & $1.83\pm0.09$ & $1.91^{+0.010}_{-0.09}$ & $1.94\pm0.09$ \\
\hline
\multicolumn{9}{c}{\sc WINE abs} \\
$\log\big(\frac{\xi_0}{\rm erg\ cm\ s^{-1}}\big)$ & $4.6 \pm 0.6$ & 4.2 $^{+0.7}_{-0.4}$ & $4.7^{+0.4}_{-0.6}$ & $3.75 \pm 0.2$ & $4.5^{+0.3}_{-0.5}$ & $4.7^{+0.5}_{-0.8}$ & $>$4.5 (6.5) $^b$ & $4.5^{+0.3}_{-0.4}$ \\
v$_0$ (c) & $0.45^{+0.03}_{-0.02}$ & 0.32 $^{+0.02}_{-0.03}$ & $0.21^{+0.01}_{-0.03}$ &  $0.37\pm0.01$ & $0.35^{+0.03}_{-0.02}$ & $0.27_{-0.03}^{+0.02}$ & $0.43_{-0.01}^{+0.02}$ & $0.35\pm0.01$ \\
N$_{\rm H}$ ($10^{24}\ {\rm cm^{-2}}$) & $8.2^{+2.8}_{-2.7}$ &  7.8$^{+4.2}_{-5.2}$ & $6.6^{+4.8}_{-3.3}$  & $5.1^{+3.8}_{-2.3}$ & $4.0^{+2.7}_{-2.8}$ & $5.1^{+2.4}_{-4.2}$ & $5.9^{+2.9}_{-2.0}$ & $5.8^{+1.8}_{-2.5}$  \\
$\log\big(\frac{n_0}{{\rm cm^{-3}}}\big)$ & $10.7 \pm 0.6$ & 11.35$^{+0.4}_{-0.7}$ & $11.0_{-0.4}^{+0.6}$ & $11.8 \pm 0.2$ & $11.0_{-0.3}^{+0.5}$ & $11.0_{-0.5}^{+0.8}$ & $<$10.9 (8.9) $^b$ & $11.0_{_0.3}^{+0.4}$ \\
\hline
$\chi^2$/dof & 126/135 & 102/127 & 131/130 & 177/190 & 432/379 & 399/392 & 378/382 & 448/401 \\
\hline
$\Dot M_{out}$& $5.0_{-1.6}^{+1.7}$ &  3.3$_{-2.2}^{+1.8}$ & $1.9_{-1.0}^{+1.4}$ & $2.5_{-1.1}^{+1.9}$ & $1.9_{-1.3}^{+1.7}$ & $1.8_{-1.5}^{+1.0}$ & $3.4_{-1.2}^{+1.7}$ & $2.8_{-1.2}^{+0.9}$ \\
($10^{25}$ g s$^{-1}$) \\
$\Dot p_{out}$ & $6.7_{-2.3}^{+2.4}$ &  3.2$^{+1.7}_{-2.1}$ & $1.2_{-0.6}^{+0.9}$ & $2.7_{-1.2}^{+2.1}$ & $2.0_{-1.4}^{+1.2}$ & $1.5_{-1.2}^{+0.8}$ & $4.4_{-1.5}^{+2.2}$ & $2.9_{-1.2}^{+0.9}$ \\
($10^{35}$ g cm s$^{-2}$) \\
$\Dot E_{out}$ & $5.3_{-1.8}^{+2.0}$ & 1.6$^{+0.9}_{-1.1}$ & $0.4_{-0.2}^{+0.3}$ & $1.7_{-0.8}^{+1.3}$ & $1.2_{-0.8}^{+0.7}$ & $0.6_{-0.5}^{+0.3}$ & $3.3_{-1.1}^{+1.6}$ & $1.7_{-0.7}^{+0.6}$ \\
($10^{45}$ erg s$^{-1}$) \\
\hline
$\Dot M_{out}$ ($M_{\odot}\ {\rm yr^{-1}}$) & $0.8\pm0.3$ & 0.5$\pm$0.3 & $0.3_{-0.1}^{+0.2}$ & $0.4_{-0.2}^{+0.3}$ & $0.3\pm0.2$ & $0.3_{-0.2}^{+0.1}$ & $0.5_{-0.2}^{+0.3}$ & $0.4_{-0.2}^{+0.1}$ \\
$\Dot p_{out}$ (${\rm L_{bol}}$/c) & $89_{-30}^{+32}$ & 62$^{+33}_{-41}$ & $20_{-11}^{+15}$ & $51_{-23}^{+38}$ & $37_{-26}^{+23}$ & $28_{-23}^{+15}$ & $77_{-26}^{+38}$ & $51_{-22}^{+16}$ \\
$\Dot E_{out}$ (${\rm L_{bol}}$) & $23.5_{-8.1}^{+8.9}$ & 10.6$^{+5.8}_{-7.3}$ & $2.2_{-1.3}^{+1.6}$ & $10.3_{-4.7}^{+7.7}$ & $7.2_{-5.1}^{+4.6}$ & $3.9_{-3.3}^{+2.1}$ & $19.0_{-6.6}^{+9.5}$ & $9.8_{-4.2}^{+3.3}$ \\
\hline
\end{tabular}
\caption{{\it Top}: Best fit values for the WINE fits and derived number density of the wind $n_0$. {\it Bottom}: mass, momentum and energy transfer rates (see Sect. \ref{discussion}), both in cgs and normalised units. Errors are reported at 90 \% c.l.. $(a)$: In units of $10^{-2}$ ph. keV$^{-1}$ cm$^{-2}$ s$^{-1}$ at 1 keV. $(b)$: Values in parentheses indicate the upper/lower bound. The 1$\sigma$ best fit values are $\log\big(\frac{\xi_0}{\rm erg\ cm\ s^{-1}}\big)=5.2_{-0.3}^{+0.8}, \log\big(\frac{n_0}{{\rm cm^{-3}}}\big)=10.1^{+0.3}_{-0.8}$.}
\label{fitWINE}
\end{table*}

\section{Results}
\subsection{Time-averaged spectra}
In order to get a zeroth-order characterisation of the wind features we first apply WINE to the time-averaged data from the first and second orbit, ranging respectively from 0 to 125 ks and from 175 to 300 ks. As before, the energy range is 2-12(3-79) keV for XMM-{\it Newton}({\it NuSTAR}) datasets. The model in {\sc Xspec} reads as:
\begin{flushleft}
 {\sc const$\times$TBabs$\times\big($zwabs$\times WINE_{abs}\times$powerlaw + 5zgauss}$\big)$ \\
\end{flushleft}
where the constant component ({\sc const}) accounts for the cross-calibration factor between pn and FPMA/B spectra, when present. 
In order to check the accuracy of WINE we also fit the data using the same model and replacing WINE with XSTAR tables. These tables are built using the same initial conditions (i.e. $\sigma_v$, $r_0, L_{ion}, \Gamma$) and spanning the same parameter range for $\xi_0, N_H$. Following the standard procedure, we use the XSTAR table redshift as a proxy for the wind velocity, again spanning the same range of velocities than the v$_0$ parameter in the WINE tables. We report in Table \ref{fit_avg} the best fit values using both WINE and XSTAR. For ease of comparison with the WINE values, we report the relativistically-corrected $N_H$ for XSTAR, obtained by correcting the best fit column density (and associated error) according to the best fit redshift-derived velocity (see \citealp{ltp20} for more details).
For each orbit, we obtain a reasonable agreement between the WINE and XSTAR fits, even though the overall fit statistic is better for the WINE fits.

\subsection{Time resolved spectra}
\label{resolved_results}
\begin{figure}
\centering
\includegraphics[width=\columnwidth]{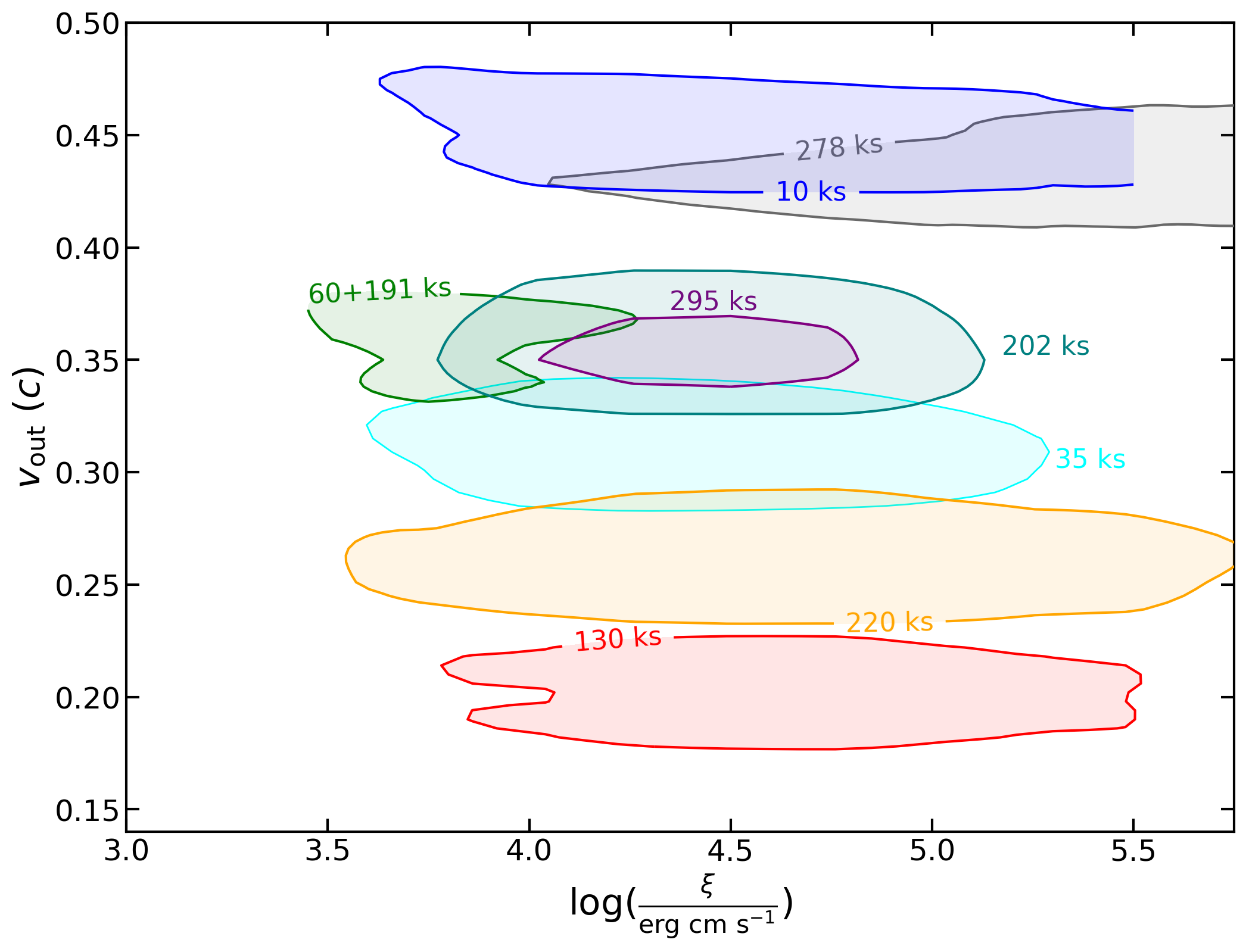} \\
\includegraphics[width=\columnwidth]{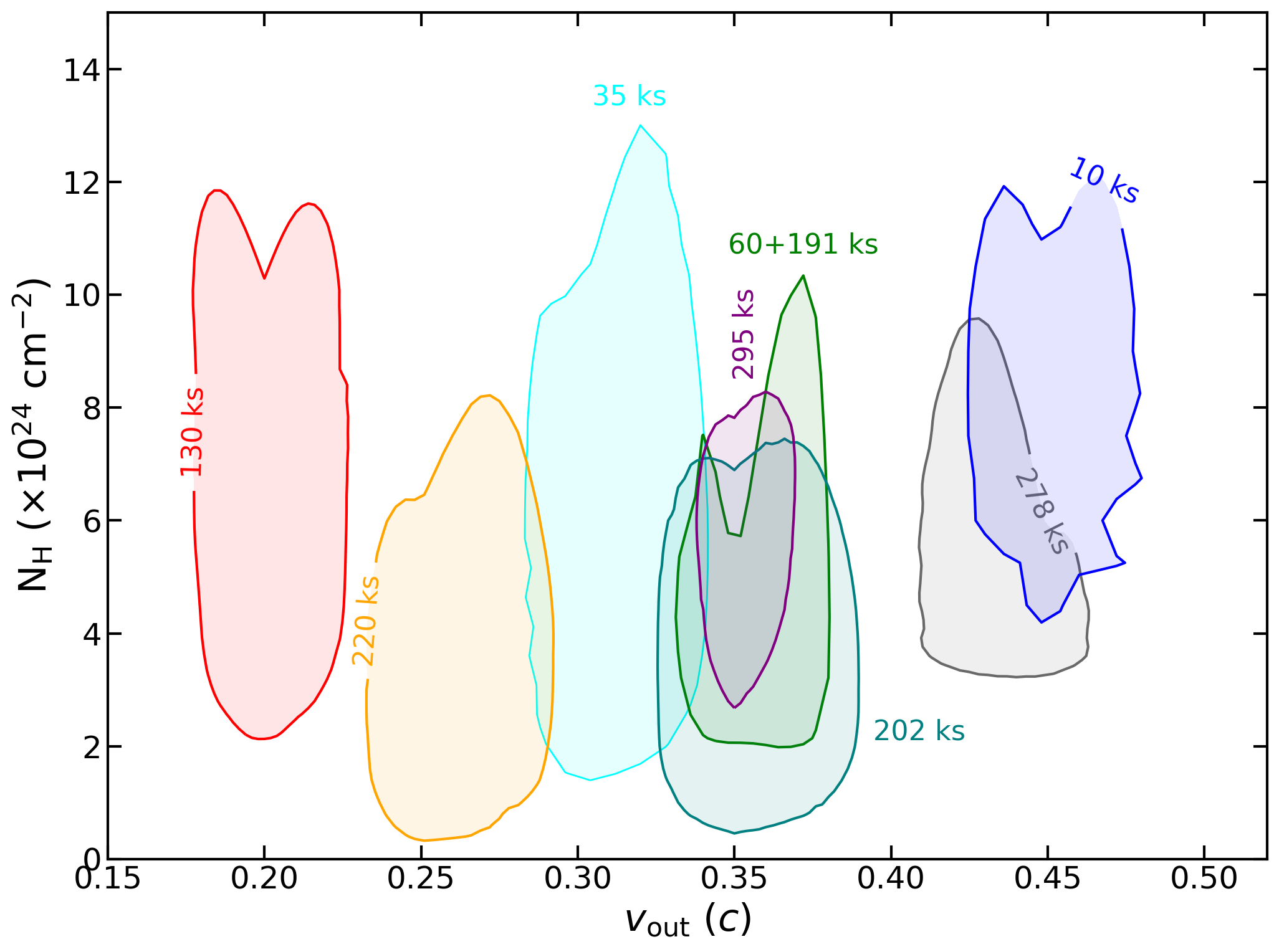}
\caption{Contour plots between the outflowing velocity of the gas and the ionisation parameter and between the column density and the outflowing velocity (top and bottom panel, respectively) obtained with WINE, for the eight analysed spectral intervals. For visual clarity, we only show curves corresponding to a 90 \% confidence level.}
\label{c_plotsWINE}
\end{figure}

We then apply the same WINE fitting model to the time slices showing absorption lines with a significance $>2\sigma$ (estimated via Monte Carlo simulations). The analysed datasets are 10, 35, 130, 191 ks (XMM-{\it Newton}) and 202, 220, 278 and 295 ks (XMM-{\it Newton}+{\it NuSTAR}). We also include the XMM-{\it Newton} coadded spectra corresponding to 60 ks and 191 ks, since their absorption features have consistent line energy according to the fit in Sect. \ref{stat}. For the 191 ks observation we obtain a reduced chi-squared $\chi^2_{\nu}=1.14$, which improves when stacking with the 60 ks one to $\chi^2_{\nu}=0.93$. Best fit values are consistent between the 191 and the stacked 60+191 fit.
Table \ref{fitWINE} reports the best fit parameters. We also indicate the wind density $n_0$, derived inverting Eq. \ref{xi_definition} and using the best fit values for $\xi_0, \rm{v}_0, \Gamma$ and $L_{ion}, r_0$ from Sect. \ref{winesection}. We note that, for a gas in photoionisation equilibrium, its ionic population, and thus the emerging spectrum, is mainly determined by the value of $\xi_0$. Observationally, once $L_{ion}, \xi_0$ are measured, it is possible to derive an estimate of $n_0 r_0^2$, but the two parameters cannot be disentangled and, thus, they both remain mostly unknown for the majority of UFOs (for further discussions see \citealp{nfp99,kne07,2022arXiv221201399L}). However, thanks to our estimate of $r_0$, we are able to break this degeneracy and, thus, to provide a value for the wind density. Fig. \ref{c_plotsWINE} shows the contour plots for all the analysed time slices, while best fit spectra and corresponding theoretical model are plotted in Fig. \ref{wine_app} in Appendix \ref{AppendixA}, where we also discuss a further absorption line detected in the 220 ks {\it NuSTAR} spectrum at E$\approx$14 keV. We note that the reduced $\chi^2$ are similar to those in Table \ref{bestfitPar}, in which absorption features are fitted with Gaussian lines. However, fit statistic shows a significant improvement with respect to the time averaged WINE fits for the 1st and 2nd orbit reported in Table \ref{fit_avg}, since we are now able to resolve the variable wind features on a 5 ks time scale and analyse them one by one. The averaged spectra are instead 125 ks long and therefore only allow for a characterisation of the average wind features.

We verified that a different model component accounting for the neutral absorption along the line of sight ({\sc ztbabs} instead of {\sc zwabs}) does not affect the best fitting parameters and the associated statistics.

\section{Discussion}
\label{discussion}
\subsection{UFO energetic and duty cycle}
\label{energetic}
The mass outflow rates associated to the UFO features can be calculated through the formula from \citet{crenshaw2012}:
\begin{equation}
\Dot{M}_{out}=4 \pi r_0 N_H \mu m_p C_f v_0
\label{mdot}
\end{equation}
where $\mu, m_p$ are the mean atomic mass per proton (set to 1.2, see \citealp{grm15}) and the proton mass, respectively, and we set $r_0=5 r_S$ following Sect. \ref{winetables}. Due to the weakness of the UFO emission features, we are not able to directly constrain the covering factor $C_f$ from the observations, so we assume a mean value of 0.4 from the detection occurrence of UFOs in AGN samples \citep{tcr10,igo20,matzeu22}. Then, we calculate the momentum transfer rate as $\Dot{P}_{out}=\Dot{M}_{out} \rm{v}_0$ and the kinetic energy according to the special relativity formula as in \citet{llt21}:
\begin{equation}
\Dot{E}_{out}= (\gamma -1)\cdot \Dot{M}_{out} c^2
\label{eout}
\end{equation}
where $\gamma= \frac{1}{\sqrt{1-\beta^2}}, \beta=\frac{\rm{v}_0}{c}$.
We report $\Dot{M}_{out}, \Dot{P}_{out}, \Dot{E}_{out}$ for each spectrum in Table \ref{fitWINE}. Errors are calculated using the standard linear propagation approximation. We do not include the error associated with the black hole mass ${\rm M_{BH}=3.0_{-1.5}^{+5.5} \cdot 10^7 M_{\odot}}$, whose normalised interval (i.e., the error interval divided by the mean value) is higher than those of the WINE parameters reported in Table \ref{fitWINE}. Including the uncertainty on ${\rm M_{BH}}$ would result in a factor $\approx 2$ and $\approx 4$ increase of the lower and upper bounds of the energetic, respectively. We also note that the commonly-used lower limit for $r_0$, built from the assumption that v$_0$ corresponds to the escape velocity of the flow, is equal to $r_{min}=4.9^{+0.4}_{-0.7}, 22.3^{+6.5}_{-2.2} {\rm r_S}$ for the fastest and slowest velocities reported in Table \ref{fitWINE}, i.e. v$_0$=0.45,0.21 c; using these values would result in generally higher values of the energetic, which linearly scales with $r_0$ (see Eq. \ref{mdot}). 

Interestingly, the fact that $r_{min} \geq r_0$ implies that the detected UFO velocities are always lower than the escape ones, therefore the wind will need additional acceleration, e.g. through radiation or magnetocentrifugal forces (e.g. \citealp{blandford82,psk00,cui20}, but see \S \ref{conclusions} for further discussion), in order to overcome the gravitational force of the central black hole. In case of insufficient acceleration, the outflow may turn into a so-called "failed wind", which is ubiquitously expected in all the radiative driving scenarios as a result of the over-ionisation of the gas closer to the black hole \citep{higginbottom14,dannen19} and, in turn, is fundamental for the shielding of the outer gas layers (see \citealp{zappacosta20} and references therein). Given the short distance from the black hole, this outflow may also be linked to the dynamics of the X-ray corona, whose physical dimension is supposed to vary according to accretion rate variations (see e.g. \citealp{kara19,alston20}); as discussed in Sect. \ref{introduction} and \ref{connection}, NGC 2992 is indeed a strongly variable source.

Our derived values for the energetic are extremely high with respect to the typical UFO ones reported in the literature; as an example, $\Dot E_{out}$ is usually found to be around 0.1 -1 times ${\rm L_{bol}}$ (see e.g. \citealp{Fiore17}), while here it is in the range 2-23 ${\rm L_{bol}}$. This is due to the very high values found for v$_0, N_H$, i.e. ${\rm \sim 0.3 c, 6 \cdot 10^{24} {\rm cm^{-2}}}$, making NGC 2992 an outlier with respect to the nearby Seyferts, which typically show v $\sim$ 0.1 c, $N_H \sim 10^{23} {\rm cm^{-2}}$ (see e.g. \citealp{tcr11}). 
We note that, as a result of the high observed velocities, the relativistic reduction of the gas opacity (and, then, of its observed column density) is particularly significant. This, together with the lower collecting area of XMM-{\it Newton} above 10 keV, makes the features with low $N_H$ more difficult to be detected, possibly resulting in a bias in our analysis toward high $N_H$ and, in turn, in an underestimate of the wind activity. This bias may explain, at least partly, the high column densities reported in Table \ref{fitWINE} with respect to the \citet{tcr11} mean values.
Moreover, as we will discuss below, NGC 2992 shows evidence of a "changing-look" activity, hinting at differences in the accretion-ejection dynamics with respect to "canonical" Seyfert galaxies. Interestingly, UFOs detected in Quasars show higher v$_0,N_H$ with respect to Seyfert galaxies; as an example, \citet{chartas21} concentrate on a sample of Quasars at $1.4 \leq z \leq 3.9$, finding an average v$_0 \approx 0.3$ c and $N_H \approx 4 \cdot 10^{23} {\rm cm^{-2}}$ (which, once relativistically corrected for 0.3 c, corresponds to $N_H^{rel} \sim 8 \cdot 10^{23} {\rm cm^{-2}}$). \citet{nrg15} and \citet{tmv15} both found similar v$_0 \approx 0.25 c, N_H^{rel} \approx 1.2 \cdot 10^{24} {\rm cm^{-2}}$ for the UFOs in PDS 456 (z=0.184) and in IRASF1119+3257 (z=0.189), respectively.

Thanks to our time-resolved analysis we are able to estimate the duty cycle of the wind, i.e. the fraction of time during which it is observed. Considering that the amount of spurious detection within the full set of time slices amounts to $\approx 2$ for our 2$\sigma$ significance threshold (see Sect. \ref{stat} and Appendix \ref{appsim}), we can conservatively assume that at least 6 of the XMM-{\it Newton} UFO detections (out of a total of 8) are not due to noise fluctuations. For a total of 50 time slices, this translates into a lower limit for the duty cycle of 6/50=12\%.
However, we caution that the observing bias discussed above may likely result in an underestimate of the duty cycle. We can derive mass and energy outflow rates representative of the total observing time as the average of the values reported in Table \ref{fitWINE} times the wind duty cycle, thus obtaining $\Dot M_{out}^{tot} = 0.05 {\rm M_{\odot} yr^{-1}}, \Dot E_{out}^{tot}=1.3{\rm L_{bol}}$.
We also note that, from an historical point of view, UFO features have been detected in high-flux observations only, which however occurred only $\approx 30 \%$ of the time from the first X-ray observation of NGC 2992 in 1978 \citep{mbb18}. Thus, the inferred duty cycle may not be regarded as representative of the global AGN lifetime.

UFO features were already detected in \citet{mbb18} in two high-flux observations of NGC 2992, the 2003 XMM-{\it Newton} and the joint 2015 {\it NuSTAR}+Swift one, with 2-10 keV luminosities $L_{2-10}=1.3 \cdot 10^{43}, 7.6 \cdot 10^{42} {\rm erg\ s^{-1}}$, respectively, comparable to that of our 2019 observations, $L_{2-10}=1.0 \cdot 10^{43} {\rm erg\ s^{-1}}$. Our derived values for $\Dot{M}_{out}, \Dot{E}_{out}$ are a factor $>10$ higher than those reported in \citet{mbb18} due to the different wind properties: our best fit values for $N_H, v_{out}$ are in the range $4-8 \cdot 10^{24}$ cm$^{-2}$ and 0.2-0.4 c, while in 2003(2015) observations they found $N_H = 2.2 \cdot 10^{23}$($1.8 \cdot 10^{22}$) cm$^{-2}$ and v$_0 \approx$ 0.2 - 0.3\ c. We note that neglecting the relativistic reduction of the wind opacity would have led to a 35-55\% lower $N_H$ (according to our range of v$_0$) with respect to the WINE-derived value.
As an additional consistency check, we fit again the 2003 XMM-{\it Newton} observations using the same fitting model of \citet{mbb18} and replacing the original wind tables, computed with the {\it Cloudy} code \citep{fcg17}, with our WINE tables. We obtain best fit values consistent with those in \citet{mbb18} and significantly lower than the present ones, further confirming the robustness of WINE when compared to different codes on one side and, on the other, the peculiarity of the wind features reported in this paper with respect to the archival ones. The high column densities inferred for the 2019 data set, easily exceeding the Compton-thickness threshold, suggest a scenario in which the observed features can be ascribed to independent, high-velocity clouds ejected from the disc, possibly embedded in a much lower $N_H$ wind. We discuss such scenario in the next subsection.

\subsection{UFO energetic according to the "cloud scenario"}

Equation \ref{mdot} is calculated under the assumption of a spherical symmetric flow (\citealp{crenshaw03,crenshaw2012}), as commonly assumed for UFOs (see e.g. \citealp{tcr12,nrg15,Fiore17,chartas21}). However, the outflow in NGC 2992 shows a very low duty cycle and a short term variation of its spectral appearance, equal or lower than the 5 ks time scale of our observations. Moreover, we do not detect any emission feature associated to the UFO, therefore we are not able to put constraints on its angular extension. Therefore, we also consider a complementary scenario, dubbed "cloud scenario" in which the UFO absorption features are due to outflowing (spherical) gas clouds passing through our line of sight for a time t$\approx$5 ks (see \citealp{bianchi09c} for a similar approach). For each time slice we derive the mass and energy outflows as follows. We compute the average radial distance of each cloud as:
\begin{equation}
    r_{avg}=r_0 + r(N_H/2) = r_0 + \frac{N_H/2}{n_0}
\end{equation}
where $r(N_H/2)$ is the radius enclosing half of the wind column density and the last term is valid for a constant density wind ($\alpha=0$, see Eq. \ref{expr_deltar}). Then, assuming that the cloud is rotating at a distance $r_{avg}$ from the black hole with a Keplerian velocity $v_{rot}$, its dimension D can be estimated as $D=t \cdot v_{rot}= t \cdot \sqrt{G M_{BH}/r_{avg}}$, where t=5 ks. Finally, the mass of the cloud is given by:
\begin{equation}
M_{out}=\frac{4}{3} \pi (D/2)^3 n_0
\end{equation}
We report $M_{out}$ in Table \ref{WINEcloud}, together with the associated energy $E_{out}=(\gamma -1) {M}_{out} c^2$, where now we use the composition between outflowing and rotational velocity, i.e. $\beta_+=\sqrt{\rm v_{rot}^2+v_0^2}\ /c$. We also calculate the average mass and energy rates as the ratio between the sum of $M_{out}, E_{out}$ for all the observations divided by the total observing time, i.e. 250 ks. With respect to $\Dot M_{out}^{tot}, \Dot E_{out}^{tot}$ computed assuming spherical symmetry we obtain lower values by a factor of $\approx 100$ and 2, respectively. Given the large number of assumptions, in Table \ref{WINEcloud} we only report the mean values, with the aim of demonstrating the importance of a proper description of the UFO geometry and dynamic for a reliable estimate of its energetic.

\begin{table}
\centering
\begin{tabular}{c c c c}
Time frame (ks) & $ M_{out}$ & $ E_{out}$ \\
 & ($10^{-6} {\rm M_{\odot}}$) & ($10^{47} {\rm erg}$) \\
\hline
10 & 0.65 & 1.71\\
35 & 7.55 & 13.46 \\
130 & 2.71 & 2.75\\
60+191 & 27.7 & 67.6\\
202 & 3.61 & 7.35\\
220 & 2.82 & 3.78\\
278 & 0.06 & 0.12\\
295 & 3.06 & 5.94\\
Total & 48.2 & 102.7 \\
\hline
& $\Dot M_{out}$ & $\Dot E_{out}$ & $\Dot E_{out}$ \\
& (${\rm M_{\odot}\ yr^{-1}}$) & (erg s$^{-1}$) & ($\rm{L_{bol}}$) \\
Time-avg. & 6.08$\cdot 10^{-3}$ & 4.11$\cdot 10^{43}$ &0.23 \\
\end{tabular}
\caption{Energetic according to the cloud scenario. From top to bottom we report $M_{out}$ and $E_{out}$ (left to right) for the single observations and their sum. Last row shows the average mass and energy rates for the total observing time.}
\label{WINEcloud}
\end{table}

\subsection{Connection with the accretion disc}
\label{connection}
We are not able to find any correlation among the UFO parameters and between these and the properties of the continuum spectrum (e.g., photon index, normalisation), both for the present and the \citet{mbb18} observations. However, we note that the 2003 and 2015 UFO features were also accompanied by a broad component of the Fe K$\alpha$ emission line, while instead both the wind and the emission are absent in the low flux XMM-{\it Newton} observations from 2010 to 2013, which have $L_{2-10}<4 \cdot 10^{42}$ erg s$^{-1}$. 

By analysing all the optical and X-ray observations from 1978 up to 2021, \citet{guolo21} found evidence for a "changing-look" behaviour of NGC 2992, which appears to be driven primarily by the intrinsic luminosity (i.e., the accretion rate), rather than by obscuration or transient events such as TDEs. Their main finding supporting this interpretation is the anti-correlation between $L_{2-10}$ and the full width at half maximum (FWHM) of the H$\alpha$ line; moreover, they also found a positive correlation between $L_{2-10}$ and i) the flux of the Fe K$\alpha$ line and ii) the flux of the H$\alpha$ line in the optical band. H$\beta$ line was also detected, albeit with a lower confidence, for the brightest observations, showing a fairly constant H$\alpha$/H$\beta$ flux ratio of $\sim 9$. The luminosity threshold for the appearance of the H$\alpha$ line is $2.6 \cdot 10^{42}$ erg s$^{-1}$, corresponding to an Eddington ratio $\approx 1 \%$. Interestingly, they suggest the possibility that, due to the low accretion rate, the accretion disc could be thin, pressure dominated at large radii (as in the standard \citealp{ss73} picture) and radiatively inefficient in the innermost regions (RIAF; \citealp{yn14}), and the high-luminosity intervals are associated to instabilities at the boundary between these two regimes. UFOs have been always observed in high-luminosity states; unless this evidence is entirely due to the low constraining power of the low-flux spectra, it represents an indication for a link between the radiative efficiency of the inner region and the ejection of matter in the form of relativistic disc winds.

\section{Conclusions}
\label{conclusions}
In this paper we analyse the UFO absorption features, at energies E$\gtrsim$ 9 keV, detected in the 2019 XMM-{\it Newton}+\textit{NuSTAR} monitoring campaign of the Seyfert galaxy NGC 2992. The time-averaged spectra shows absorption lines at a significance $>3 \sigma$, estimated via a set of 1000 Monte Carlo simulations.
Moreover, the high flux of these observations allows us to perform a time-resolved spectroscopic analysis of the absorption features  with a temporal resolution of $\sim$5 ks. We obtain a Monte-Carlo derived significance $> 2 \sigma$ for 4 time slices of the first XMM-{\it Newton} orbit and 4 of the joint XMM-{\it Newton}+{\it NuSTAR} observations.

We fit these spectra with the novel photoionisation and spectroscopic model WINE (Wind in the Ionised Nuclear Environment; \citealp{lpt18,ltp20,llt21}, Luminari et al., in prep.), which self-consistently calculates wind absorption and emission profiles using a realistic, physically-motivated dynamical and geometrical representation of the outflows for AGN and compact sources (Sect. \ref{winesection}). Notably, WINE also includes the special relativity effects on the gas opacity (as discussed in \citealp{ltp20}), which are particularly important given the high detected velocities, between v$_0=0.21$ and 0.45 c, resulting in an increase of the intrinsic wind column $N_H$ with respect to the observed (i.e. apparent) one by a factor between 35\% and 55\%, respectively. 

\noindent Our main findings can be summarized as follows:
\begin{itemize}
\item We detect fast, massive and ionised outflows, with average v$_0=0.35 c, N_H=5.8 \cdot 10^{24} {\rm cm^{-2}}, \log\big(\frac{\xi_0}{\rm erg\ cm\ s^{-1}}\big)=4.5$. Notably, these values are far higher than those typically found for UFOs in nearby Seyfert galaxies and resemble those observed in highly-accreting Quasars. On the basis of geometrical and dynamical considerations we suggest the wind is launched from a short distance to the black hole, of the order of 5 $\rm{r_S}$. Interestingly, this value is in rough agreement with the dynamical length obtained as the product between the timescale of the UFO appearance ($\approx$ 5 ks) and a typical flow velocity of 0.4 c. Through $r_0, \xi_0$ we are also able to provide an estimate of the wind density $n_0 \approx 10^{11} {\rm cm^{-3}}$.
We estimate a UFO duty cycle of $\approx$12\% as the fraction of time slices in which we obtained a detection with a $> 2 \sigma$ significance of the corresponding absorption features, corrected by the possible contribution from noise fluctuations. However, this value likely represents a lower limit, since both relativistic effects and the limited XMM-{\it Newton} effective area above 10 keV may result in a lower detection rate of the features with lower $N_H$. The wind best fit values of the time averaged spectra are consistent with the time resolved ones, further confirming the robustness of our analysis.

\item The "instantaneous" momentum outflow rate, $\Dot p_{out}$, for the analysed time slices lies in the range $20-89 L_{bol}/c$, strongly suggesting the presence of additional launching mechanisms at work beside radiation pressure, such as magneto-hydrodynamic (MHD) acceleration, especially given the low bolometric luminosity during these observations, around 4\% the Eddington value \citep{fkc10,gaspari17,cui20,lne21}. $\Dot p_{out}$ has been computed assuming a spherical symmetric outflow, as usually done for UFOs (see Eq. \ref{mdot}); similarly, we compute "instantaneous" mass and energy outflow rates, of the order of $\Dot M_{out} \approx 0.5 M_{\odot} yr^{-1}, \Dot E_{out} \approx 10 L_{bol}$, respectively.

Using the wind duty cycle we are able to derive "average" mass and energy outflow rates representative of the total observing time, i.e. $\Dot M_{out}^{tot} =$0.05${\rm M_{\odot} yr^{-1}}, \Dot E_{out}^{tot}=$1.3${\rm L_{bol}}$. The energy outflow rate is of the same order of the theoretical threshold (between 0.5 - 5  \% ${\rm L_{bol}}$, \citealp{dsh05,he10}) required to switch on feedback effects in the host galaxy. However, we caution that the low accretion rate of the central black hole will probably prevent the nuclear wind to develop a fully energy conserving galactic outflow (see e.g. \citealp{Faucher12,King15,torrey20}); therefore, we only expect a moderate outflow activity in the host galaxy.

We also caution that for an MHD-driven wind the conservation of angular momentum implies $\dot M_{in} = (r_A/r_0)^2 \dot M_{out}$, where $\dot M_{in}$ is the rate of accreting mass and $r_A/r_0$ is the ratio between the Alfven and the launching radii of the wind. Such ratio is usually estimated in the interval 1 - 10 (see e.g. \citealp{pudritz07,cui20,fiore23}), thus predicting $\dot M_{out}=0.1 - 1 \dot M_{in}$. Our "instantaneous" mass outflow rates are $\approx 10 -20 \dot M_{in}$ and, therefore, in tension with the MHD prediction. Such high rates would remove more angular momentum than that of the accreting mass, leading to accretion "bursts" and, thus, significant X-ray variability, as indeed observed in NGC2992 (see \S \ref{introduction}). The average mass outflow rate is $\dot M_{out}^{tot} \approx 1.7 \dot M_{in}$, possibly indicating a longer-term equilibrium between accretion and ejection. However, additional observations of NGC2992 are needed to carefully investigate this hypothesis.

\item We propose the alternative scenario in which the UFOs are associated to a series of clouds passing through our line of sight with a 5 ks time scale. The associated mass and energy rates, $\approx$ 6 $\cdot 10^{-3} \rm{M_{\odot}\ yr^{-1}}, 0.2\ \rm{L_{bol}}$ respectively, are significantly lower than the above ones. Even though we are not able to provide precise measurements due to the large number of assumptions, this exercise demonstrates the importance of a proper physical setting when calculating the wind energetic. 
\item Disc winds have been observed in NGC 2992 only in three high luminosity observations, i.e. the 2019 campaign analysed here and two previous observations from 2003 and 2015 \citep{mbb18}, where also a broad Fe K$\alpha$ emission line was present. Interestingly, all these observations caught the source with a luminosity above the threshold of $L_{2-10}=2.6 \cdot 10^{42}$ erg s$^{-1}$ (corresponding to 1\% ${\rm L_{Edd}}$) identified by \citet{guolo21}, over which broad H$\alpha$ and H$\beta$ lines are detected in the optical band. This evidence may suggest a link between the accretion disc activity and the presence of ultra fast outflows.
\item Outflows in the optical band have been observed from the Broad Line Region up to galactic scales by several authors \citep{veilleux01,irwin17,Mingozzi19,Pereira21}, suggesting a link between nuclear and galactic scales. A forthcoming paper (Zanchettin et al., subm.) will analyse further ALMA and MUSE observations probing the complex interplay between the different gas phases (cold molecular, warm ionised, and the radio jet), in order to investigate the relation between AGN, winds, jet and the galactic disc.
\end{itemize}
This is the third paper of a series devoted to the 2019 observation campaign of NGC 2992, after \cite{mbb20,middei20}. As a further step, we plan to apply WINE to the 2003 and 2015 high flux observations in order to build an homogeneous census of the UFO features in NGC 2992 and assess in detail the relation between disc winds and accretion disc.

\emph{Acknowledgments.} We thank the referee for their valuable and interesting comments. AM, SB, GM, EN, EP, SP acknowledge support from PRIN MIUR project "Black Hole winds and the Baryon Life Cycle of Galaxies: the stone-guest at the galaxy evolution supper",  contract no. 2017PH3WAT. AL acknowledges support from the HORIZON-2020 grant “Integrated Activities for the High Energy Astrophysics Domain" (AHEAD-2020), G.A. 871158. BDM acknowledges support via Ramón y Cajal Fellowship (RYC2018-025950-I), the Spanish MINECO grant PID2020-117252GB-I00, and the AGAUR/Generalitat de Catalunya grant SGR-386/2021. RM acknowledges the financial support of INAF (Istituto Nazionale di Astrofisica), Osservatorio Astronomico di Roma, ASI (Agenzia Spaziale Italiana) under contract to INAF: ASI 2014-049-R.0 dedicated to SSDC. SB, EP acknowledge financial support from ASI under grants ASI-INAF I/037/12/1 and n. 2017-14-H.O. We used {\sc Astropy}, a community-developed core {\sc Python} package for Astronomy \citep{astropy13, astropy18}, {\sc numpy} \citep{numpy} and {\sc matplotlib} \citep{h07}. This research has made use of the {\it NuSTAR} Data Analysis Software (NuSTARDAS) jointly developed by the ASI Science Data Center (ASDC, Italy) and the California Institute of Technology (USA).

\appendix

\section{EPIC pn background and calibration}
\label{AppendixBACK}
\begin{figure*}
\centering
\includegraphics[width=0.2 \paperwidth]{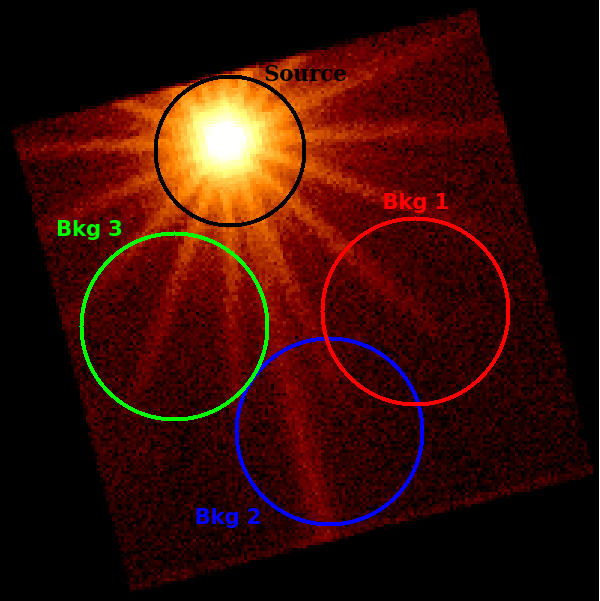}
\includegraphics[width=0.3 \paperwidth]{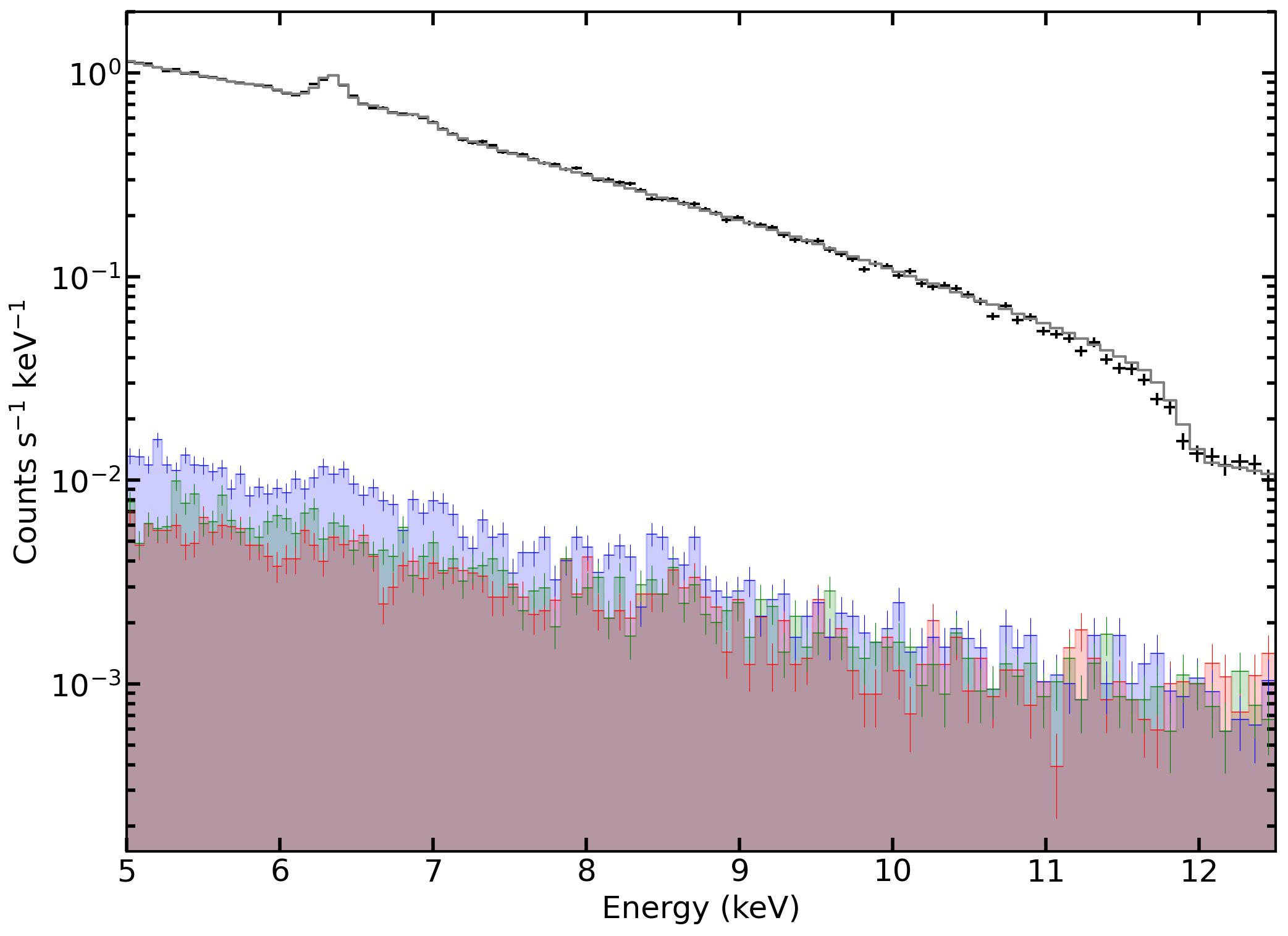} \\
\includegraphics[width=0.2 \paperwidth]{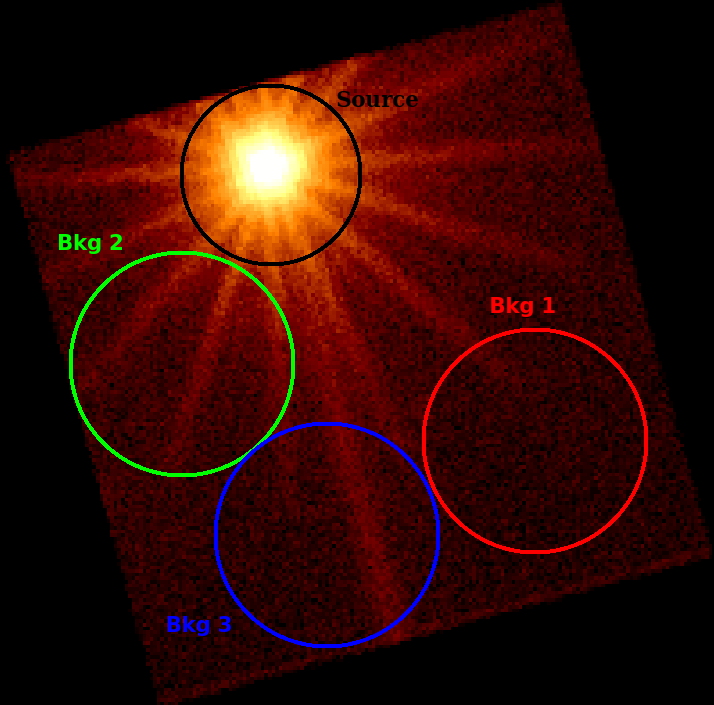}
\includegraphics[width=0.3 \paperwidth]{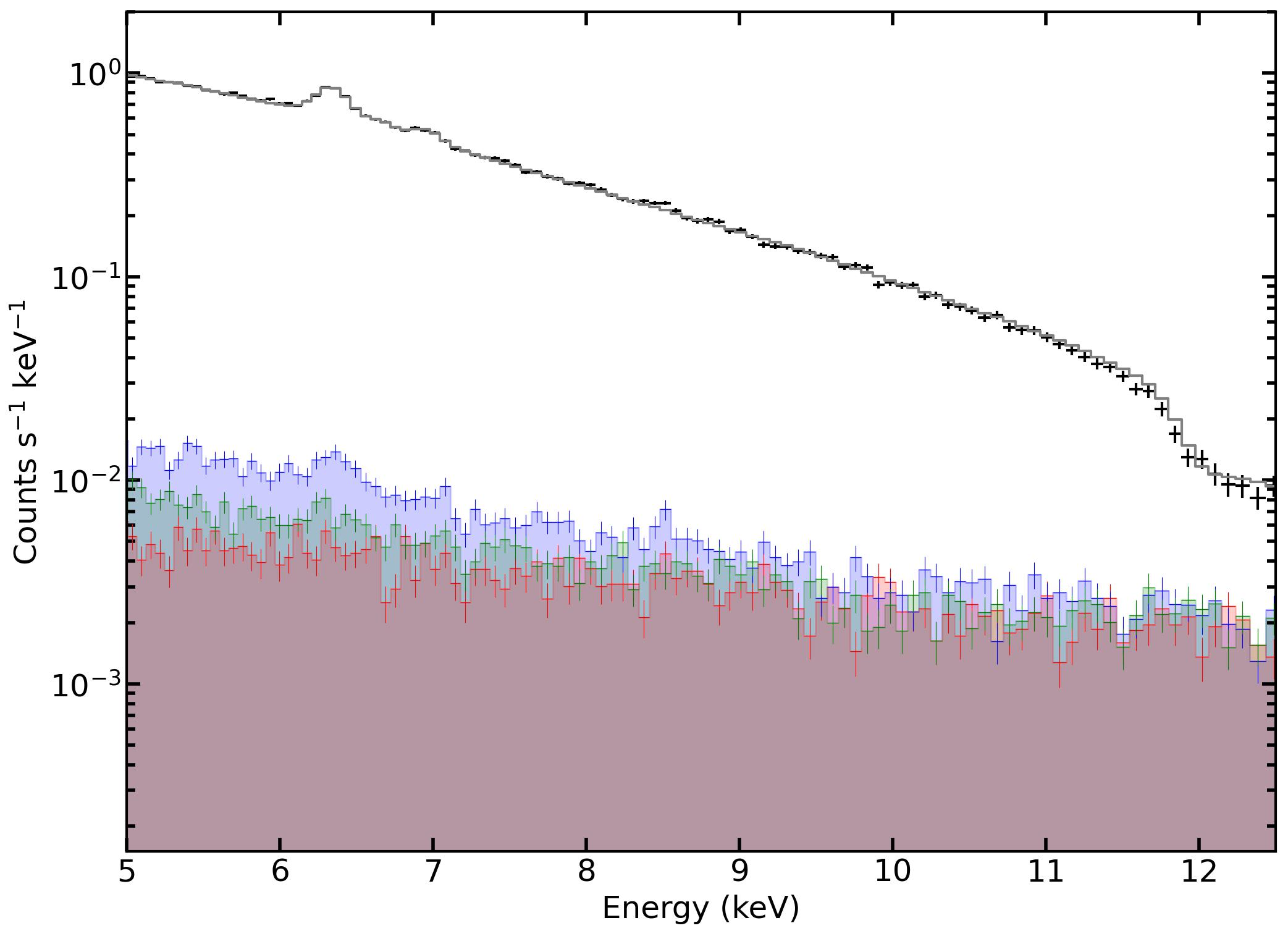}
\caption{{\it Left panels:} EPIC pn filtered event files for orbits 1 (top) and 2 (down). The source spectrum is a circular region with 40'' radius, while different circular regions with 50'' radii are tested as background regions. {\it Right panels:} time-averaged spectra, in black, for orbit 1 (92.6 ks long, top figure) and 2 (92.8 ks long, bottom figure). Shaded regions indicate different background spectra; red ones correspond to those used throughout the paper. Best fit models are shown as gray solid lines.}
\label{xmm_back}
\end{figure*}

The 2019 flux levels of the source were remarkably high compared to previous observations (\citealp{mkt07}). However, the observed background can still be relevant above 10 keV, where the EPIC pn effective area has a significant drop. In this Appendix we try different extraction regions for the time-averaged background spectra of the two XMM orbits and compare them with the ones used throughout our previous analysis. Fig. \ref{xmm_back} (left panels) shows the three circular regions (with 50'' radii) from which spectra are extracted. Red circles indicate the background regions used in Sect. \ref{stat}. In the 10-12.5 keV energy band the flux of the background is 3.7\% and 7.9\% of the source flux for the first and second orbit, respectively. No significant statistical variations are found in the best fit values and in the overall $\chi^2$/dof for the time-averaged spectra when using the three different background regions.

\begin{figure}
\centering
\includegraphics[width=0.5\paperwidth]{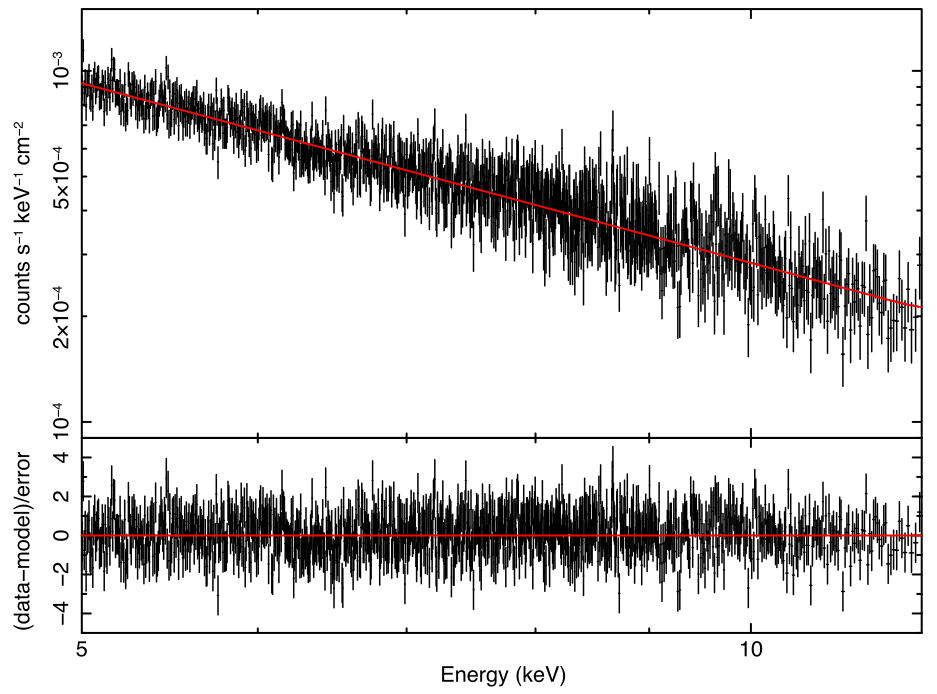}
\caption{The EPIC-pn spectrum of 3C 273 and its best fit powerlaw (top) and the associated residuals (bottom).}
\label{3c273}
\end{figure}

As an additional test we reduced the XMM-Newton pn data set for the Blazar 3C 273, which is well known for showing a smooth broadband continuum at hard X-ray energies \citep{madsen15}. The observation was performed on July 2018 in Small Window Mode (ObsID 0414191401), i.e. less than a year before our observations of NGC 2992 and with the same operational settings. The 3C 273 data are not affected by background flares and have a net exposure time of 44.3 ks. The flux level of the source (F$_{2-10}$=6 $\cdot10^{-11}$ erg cm$^{-2}$ s$^{-1}$) is comparable with that of NGC 2992 during the second orbit. The extracted spectrum is shown in Fig. \ref{3c273} with its best-fit power law over the range 5-12 keV ($\Gamma = 1.70$) and the relative residuals (bottom panel). It is worth noting that the spectrum of 3C 273 does not show any deviation from a simple power law up to 12 keV (compare with Fig. \ref{ratios} for NGC 2992), implying that there are no calibration issues above 10 keV and that the ancillary response files of the EPIC pn detector are fully reliable in this spectral region in spite of the drop of the effective area. We replaced the 3C 273 pn background with the NGC 2992 one, we modelled the continuum with an absorbed power law and applied the same blind scan of Sect. \ref{stat} to search for absorption lines in the 6-12 keV energy range. No absorption lines are detected at 99\% c.l., further demonstrating that the lines detected in the NGC 2992 observations are not due to background artefacts.

\section{Monte Carlo simulations}
\label{appsim}
\begin{figure*}
\centering
\includegraphics[width=0.75\paperwidth]{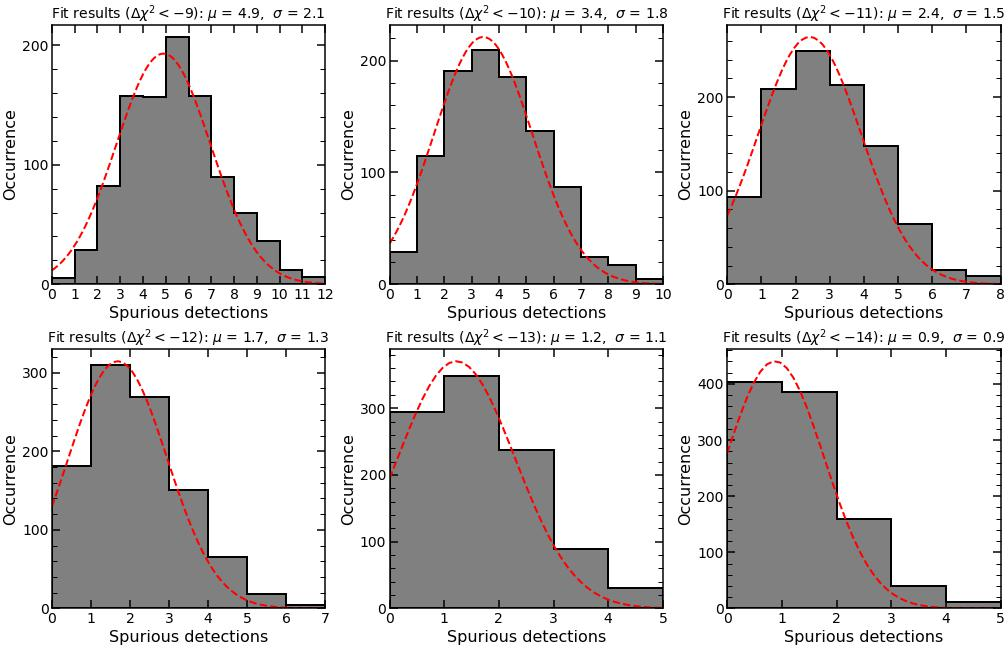}
\caption{Histograms of the absorption lines detected in the 1000 simulated data sets. Red dashed lines represent the corresponding best fit Gaussian distributions, whose mean $\mu$ and standard deviation $\sigma$ are reported on the top of each box. From left to right and top to bottom the statistical threshold increases (from $\Delta \chi^2 <-9$ to <-14).}
\label{setsim}
\end{figure*}
Fig. \ref{setsim} reports the results of the 1000 Monte Carlo simulations of the entire set of 50 XMM-{\it Newton} time slices. From left to right and top to bottom, each panel reports, for increasing significance, the distribution of the spurious Gaussian absorption components detected in each set. Each distribution is fitted with a Gaussian profile to derive its mean $\mu$ and standard deviation $\sigma$, which are reported on top of each plot. The mean number of spurious detections ranges from $\mu=4.9$ for $\Delta \chi^2<-9$ to 0.9 for $\Delta \chi^2<-14$.

\section{{\sc WINE} best fits}
\label{AppendixA}
\begin{figure*}
\begin{center}
\includegraphics[width=0.3\columnwidth]{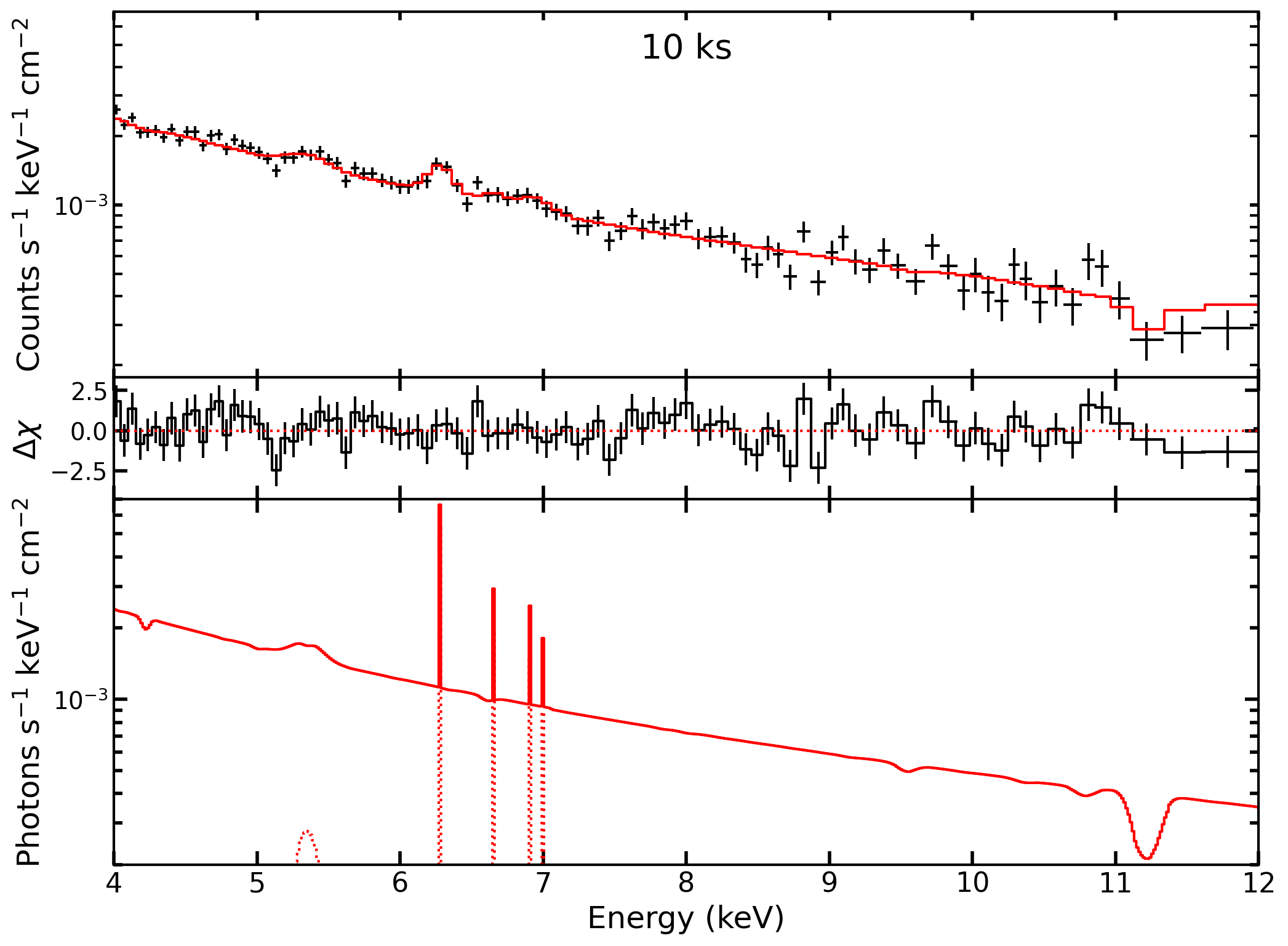}
\includegraphics[width=0.3\columnwidth]{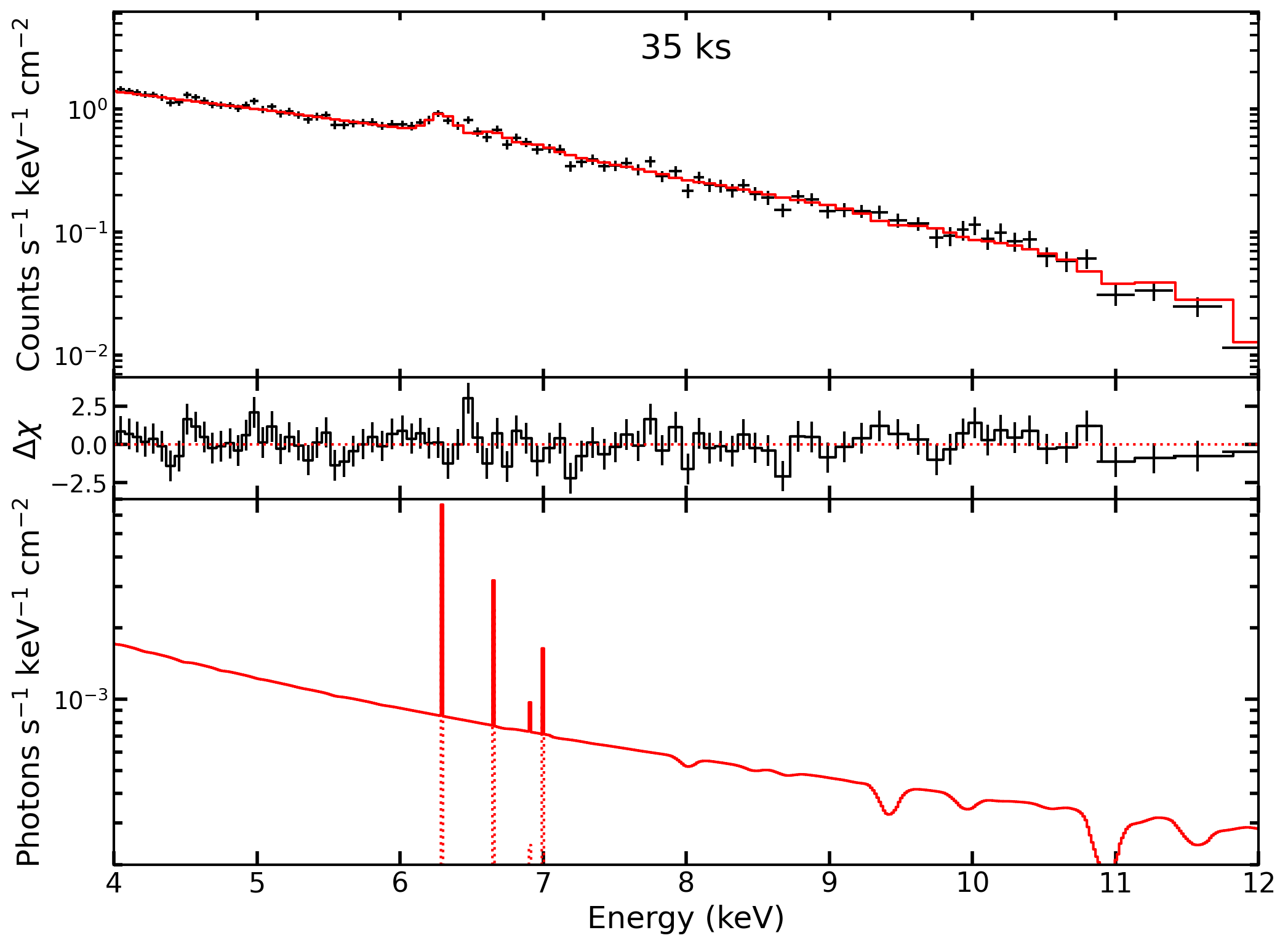}
\includegraphics[width=0.3\columnwidth]{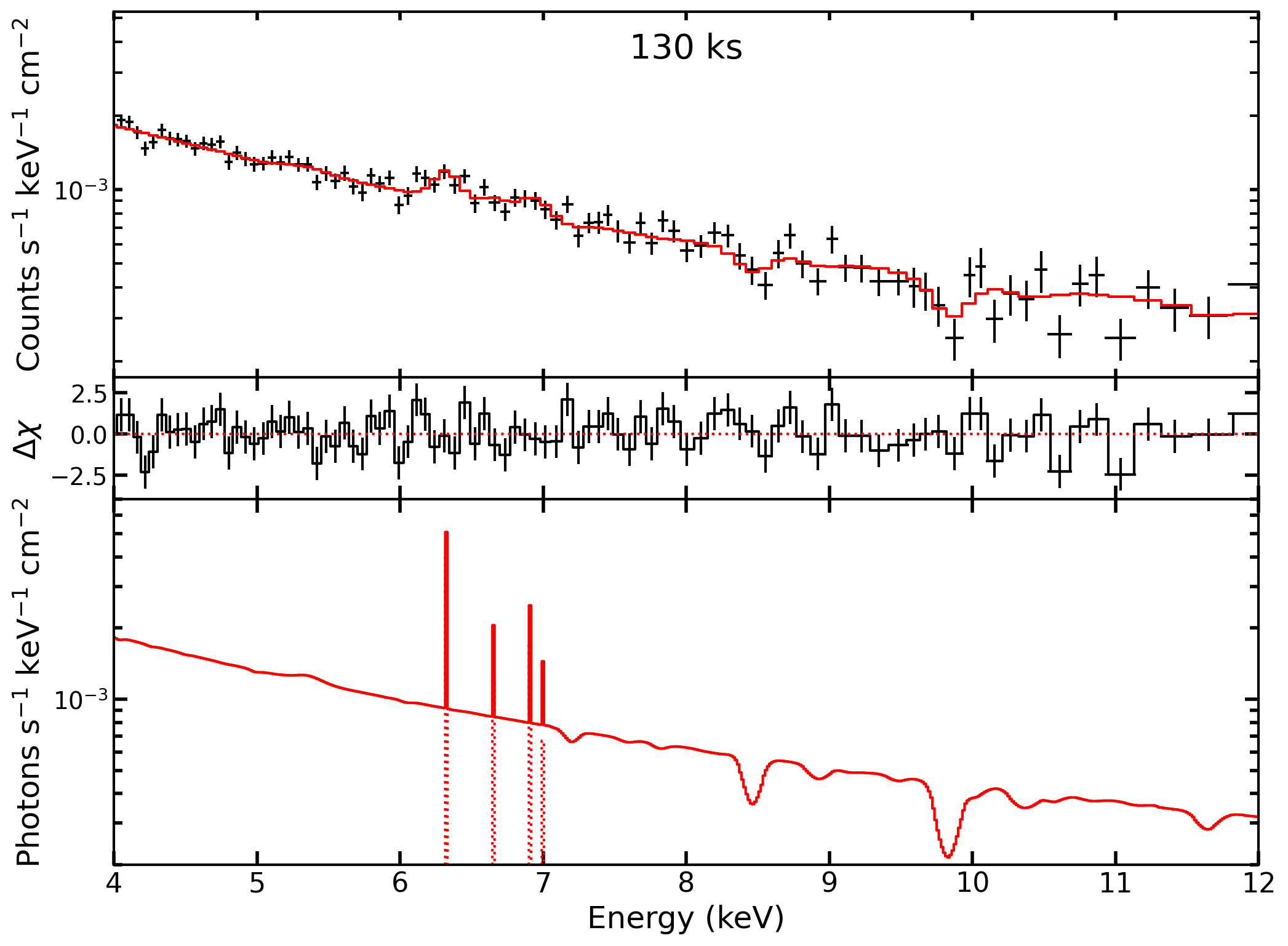}
\includegraphics[width=0.3\columnwidth]{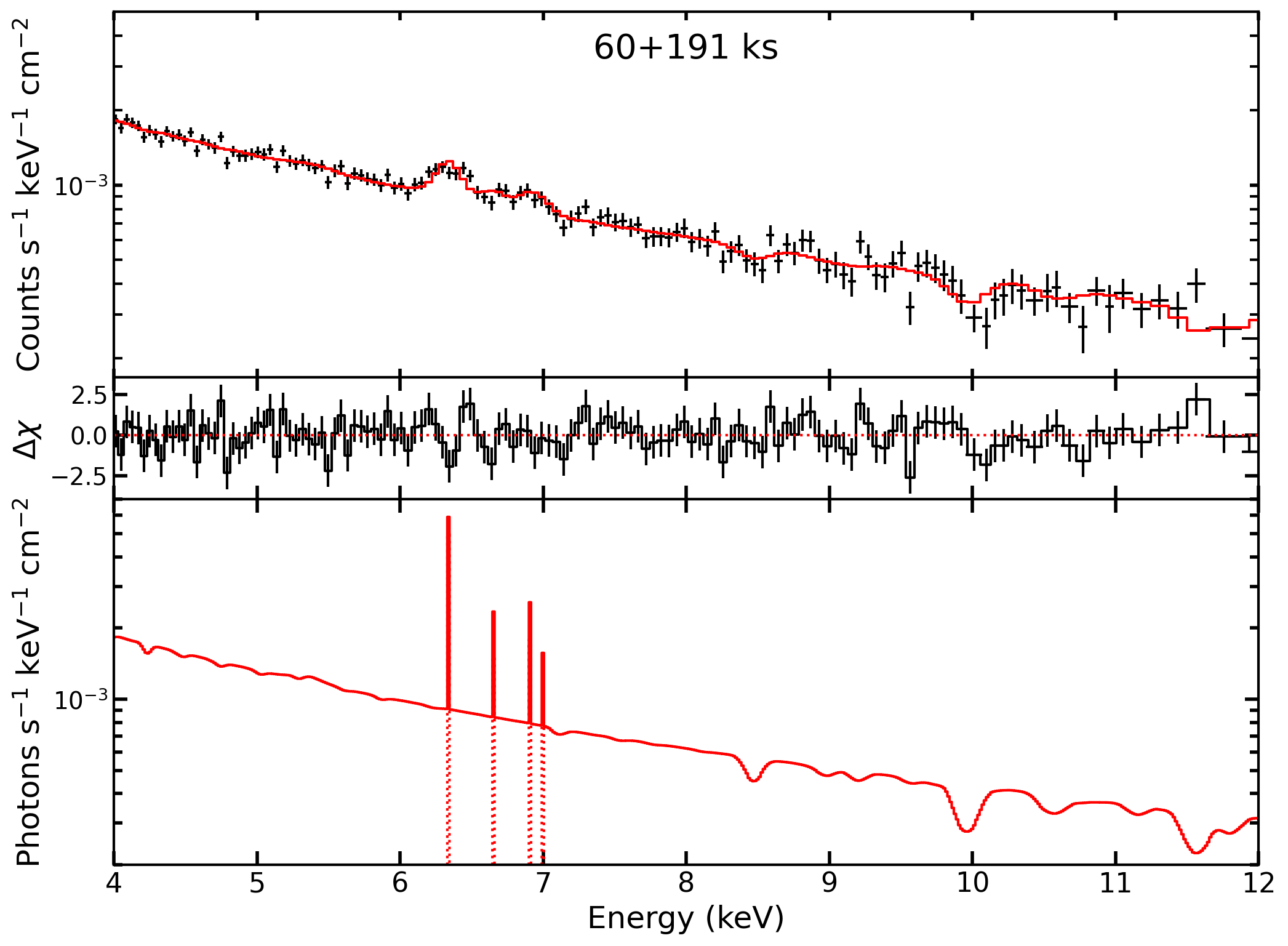}
\includegraphics[width=0.3\columnwidth]{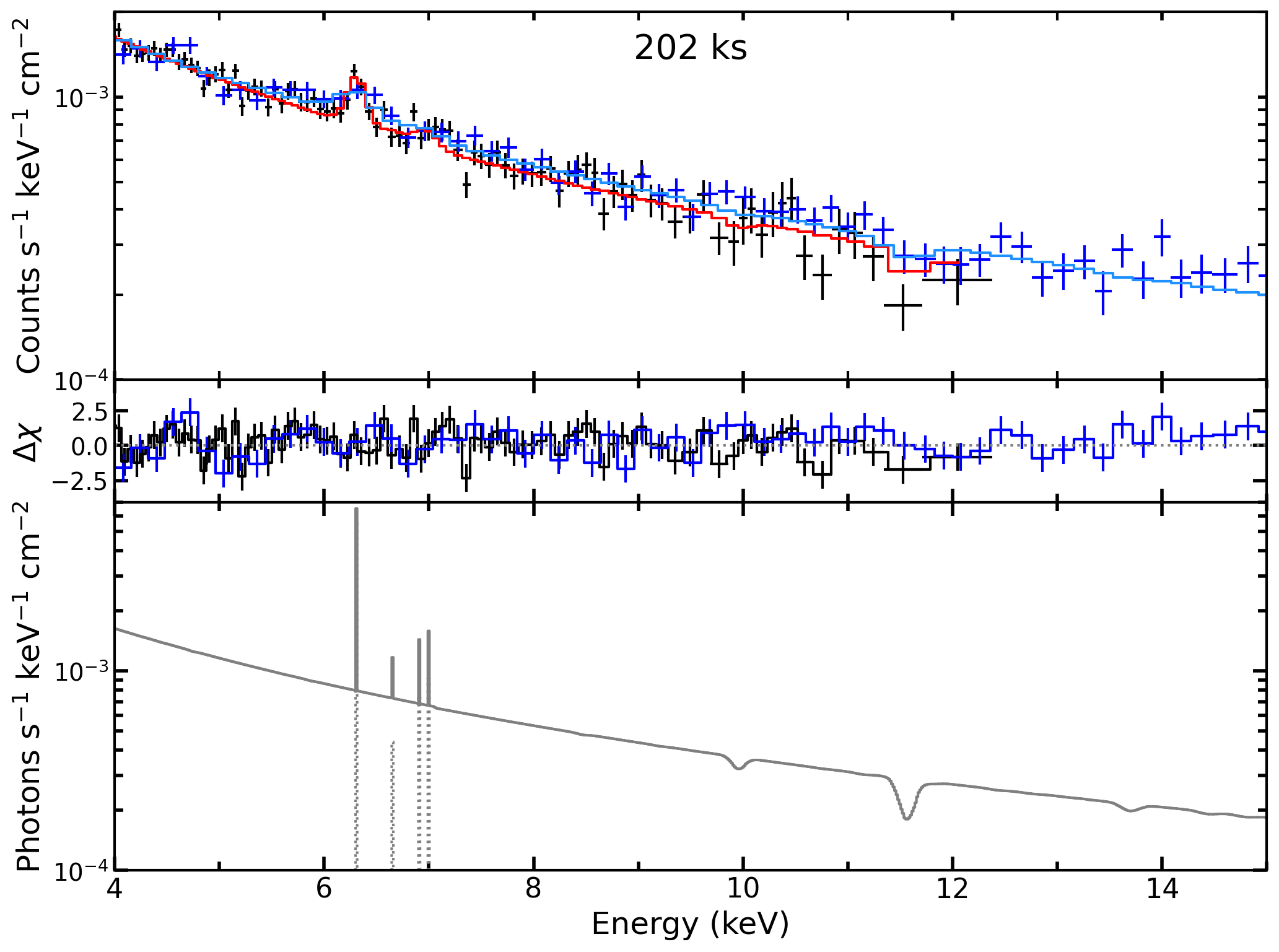}
\includegraphics[width=0.3\columnwidth]{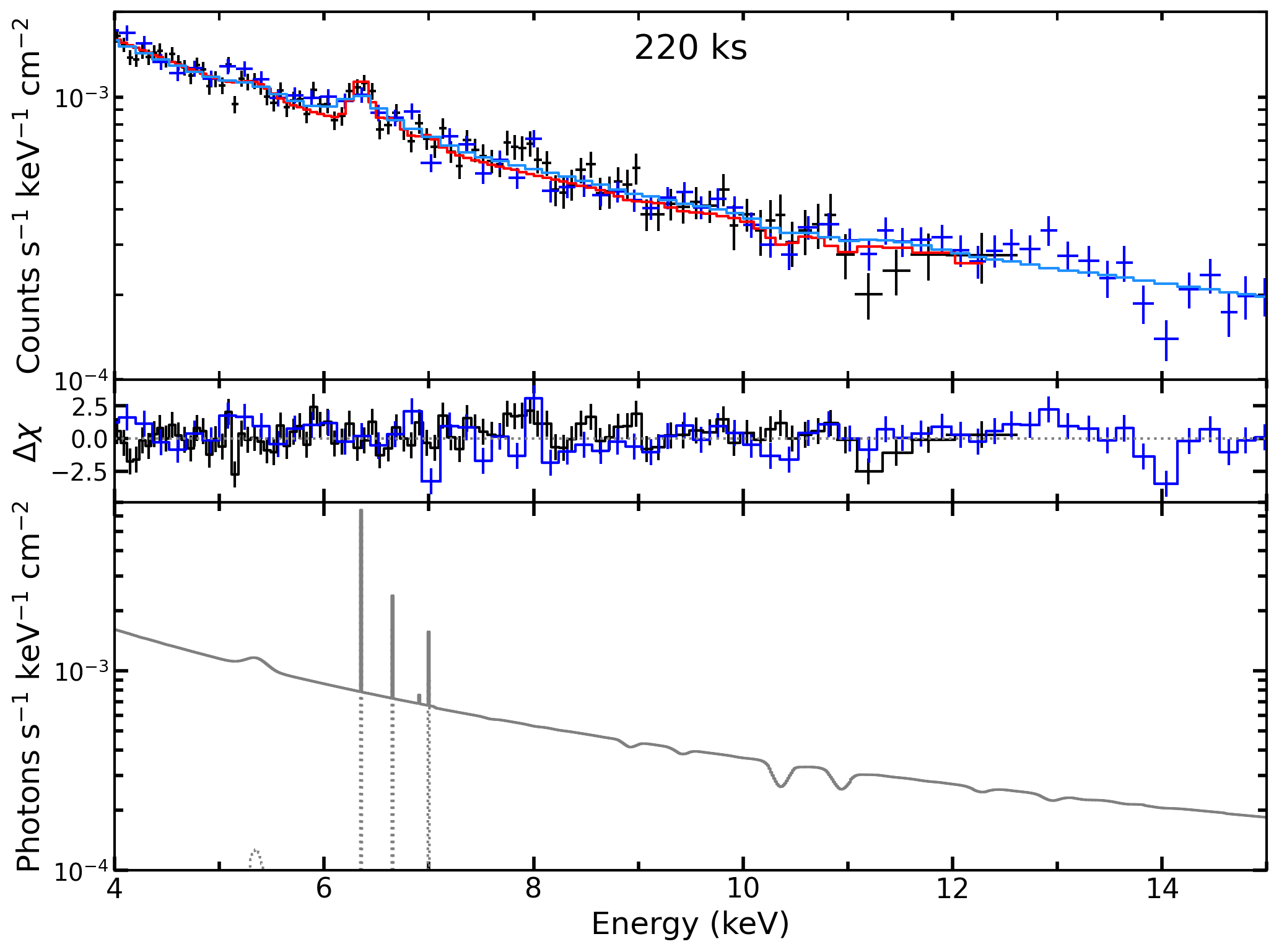}
\includegraphics[width=0.3\columnwidth]{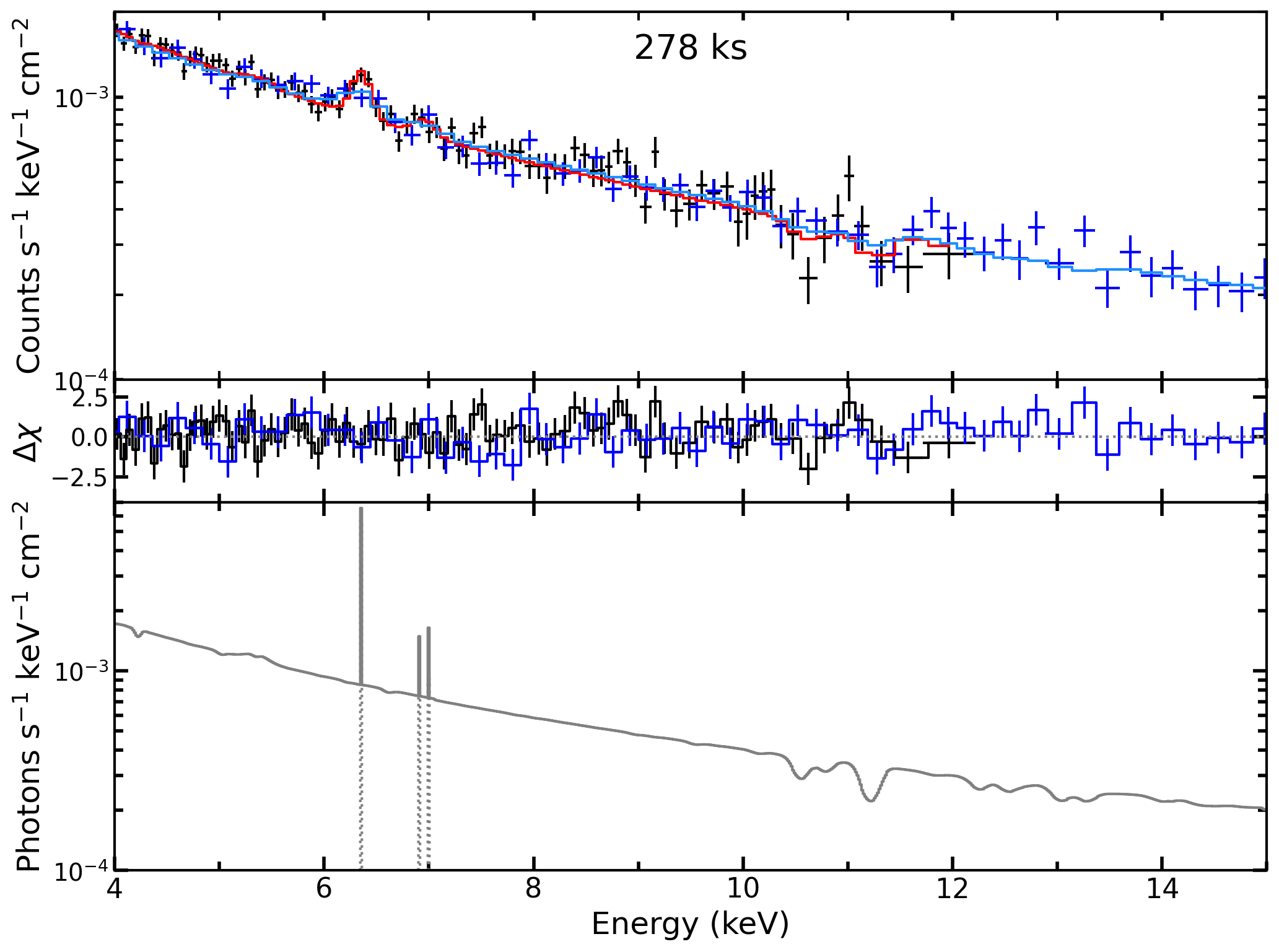}
\includegraphics[width=0.3\columnwidth]{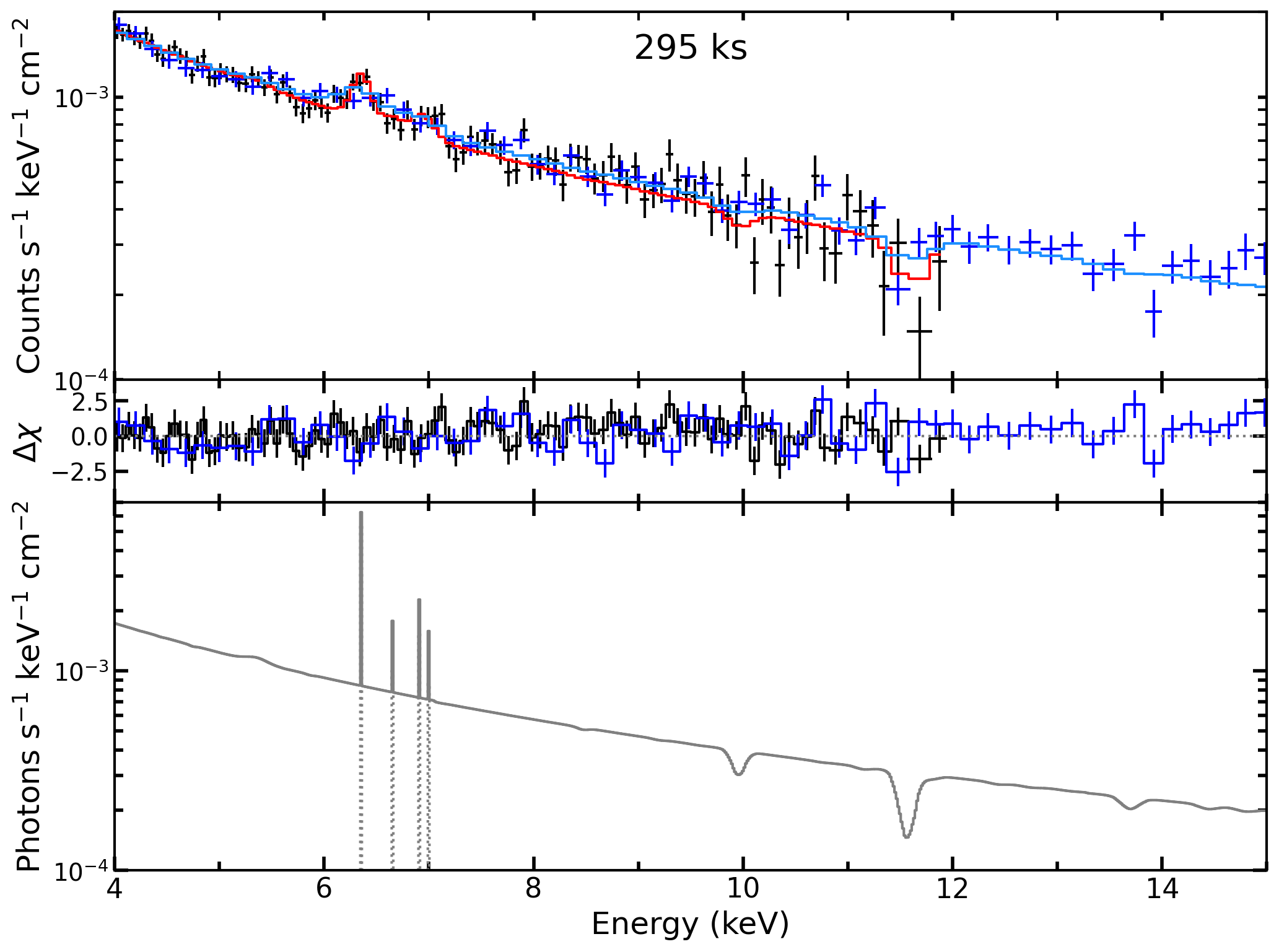}
\end{center}
\caption{Each box reports the best fit and the model for the observations fitted with WINE. {\it Top panels:} spectrum and best fit model, folded with the instrument response. {\it Middle panels:} residuals in units of $\chi^2$. {\it Bottom panels:} best fit model (unfolded). Datapoints are plotted in black(blue) and models in red(lightblue) for XMM-{\it Newton}({\it NuSTAR}).}
\label{wine_app}
\end{figure*}

Fig. \ref{wine_app} shows the best fits obtained with WINE. For each time slice we show: the datasets and the best fit model, folded with the instrumental response (top), the residuals (middle), the (unfolded) best fit model (bottom). Dotted lines indicate the Gaussian lines associated to the accretion disc emission (see Sect. \ref{stat}). 

\begin{figure*}
\centering
\includegraphics[width=0.6\paperwidth]{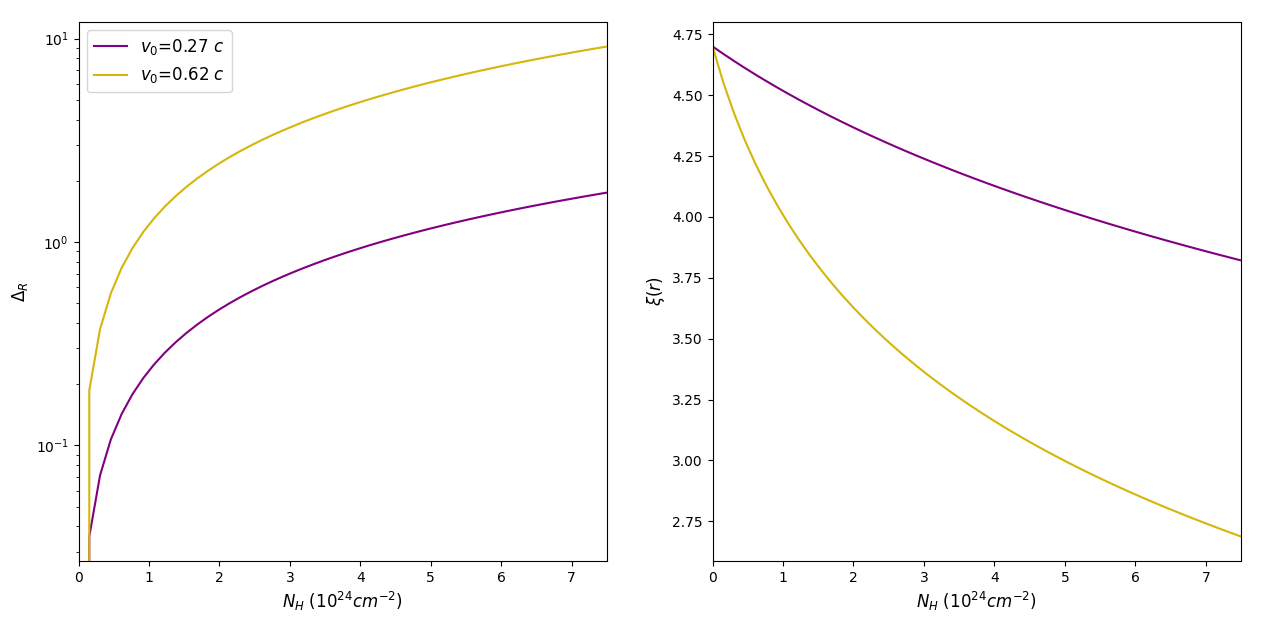}
\caption{Scalings of the wind radial thickness and ionisation parameter (left and right panel, respectively) for $\log\big(\frac{\xi_0}{\rm erg\ cm\ s^{-1}}\big)=4.7$ and v$_0$=0.27 c, =0.62 c, i.e. the best fit velocity and the velocity inferred for the E=14 kev line (purple and yellow line, respectively).}
\label{220ks_analitic}
\end{figure*}

We note that the E $\approx$ 14 keV absorption line detected by {\it NuSTAR} in the 220 ks spectrum, with a Monte Carlo-estimated significance of $2.07 \sigma$ (see Table \ref{bestfitPar}), would imply an outflow velocity $\sim$0.62 c if ascribed to Fe XXV He$\alpha$ or Fe XXVI Ly$\alpha$. We try to fit this line with WINE using an updated set of tables, spanning velocities up to v$_0$=0.70 c. However, we are not able to reproduce such a feature within our model, since for such high velocities the radial thickness $\Delta_r$ dramatically increases and, as a result, the ionisation parameter $\xi(r)$ strongly decreases along the wind column due to the geometric dilution of the ionising flux. For comparison, we show in Fig. \ref{220ks_analitic} a plot of $\Delta_r, \xi(r)$ (top and bottom panel, respectively) for the best fit values of the 220 ks observation, $\log\big(\frac{\xi_0}{\rm erg\ cm\ s^{-1}}\big)=4.7, \rm{v}_0$=0.27 c (purple line, see Table \ref{fitWINE}), together with a solution with same $\xi_0$ but v$_0$=0.62 c, where the effects of the higher velocity are clearly visible.

\section{Analytical constrains for $r_0$}
\label{r0_determination}
In Sect. \ref{winetables} we set a constrain for the launching radius of the wind $r_0$ starting from geometrical considerations. In particular, we required a small radial thickness of the wind $\Delta_r$ in order to obtain a reduced geometric dilution of the ionisation parameter $\xi(r)$ along the wind column. Using the best fit values for the time resolved spectra (see Table \ref{fitWINE}), we are now able to put updated constraints on $r_0$. For each time slice, the uncertainty associated to $\xi_0$ \footnote{We recall that $\xi_0$ is defined as the ionisation parameter $\xi(r)$ for $r=r_0$, i.e. at the inner boundary of the wind.} is given by its error bar; we define $\xi_{up},\xi_{low}$ as the maximum and minimum values at 90 \% c.l. (as an example, for the 10 ks time slice $\log\big(\frac{\xi_0}{\rm erg\ cm\ s^{-1}}\big)=4.6 \pm 0.6$, therefore $\log(\xi_{up})=5.2,\log(\xi_{low})=4.0$). 

Each value $\xi_{low} \leq \xi_0 \leq \xi_{up}$ (together with the best fit values $N_H, \rm{v}_0, \Gamma$) implies a different wind thickness. In fact, the gas density $n$ can be derived as $n=L'_{ion} /( \xi_0 r_0^2)= \Big(\frac{1-\rm{v}_0}{1+\rm{v}_0} \Big)^{\frac{2+\Gamma}{2}} \cdot L_{ion} /( \xi_0 r_0^2)$ (see Sect. \ref{xi_definition}), where $L_{ion}=2.67 \cdot 10^{43}$ erg s$^{-1}$, as in the rest of the paper. Then, the wind outer boundary (i.e. the final radius) can be determined as $r_{N_H} \equiv r(N_H)=r_0+N_H/n$, once assuming $n$ as constant throughout the wind (see Sect. \ref{winetables}). Thus, as a result of the geometric dilution of the ionising flux, the ionisation parameter will range from $\xi_0$ to $\xi (r_{N_H})$.

In order to obtain an order-of-magnitude estimate, we require the radial variation of $\xi(r)$ to be comprised within the interval $[\xi_{low}, \xi_{up}]$. This can be expressed mathematically by requiring that when $\xi_0=\xi_{up}$, then $\xi(r_{N_H}) \geq \xi_{low}$:
\begin{equation}
\begin{split}
\xi(r_{N_H}) &= \frac{L'_{ion}}{n (r_{N_H})^2} = \xi_{up}  \Big( \frac{r_0}{r_{N_H}} \Big)^2 \\ 
& = \xi_{up}  \Big( \frac{r_0}{r_0+r_0^2 N_H \xi_{up} /L'_{ion}} \Big)^2 \geq \xi_{low}
\end{split}
\end{equation}
which can be translated in the following condition for $r_0$:
\begin{equation}
    r_0 \leq r_0^{max} = \frac{\sqrt{\xi_{up}/\xi_{low}}-1}{N_H \xi_{up}/L'_{ion}}
\end{equation}
We report $r_0^{max}$ for the different time slices in Table \ref{rmax}. We note that the average value is very close to our assumed $r_0= 5 {\rm r_S}$, representing an important confirmation of our fitting strategy.

\begin{table}
\centering
\begin{tabular}{c c}
Time frame (ks) & $r_0^{max}$ (${\rm r_S}$) \\
\hline
10 & 1.2\\
35 & 3.2 \\
130 & 3.3\\
60+191 & 11.1\\
202 & 4.9\\
220 & 4.9\\
278 & 0.2\\
295 & 1.5\\
\hline
Mean & 3.4 \\
Median & 3.2 \\
\hline
\end{tabular}
\caption{Maximum launching radius $r_0^{max}$ for each given time slice. Last two lines report the mean and the median values, respectively}
\label{rmax}
\end{table}

\bibliography{sample631}{}

\begin{thebibliography}{}
\expandafter\ifx\csname natexlab\endcsname\relax\def\natexlab#1{#1}\fi
\providecommand{\url}[1]{\href{#1}{#1}}
\providecommand{\dodoi}[1]{doi:~\href{http://doi.org/#1}{\nolinkurl{#1}}}
\providecommand{\doeprint}[1]{\href{http://ascl.net/#1}{\nolinkurl{http://ascl.net/#1}}}
\providecommand{\doarXiv}[1]{\href{https://arxiv.org/abs/#1}{\nolinkurl{https://arxiv.org/abs/#1}}}

\bibitem[{{Alston} {et~al.}(2020){Alston}, {Fabian}, {Kara}, {Parker},
  {Dovciak}, {Pinto}, {Jiang}, {Middleton}, {Miniutti}, {Walton}, {Wilkins},
  {Buisson}, {Caballero-Garcia}, {Cackett}, {De Marco}, {Gallo}, {Lohfink},
  {Reynolds}, {Uttley}, {Young}, \& {Zogbhi}}]{alston20}
{Alston}, W.~N., {Fabian}, A.~C., {Kara}, E., {et~al.} 2020, Nature Astronomy,
  4, 597, \dodoi{10.1038/s41550-019-1002-x}

\bibitem[{{Arnaud}(1996)}]{Xspec}
{Arnaud}, K.~A. 1996, in ASP Conf. Ser. 101: Astronomical Data Analysis
  Software and Systems V, 17

\bibitem[{{Astropy Collaboration} {et~al.}(2013){Astropy Collaboration},
  {Robitaille}, {Tollerud}, {Greenfield}, {Droettboom}, {Bray}, {Aldcroft},
  {Davis}, {Ginsburg}, {Price-Whelan}, {Kerzendorf}, {Conley}, {Crighton},
  {Barbary}, {Muna}, {Ferguson}, {Grollier}, {Parikh}, {Nair}, {Unther},
  {Deil}, {Woillez}, {Conseil}, {Kramer}, {Turner}, {Singer}, {Fox}, {Weaver},
  {Zabalza}, {Edwards}, {Azalee Bostroem}, {Burke}, {Casey}, {Crawford},
  {Dencheva}, {Ely}, {Jenness}, {Labrie}, {Lim}, {Pierfederici}, {Pontzen},
  {Ptak}, {Refsdal}, {Servillat}, \& {Streicher}}]{astropy13}
{Astropy Collaboration}, {Robitaille}, T.~P., {Tollerud}, E.~J., {et~al.} 2013,
  \aap, 558, A33, \dodoi{10.1051/0004-6361/201322068}

\bibitem[{{Astropy Collaboration} {et~al.}(2018){Astropy Collaboration},
  {Price-Whelan}, {Sip{\H{o}}cz}, {G{\"u}nther}, {Lim}, {Crawford}, {Conseil},
  {Shupe}, {Craig}, {Dencheva}, {Ginsburg}, {Vand erPlas}, {Bradley},
  {P{\'e}rez-Su{\'a}rez}, {de Val-Borro}, {Aldcroft}, {Cruz}, {Robitaille},
  {Tollerud}, {Ardelean}, {Babej}, {Bach}, {Bachetti}, {Bakanov}, {Bamford},
  {Barentsen}, {Barmby}, {Baumbach}, {Berry}, {Biscani}, {Boquien}, {Bostroem},
  {Bouma}, {Brammer}, {Bray}, {Breytenbach}, {Buddelmeijer}, {Burke},
  {Calderone}, {Cano Rodr{\'\i}guez}, {Cara}, {Cardoso}, {Cheedella}, {Copin},
  {Corrales}, {Crichton}, {D'Avella}, {Deil}, {Depagne}, {Dietrich}, {Donath},
  {Droettboom}, {Earl}, {Erben}, {Fabbro}, {Ferreira}, {Finethy}, {Fox},
  {Garrison}, {Gibbons}, {Goldstein}, {Gommers}, {Greco}, {Greenfield},
  {Groener}, {Grollier}, {Hagen}, {Hirst}, {Homeier}, {Horton}, {Hosseinzadeh},
  {Hu}, {Hunkeler}, {Ivezi{\'c}}, {Jain}, {Jenness}, {Kanarek}, {Kendrew},
  {Kern}, {Kerzendorf}, {Khvalko}, {King}, {Kirkby}, {Kulkarni}, {Kumar},
  {Lee}, {Lenz}, {Littlefair}, {Ma}, {Macleod}, {Mastropietro}, {McCully},
  {Montagnac}, {Morris}, {Mueller}, {Mumford}, {Muna}, {Murphy}, {Nelson},
  {Nguyen}, {Ninan}, {N{\"o}the}, {Ogaz}, {Oh}, {Parejko}, {Parley}, {Pascual},
  {Patil}, {Patil}, {Plunkett}, {Prochaska}, {Rastogi}, {Reddy Janga},
  {Sabater}, {Sakurikar}, {Seifert}, {Sherbert}, {Sherwood-Taylor}, {Shih},
  {Sick}, {Silbiger}, {Singanamalla}, {Singer}, {Sladen}, {Sooley},
  {Sornarajah}, {Streicher}, {Teuben}, {Thomas}, {Tremblay}, {Turner},
  {Terr{\'o}n}, {van Kerkwijk}, {de la Vega}, {Watkins}, {Weaver}, {Whitmore},
  {Woillez}, {Zabalza}, \& {Astropy Contributors}}]{astropy18}
{Astropy Collaboration}, {Price-Whelan}, A.~M., {Sip{\H{o}}cz}, B.~M., {et~al.}
  2018, \aj, 156, 123, \dodoi{10.3847/1538-3881/aabc4f}

\bibitem[{{Bianchi} {et~al.}(2009){Bianchi}, {Piconcelli}, {Chiaberge},
  {Bail{\'o}n}, {Matt}, \& {Fiore}}]{bianchi09c}
{Bianchi}, S., {Piconcelli}, E., {Chiaberge}, M., {et~al.} 2009, \apj, 695,
  781, \dodoi{10.1088/0004-637X/695/1/781}

\bibitem[{{Blandford} \& {Payne}(1982)}]{blandford82}
{Blandford}, R.~D., \& {Payne}, D.~G. 1982, \mnras, 199, 883,
  \dodoi{10.1093/mnras/199.4.883}

\bibitem[{{Braito} {et~al.}(2021){Braito}, {Reeves}, {Severgnini}, {Della
  Ceca}, {Ballo}, {Cicone}, {Matzeu}, {Serafinelli}, \& {Sirressi}}]{brs21}
{Braito}, V., {Reeves}, J.~N., {Severgnini}, P., {et~al.} 2021, \mnras, 500,
  291, \dodoi{10.1093/mnras/staa3264}

\bibitem[{{Cash}(1976)}]{cash76}
{Cash}, W. 1976, \aap, 52, 307

\bibitem[{{Chartas} {et~al.}(2021){Chartas}, {Cappi}, {Vignali}, {Dadina},
  {James}, {Lanzuisi}, {Giustini}, {Gaspari}, {Strickland}, \&
  {Bertola}}]{chartas21}
{Chartas}, G., {Cappi}, M., {Vignali}, C., {et~al.} 2021, \apj, 920, 24,
  \dodoi{10.3847/1538-4357/ac0ef2}

\bibitem[{{Costanzo} {et~al.}(2021){Costanzo}, {Dadina}, {Vignali}, {De Marco},
  {Cappi}, {Petrucci}, {Bianchi}, {Kriss}, {Kaastra}, {Mehdipour}, {Behar}, \&
  {Matzeu}}]{cdv21}
{Costanzo}, D., {Dadina}, M., {Vignali}, C., {et~al.} 2021, arXiv e-prints,
  arXiv:2112.09096.
\newblock \doarXiv{2112.09096}

\bibitem[{{Crenshaw} \& {Kraemer}(2012)}]{crenshaw2012}
{Crenshaw}, D.~M., \& {Kraemer}, S.~B. 2012, \apj, 753, 75,
  \dodoi{10.1088/0004-637X/753/1/75}

\bibitem[{{Crenshaw} {et~al.}(2003){Crenshaw}, {Kraemer}, \&
  {George}}]{crenshaw03}
{Crenshaw}, D.~M., {Kraemer}, S.~B., \& {George}, I.~M. 2003, \araa, 41, 117,
  \dodoi{10.1146/annurev.astro.41.082801.100328}

\bibitem[{{Cui} \& {Yuan}(2020)}]{cui20}
{Cui}, C., \& {Yuan}, F. 2020, \apj, 890, 81, \dodoi{10.3847/1538-4357/ab6e6f}

\bibitem[{{Dannen} {et~al.}(2019){Dannen}, {Proga}, {Kallman}, \&
  {Waters}}]{dannen19}
{Dannen}, R.~C., {Proga}, D., {Kallman}, T.~R., \& {Waters}, T. 2019, \apj,
  882, 99, \dodoi{10.3847/1538-4357/ab340b}

\bibitem[{{Di Matteo} {et~al.}(2005){Di Matteo}, {Springel}, \&
  {Hernquist}}]{dsh05}
{Di Matteo}, T., {Springel}, V., \& {Hernquist}, L. 2005, \nat, 433, 604,
  \dodoi{10.1038/nature03335}

\bibitem[{{Faucher-Gigu{\`e}re} \& {Quataert}(2012)}]{Faucher12}
{Faucher-Gigu{\`e}re}, C.-A., \& {Quataert}, E. 2012, \mnras, 425, 605,
  \dodoi{10.1111/j.1365-2966.2012.21512.x}

\bibitem[{{Ferland} {et~al.}(2017){Ferland}, {Chatzikos}, {Guzm{\'a}n},
  {Lykins}, {van Hoof}, {Williams}, {Abel}, {Badnell}, {Keenan}, {Porter}, \&
  {Stancil}}]{fcg17}
{Ferland}, G.~J., {Chatzikos}, M., {Guzm{\'a}n}, F., {et~al.} 2017, \rmxaa, 53,
  385.
\newblock \doarXiv{1705.10877}

\bibitem[{{Fiore} {et~al.}(2023){Fiore}, {Gaspari}, {Luminari}, {Tozzi}, \& {De
  Arcangelis}}]{fiore23}
{Fiore}, F., {Gaspari}, M., {Luminari}, A., {Tozzi}, P., \& {De Arcangelis}, L.
  2023, arXiv e-prints, arXiv:2304.12696, \dodoi{10.48550/arXiv.2304.12696}

\bibitem[{{Fiore} {et~al.}(2017){Fiore}, {Feruglio}, {Shankar}, {Bischetti},
  {Bongiorno}, {Brusa}, {Carniani}, {Cicone}, {Duras}, {Lamastra}, {Mainieri},
  {Marconi}, {Menci}, {Maiolino}, {Piconcelli}, {Vietri}, \&
  {Zappacosta}}]{Fiore17}
{Fiore}, F., {Feruglio}, C., {Shankar}, F., {et~al.} 2017, \aap, 601, A143,
  \dodoi{10.1051/0004-6361/201629478}

\bibitem[{{Fukumura} {et~al.}(2010){Fukumura}, {Kazanas}, {Contopoulos}, \&
  {Behar}}]{fkc10}
{Fukumura}, K., {Kazanas}, D., {Contopoulos}, I., \& {Behar}, E. 2010, \apj,
  715, 636, \dodoi{10.1088/0004-637X/715/1/636}

\bibitem[{{Gaspari} \& {Sadowski}(2017)}]{gaspari17}
{Gaspari}, M., \& {Sadowski}, A. 2017, \apj, 837, 149,
  \dodoi{10.3847/1538-4357/aa61a3}

\bibitem[{{Gofford} {et~al.}(2015){Gofford}, {Reeves}, {McLaughlin}, {Braito},
  {Turner}, {Tombesi}, \& {Cappi}}]{grm15}
{Gofford}, J., {Reeves}, J.~N., {McLaughlin}, D.~E., {et~al.} 2015, \mnras,
  451, 4169, \dodoi{10.1093/mnras/stv1207}

\bibitem[{{Gofford} {et~al.}(2013){Gofford}, {Reeves}, {Tombesi}, {Braito},
  {Turner}, {Miller}, \& {Cappi}}]{grt13}
{Gofford}, J., {Reeves}, J.~N., {Tombesi}, F., {et~al.} 2013, \mnras, 430, 60,
  \dodoi{10.1093/mnras/sts481}

\bibitem[{{Guolo} {et~al.}(2021){Guolo}, {Ruschel-Dutra}, {Grupe}, {Peterson},
  {Storchi-Bergmann}, {Schimoia}, {Nemmen}, \& {Robinson}}]{guolo21}
{Guolo}, M., {Ruschel-Dutra}, D., {Grupe}, D., {et~al.} 2021, \mnras, 508, 144,
  \dodoi{10.1093/mnras/stab2550}

\bibitem[{{Guolo-Pereira} {et~al.}(2021){Guolo-Pereira}, {Ruschel-Dutra},
  {Storchi-Bergmann}, {Schnorr-M{\"u}ller}, {Cid Fernandes}, {Couto},
  {Dametto}, \& {Hernandez-Jimenez}}]{Pereira21}
{Guolo-Pereira}, M., {Ruschel-Dutra}, D., {Storchi-Bergmann}, T., {et~al.}
  2021, \mnras, 502, 3618, \dodoi{10.1093/mnras/stab245}

\bibitem[{Harris {et~al.}(2020)Harris, Millman, van~der Walt, Gommers,
  Virtanen, Cournapeau, Wieser, Taylor, Berg, Smith, Kern, Picus, Hoyer, van
  Kerkwijk, Brett, Haldane, del R{\'{i}}o, Wiebe, Peterson,
  G{\'{e}}rard-Marchant, Sheppard, Reddy, Weckesser, Abbasi, Gohlke, \&
  Oliphant}]{numpy}
Harris, C.~R., Millman, K.~J., van~der Walt, S.~J., {et~al.} 2020, Nature, 585,
  357, \dodoi{10.1038/s41586-020-2649-2}

\bibitem[{{Higginbottom} {et~al.}(2014){Higginbottom}, {Proga}, {Knigge},
  {Long}, {Matthews}, \& {Sim}}]{higginbottom14}
{Higginbottom}, N., {Proga}, D., {Knigge}, C., {et~al.} 2014, \apj, 789, 19,
  \dodoi{10.1088/0004-637X/789/1/19}

\bibitem[{{Hopkins} \& {Elvis}(2010)}]{he10}
{Hopkins}, P.~F., \& {Elvis}, M. 2010, \mnras, 401, 7,
  \dodoi{10.1111/j.1365-2966.2009.15643.x}

\bibitem[{Hunter(2007)}]{h07}
Hunter, J.~D. 2007, Computing in Science \& Engineering, 9, 90,
  \dodoi{10.1109/MCSE.2007.55}

\bibitem[{{Igo} {et~al.}(2020){Igo}, {Parker}, {Matzeu}, {Alston}, {Alvarez
  Crespo}, {F{\"u}rst}, {Buisson}, {Lobban}, {Joyce}, {Mallick}, {Schartel}, \&
  {Santos-Lle{\'o}}}]{igo20}
{Igo}, Z., {Parker}, M.~L., {Matzeu}, G.~A., {et~al.} 2020, \mnras, 493, 1088,
  \dodoi{10.1093/mnras/staa265}

\bibitem[{{Irwin} {et~al.}(2017){Irwin}, {Schmidt}, {Damas-Segovia}, {Beck},
  {English}, {Heald}, {Henriksen}, {Krause}, {Li}, {Rand}, {Wang}, {Wiegert},
  {Kamieneski}, {Par{\'e}}, \& {Sullivan}}]{irwin17}
{Irwin}, J.~A., {Schmidt}, P., {Damas-Segovia}, A., {et~al.} 2017, \mnras, 464,
  1333, \dodoi{10.1093/mnras/stw2414}

\bibitem[{{Kaastra} {et~al.}(1996){Kaastra}, {Mewe}, \&
  {Nieuwenhuijzen}}]{spex}
{Kaastra}, J.~S., {Mewe}, R., \& {Nieuwenhuijzen}, H. 1996, in UV and X-ray
  Spectroscopy of Astrophysical and Laboratory Plasmas, 411--414

\bibitem[{{Kalberla} {et~al.}(2005){Kalberla}, {Burton}, {Hartmann}, {Arnal},
  {Bajaja}, {Morras}, \& {P{\"o}ppel}}]{kalberla05}
{Kalberla}, P.~M.~W., {Burton}, W.~B., {Hartmann}, D., {et~al.} 2005, \aap,
  440, 775, \dodoi{10.1051/0004-6361:20041864}

\bibitem[{{Kallman} \& {Bautista}(2001)}]{xstar}
{Kallman}, T., \& {Bautista}, M. 2001, \apjs, 133, 221, \dodoi{10.1086/319184}

\bibitem[{{Kara} {et~al.}(2019){Kara}, {Steiner}, {Fabian}, {Cackett},
  {Uttley}, {Remillard}, {Gendreau}, {Arzoumanian}, {Altamirano}, {Eikenberry},
  {Enoto}, {Homan}, {Neilsen}, \& {Stevens}}]{kara19}
{Kara}, E., {Steiner}, J.~F., {Fabian}, A.~C., {et~al.} 2019, \nat, 565, 198,
  \dodoi{10.1038/s41586-018-0803-x}

\bibitem[{{Keel}(1996)}]{keel96}
{Keel}, W.~C. 1996, \apjs, 106, 27, \dodoi{10.1086/192326}

\bibitem[{{King} \& {Pounds}(2015)}]{King15}
{King}, A., \& {Pounds}, K. 2015, \araa, 53, 115,
  \dodoi{10.1146/annurev-astro-082214-122316}

\bibitem[{{Krongold} {et~al.}(2007){Krongold}, {Nicastro}, {Elvis},
  {Brickhouse}, {Binette}, {Mathur}, \& {Jim{\'e}nez-Bail{\'o}n}}]{kne07}
{Krongold}, Y., {Nicastro}, F., {Elvis}, M., {et~al.} 2007, \apj, 659, 1022,
  \dodoi{10.1086/512476}

\bibitem[{{Laurenti} {et~al.}(2021){Laurenti}, {Luminari}, {Tombesi},
  {Vagnetti}, {Middei}, \& {Piconcelli}}]{llt21}
{Laurenti}, M., {Luminari}, A., {Tombesi}, F., {et~al.} 2021, \aap, 645, A118,
  \dodoi{10.1051/0004-6361/202039409}

\bibitem[{{Luminari} {et~al.}(2021){Luminari}, {Nicastro}, {Elvis},
  {Piconcelli}, {Tombesi}, {Zappacosta}, \& {Fiore}}]{lne21}
{Luminari}, A., {Nicastro}, F., {Elvis}, M., {et~al.} 2021, \aap, 646, A111,
  \dodoi{10.1051/0004-6361/202039396}

\bibitem[{{Luminari} {et~al.}(2022){Luminari}, {Nicastro}, {Krongold}, {Piro},
  \& {Linesh Thakur}}]{2022arXiv221201399L}
{Luminari}, A., {Nicastro}, F., {Krongold}, Y., {Piro}, L., \& {Linesh Thakur},
  A. 2022, arXiv e-prints, arXiv:2212.01399, \dodoi{10.48550/arXiv.2212.01399}

\bibitem[{{Luminari} {et~al.}(2018){Luminari}, {Piconcelli}, {Tombesi},
  {Zappacosta}, {Fiore}, {Piro}, \& {Vagnetti}}]{lpt18}
{Luminari}, A., {Piconcelli}, E., {Tombesi}, F., {et~al.} 2018, \aap, 619,
  A149, \dodoi{10.1051/0004-6361/201833623}

\bibitem[{{Luminari} {et~al.}(2020){Luminari}, {Tombesi}, {Piconcelli},
  {Nicastro}, {Fukumura}, {Kazanas}, {Fiore}, \& {Zappacosta}}]{ltp20}
{Luminari}, A., {Tombesi}, F., {Piconcelli}, E., {et~al.} 2020, \aap, 633, A55,
  \dodoi{10.1051/0004-6361/201936797}

\bibitem[{{Madsen} {et~al.}(2015){Madsen}, {F{\"u}rst}, {Walton}, {Harrison},
  {Nalewajko}, {Ballantyne}, {Boggs}, {Brenneman}, {Christensen}, {Craig},
  {Fabian}, {Forster}, {Grefenstette}, {Guainazzi}, {Hailey}, {Madejski},
  {Matt}, {Stern}, {Walter}, \& {Zhang}}]{madsen15}
{Madsen}, K.~K., {F{\"u}rst}, F., {Walton}, D.~J., {et~al.} 2015, \apj, 812,
  14, \dodoi{10.1088/0004-637X/812/1/14}

\bibitem[{{Marinucci} {et~al.}(2020){Marinucci}, {Bianchi}, {Braito}, {De
  Marco}, {Matt}, {Middei}, {Nardini}, \& {Reeves}}]{mbb20}
{Marinucci}, A., {Bianchi}, S., {Braito}, V., {et~al.} 2020, \mnras, 496, 3412,
  \dodoi{10.1093/mnras/staa1683}

\bibitem[{{Marinucci} {et~al.}(2018){Marinucci}, {Bianchi}, {Braito}, {Matt},
  {Nardini}, \& {Reeves}}]{mbb18}
---. 2018, \mnras, 478, 5638, \dodoi{10.1093/mnras/sty1436}

\bibitem[{{Matzeu} {et~al.}(2022){Matzeu}, {Brusa}, {Lanzuisi}, {Dadina},
  {Bianchi}, {Kriss}, {Mehdipour}, {Nardini}, {Chartas}, {Middei},
  {Piconcelli}, {Gianolli}, {Comastri}, {Longinotti}, {Krongold}, {Ricci},
  {Petrucci}, {Tombesi}, {Luminari}, {Zappacosta}, {Miniutti}, {Gaspari},
  {Behar}, {Bischetti}, {Mathur}, {Perna}, {Giustini}, {Grandi}, {Torresi},
  {Vignali}, {Bruni}, {Cappi}, {Costantini}, {Cresci}, {De Marco}, {De Rosa},
  {Gilli}, {Guainazzi}, {Kaastra}, {Kraemer}, {La Franca}, {Marconi},
  {Panessa}, {Ponti}, {Proga}, {Ursini}, {Fiore}, {King}, {Maiolino}, {Matt},
  \& {Merloni}}]{matzeu22}
{Matzeu}, G.~A., {Brusa}, M., {Lanzuisi}, G., {et~al.} 2022, arXiv e-prints,
  arXiv:2212.02960, \dodoi{10.48550/arXiv.2212.02960}

\bibitem[{{Middei} {et~al.}(2022){Middei}, {Marinucci}, {Braito}, {Bianchi},
  {De Marco}, {Luminari}, {Matt}, {Nardini}, {Perri}, {Reeves}, \&
  {Vagnetti}}]{middei20}
{Middei}, R., {Marinucci}, A., {Braito}, V., {et~al.} 2022, \mnras, 514, 2974,
  \dodoi{10.1093/mnras/stac1381}

\bibitem[{{Mingozzi} {et~al.}(2019){Mingozzi}, {Cresci}, {Venturi}, {Marconi},
  {Mannucci}, {Perna}, {Belfiore}, {Carniani}, {Balmaverde}, {Brusa}, {Cicone},
  {Feruglio}, {Gallazzi}, {Mainieri}, {Maiolino}, {Nagao}, {Nardini}, {Sani},
  {Tozzi}, \& {Zibetti}}]{Mingozzi19}
{Mingozzi}, M., {Cresci}, G., {Venturi}, G., {et~al.} 2019, \aap, 622, A146,
  \dodoi{10.1051/0004-6361/201834372}

\bibitem[{{Murphy} {et~al.}(2007){Murphy}, {Yaqoob}, \& {Terashima}}]{mkt07}
{Murphy}, K.~D., {Yaqoob}, T., \& {Terashima}, Y. 2007, \apj, 666, 96,
  \dodoi{10.1086/520039}

\bibitem[{{Nardini} {et~al.}(2015){Nardini}, {Reeves}, {Gofford}, {Harrison},
  {Risaliti}, {Braito}, {Costa}, {Matzeu}, {Walton}, {Behar}, {Boggs},
  {Christensen}, {Craig}, {Hailey}, {Matt}, {Miller}, {O'Brien}, {Stern},
  {Turner}, \& {Ward}}]{nrg15}
{Nardini}, E., {Reeves}, J.~N., {Gofford}, J., {et~al.} 2015, Science, 347,
  860, \dodoi{10.1126/science.1259202}

\bibitem[{{Nicastro} {et~al.}(1999){Nicastro}, {Fiore}, {Perola}, \&
  {Elvis}}]{nfp99}
{Nicastro}, F., {Fiore}, F., {Perola}, G.~C., \& {Elvis}, M. 1999, \apj, 512,
  184, \dodoi{10.1086/306736}

\bibitem[{{Parker} {et~al.}(2017){Parker}, {Pinto}, {Fabian}, {Lohfink},
  {Buisson}, {Alston}, {Kara}, {Cackett}, {Chiang}, {Dauser}, {De Marco},
  {Gallo}, {Garcia}, {Harrison}, {King}, {Middleton}, {Miller}, {Miniutti},
  {Reynolds}, {Uttley}, {Vasudevan}, {Walton}, {Wilkins}, \& {Zoghbi}}]{ppf17}
{Parker}, M.~L., {Pinto}, C., {Fabian}, A.~C., {et~al.} 2017, \nat, 543, 83,
  \dodoi{10.1038/nature21385}

\bibitem[{{Parker} {et~al.}(2020){Parker}, {Matzeu}, {Alston}, {Fabian},
  {Lobban}, {Miniutti}, {Pinto}, {Santos-Lle{\'o}}, \& {Schartel}}]{pma20}
{Parker}, M.~L., {Matzeu}, G.~A., {Alston}, W.~N., {et~al.} 2020, \mnras, 498,
  L140, \dodoi{10.1093/mnrasl/slaa144}

\bibitem[{{Proga} {et~al.}(2000){Proga}, {Stone}, \& {Kallman}}]{psk00}
{Proga}, D., {Stone}, J.~M., \& {Kallman}, T.~R. 2000, \apj, 543, 686,
  \dodoi{10.1086/317154}

\bibitem[{{Protassov} {et~al.}(2002){Protassov}, {van Dyk}, {Connors},
  {Kashyap}, \& {Siemiginowska}}]{prot02}
{Protassov}, R., {van Dyk}, D.~A., {Connors}, A., {Kashyap}, V.~L., \&
  {Siemiginowska}, A. 2002, \apj, 571, 545, \dodoi{10.1086/339856}

\bibitem[{{Pudritz} {et~al.}(2007){Pudritz}, {Ouyed}, {Fendt}, \&
  {Brandenburg}}]{pudritz07}
{Pudritz}, R.~E., {Ouyed}, R., {Fendt}, C., \& {Brandenburg}, A. 2007, in
  Protostars and Planets V, ed. B.~{Reipurth}, D.~{Jewitt}, \& K.~{Keil}, 277,
  \dodoi{10.48550/arXiv.astro-ph/0603592}

\bibitem[{{Reeves} {et~al.}(2018){Reeves}, {Braito}, {Nardini}, {Lobban},
  {Matzeu}, \& {Costa}}]{rbn18}
{Reeves}, J.~N., {Braito}, V., {Nardini}, E., {et~al.} 2018, \apjl, 854, L8,
  \dodoi{10.3847/2041-8213/aaaae1}

\bibitem[{{Risaliti} {et~al.}(2005){Risaliti}, {Elvis}, {Fabbiano}, {Baldi}, \&
  {Zezas}}]{ris05}
{Risaliti}, G., {Elvis}, M., {Fabbiano}, G., {Baldi}, A., \& {Zezas}, A. 2005,
  \apjl, 623, L93, \dodoi{10.1086/430252}

\bibitem[{{Shakura} \& {Sunyaev}(1973)}]{ss73}
{Shakura}, N.~I., \& {Sunyaev}, R.~A. 1973, \aap, 24, 337

\bibitem[{{Tombesi} {et~al.}(2012){Tombesi}, {Cappi}, {Reeves}, \&
  {Braito}}]{tcr12}
{Tombesi}, F., {Cappi}, M., {Reeves}, J.~N., \& {Braito}, V. 2012, \mnras, 422,
  L1, \dodoi{10.1111/j.1745-3933.2012.01221.x}

\bibitem[{{Tombesi} {et~al.}(2011){Tombesi}, {Cappi}, {Reeves}, {Palumbo},
  {Braito}, \& {Dadina}}]{tcr11}
{Tombesi}, F., {Cappi}, M., {Reeves}, J.~N., {et~al.} 2011, \apj, 742, 44,
  \dodoi{10.1088/0004-637X/742/1/44}

\bibitem[{{Tombesi} {et~al.}(2010){Tombesi}, {Cappi}, {Reeves}, {Palumbo},
  {Yaqoob}, {Braito}, \& {Dadina}}]{tcr10}
---. 2010, \aap, 521, A57, \dodoi{10.1051/0004-6361/200913440}

\bibitem[{{Tombesi} {et~al.}(2015){Tombesi}, {Mel{\'e}ndez}, {Veilleux},
  {Reeves}, {Gonz{\'a}lez-Alfonso}, \& {Reynolds}}]{tmv15}
{Tombesi}, F., {Mel{\'e}ndez}, M., {Veilleux}, S., {et~al.} 2015, \nat, 519,
  436, \dodoi{10.1038/nature14261}

\bibitem[{{Torrey} {et~al.}(2020){Torrey}, {Hopkins}, {Faucher-Gigu{\`e}re},
  {Angl{\'e}s-Alc{\'a}zar}, {Quataert}, {Ma}, {Feldmann}, {Keres}, \&
  {Murray}}]{torrey20}
{Torrey}, P., {Hopkins}, P.~F., {Faucher-Gigu{\`e}re}, C.-A., {et~al.} 2020,
  \mnras, 497, 5292, \dodoi{10.1093/mnras/staa2222}

\bibitem[{{Trippe} {et~al.}(2008){Trippe}, {Crenshaw}, {Deo}, \&
  {Dietrich}}]{trippe08}
{Trippe}, M.~L., {Crenshaw}, D.~M., {Deo}, R., \& {Dietrich}, M. 2008, \aj,
  135, 2048, \dodoi{10.1088/0004-6256/135/6/2048}

\bibitem[{{Veilleux} {et~al.}(2001){Veilleux}, {Shopbell}, \&
  {Miller}}]{veilleux01}
{Veilleux}, S., {Shopbell}, P.~L., \& {Miller}, S.~T. 2001, \aj, 121, 198,
  \dodoi{10.1086/318046}

\bibitem[{{Walton} {et~al.}(2014){Walton}, {Risaliti}, {Harrison}, {Fabian},
  {Miller}, {Arevalo}, {Ballantyne}, {Boggs}, {Brenneman}, {Christensen},
  {Craig}, {Elvis}, {Fuerst}, {Gandhi}, {Grefenstette}, {Hailey}, {Kara},
  {Luo}, {Madsen}, {Marinucci}, {Matt}, {Parker}, {Reynolds}, {Rivers}, {Ross},
  {Stern}, \& {Zhang}}]{wrh14}
{Walton}, D.~J., {Risaliti}, G., {Harrison}, F.~A., {et~al.} 2014, \apj, 788,
  76, \dodoi{10.1088/0004-637X/788/1/76}

\bibitem[{{Walton} {et~al.}(2016){Walton}, {Middleton}, {Pinto}, {Fabian},
  {Bachetti}, {Barret}, {Brightman}, {Fuerst}, {Harrison}, {Miller}, \&
  {Stern}}]{wmp16}
{Walton}, D.~J., {Middleton}, M.~J., {Pinto}, C., {et~al.} 2016, \apjl, 826,
  L26, \dodoi{10.3847/2041-8205/826/2/L26}

\bibitem[{{Ward} {et~al.}(1978){Ward}, {Wilson}, {Penston}, {Elvis},
  {Maccacaro}, \& {Tritton}}]{ward78}
{Ward}, M.~J., {Wilson}, A.~S., {Penston}, M.~V., {et~al.} 1978, \apj, 223,
  788, \dodoi{10.1086/156311}

\bibitem[{{Yuan} \& {Narayan}(2014)}]{yn14}
{Yuan}, F., \& {Narayan}, R. 2014, \araa, 52, 529,
  \dodoi{10.1146/annurev-astro-082812-141003}

\bibitem[{{Zappacosta} {et~al.}(2020){Zappacosta}, {Piconcelli}, {Giustini},
  {Vietri}, {Duras}, {Miniutti}, {Bischetti}, {Bongiorno}, {Brusa},
  {Chiaberge}, {Comastri}, {Feruglio}, {Luminari}, {Marconi}, {Ricci},
  {Vignali}, \& {Fiore}}]{zappacosta20}
{Zappacosta}, L., {Piconcelli}, E., {Giustini}, M., {et~al.} 2020, \aap, 635,
  L5, \dodoi{10.1051/0004-6361/201937292}

\bibitem[{{Zubovas} \& {Nardini}(2020)}]{zn20}
{Zubovas}, K., \& {Nardini}, E. 2020, \mnras, 498, 3633,
  \dodoi{10.1093/mnras/staa2652}

\end{thebibliography}
\bibliographystyle{aasjournal}

\end{document}